\documentclass[useAMS,usenatbib]{mn2e}
\usepackage[dvips]{graphicx}
\usepackage{epsfig}
\usepackage{rotating}
\usepackage{natbib}
\usepackage{color}
\usepackage{mathrsfs}
\usepackage[caption=false]{subfig}
    
\usepackage{amssymb}

\newcommand{\Msun}{\, {\rm M_{\odot}}}

\newcommand{\Mpc}{\, {\rm Mpc}}

\title[Formation of a hot halo]{The formation of hot gaseous haloes around galaxies}
\author[C.A.~Correa et al.]
 {Camila A.~Correa$^{1,2}$\thanks{E-mail: correa@strw.leidenuniv.nl}, Joop~Schaye$^1$, J. Stuart B.~Wyithe$^2$, Alan R.~Duffy$^{2,3}$,
 \newauthor Tom Theuns$^4$, Robert A. Crain$^5$ and Richard G. Bower$^4$\\
 $^1$ Leiden Observatory, Leiden University, P.O. Box 9513, 2300 RA Leiden, The Netherlands\\
 $^2$ School of Physics, University of Melbourne, Parkville, Victoria 3010, Australia\\
 $^3$ Centre for Astrophysics and Supercomputing, Swinburne University of Technology, Melbourne, Victoria 3122, Australia\\
 $^4$ Institute for Computational Cosmology, Physics Department, University of Durham, South Road, Durham DH1 3LE, UK\\
 $^5$ Astrophysics Research Institute, Liverpool John Moores University, 146 Brownlow Hill, Liverpool L3 5RF, UK\\
 }
\date{\today}

\pagerange{\pageref{firstpage}--\pageref{lastpage}} \pubyear{2013}

\def\LaTeX{L\kern-.36em\raise.3ex\hbox{a}\kern-.15em
    T\kern-.1667em\lower.7ex\hbox{E}\kern-.125emX}

\begin{document}
\maketitle

\begin{abstract}
We use a suite of hydrodynamical cosmological simulations from the Evolution and Assembly of GaLaxies and their Environments (EAGLE) project to investigate the formation of hot hydrostatic haloes and their dependence on feedback mechanisms. We find that the appearance of a strong bimodality in the probability density function (PDF) of the ratio of the radiative cooling and dynamical times for halo gas provides a clear signature of the formation of a hot corona. Haloes of total mass $10^{11.5}-10^{12}\Msun$ develop a hot corona independent of redshift, at least in the interval $z=0-4$ where the simulation has sufficiently good statistics. We analyse the build up of the hot gas mass in the halo, $M_{\rm{hot}}$, as a function of halo mass and redshift and find that while more energetic galactic winds powered by SNe increases $M_{\rm{hot}}$, AGN feedback reduces it by ejecting gas from the halo. We also study the thermal properties of gas accreting onto haloes and measure the fraction of shock-heated gas as a function of redshift and halo mass. We develop analytic and semianalytic approaches to estimate a `critical halo mass', $M_{\rm{crit}}$, for hot halo formation. We find that the mass for which the heating rate produced by accretion shocks equals the radiative cooling rate, reproduces the mass above which haloes develop a significant hot atmosphere. This yields a mass estimate of $M_{\rm{crit}} \approx 10^{11.7}\Msun$ at $z=0$, which agrees with the simulation results. The value of $M_{\rm{crit}}$ depends more strongly on the cooling rate than on any of the feedback parameters.
\end{abstract} 


\section{Introduction}


One of the major goals of modern galaxy formation theory is to understand the physical mechanisms that halt the star formation process, by either removing, heating or preventing the infall of cold gas onto the galactic disc. X-ray observations suggest that for haloes hosting massive galaxies the majority of baryonic matter resides not in the galaxies, but in the halo in the form of virialized hot gas (e.g. \citealt{Lin03,Crain10a,Anderson11}). This work investigates the formation of the hot gaseous corona (also refereed to as `hot halo' or `hot atmosphere') around galaxies, that may help reduce the rate of infall of gas onto galaxies, and has been suggested to explain the observed galaxy bimodality (\citealt{Dekel}).

The hot gaseous corona is produced as a result of an important heating process, that was initially discussed by \citet{Rees}, \citet{Silk77}, \citet{Binney} and \citet{White}, and later in the context of the cold dark matter paradigm by e.g. \cite{White91}, in an attempt to explain the reduced efficiency of star formation within massive haloes. They proposed that while a dark matter halo relaxes to virial equilibrium, gas falling into it experiences a shock, and determined the cooling time of gas behind the shock. As long as the cooling time is shorter than the dynamical time, the infalling gas cools (inside the current `cooling radius') and settles onto the galaxy. If, on the other hand, the cooling time exceeds the dynamical time, the gas is not able to radiate away the thermal energy that supports it. It therefore adjusts its density and temperature quasi statically, forming a hot hydrostatic halo atmosphere, pressure supported against gravitational collapse. Over the past decade, the works of \citet{Birnboim} and \citet[hereafter DB06]{Dekel} investigated the stability of accretion shocks around galaxies, and concluded that a hot atmosphere forms when the compression time of shocked gas is larger than its cooling time, occurring when haloes reach a mass of about $10^{11.7}\Msun$. 

Numerical simulations have shown, however, that cold gas accreting through filaments does not necessarily experience a shock when crossing the virial radius, even if the spherically averaged cooling radius is smaller than the virial radius. Many groups have concluded that there are two modes of gas accretion, named as hot and cold accretion, that are able to coexist in high-mass haloes at high redshift (e.g. \citealt{Keres05,Dekel,Ocvirk,Dekel09,vandeVoort11,Faucher,vandeVoort12,Nelson13}). The hot mode of accretion refers to the accreted gas that shock-heats to the halo virial temperature. The cold mode refers to gas that flows along dark matter filaments and is accreted onto the central galaxy without being shock heated near the virial radius. It has been found that the cold streams end up being the dominant mode of accretion onto galaxies at high redshift (e.g. \citealt{Dekel09}). However, it has also been found that most of the cold gas from filaments does experience significant heating when accreted by the galaxy at radii much smaller than the virial radius (\citealt{Nelson13}). 


Besides the rate of gas accretion, the hot halo can be influenced by feedback mechanisms and photoionization from local sources. Feedback mechanisms can suppress cooling from the hot halo, modify the distribution of hot gas in the halo (\citealt{vandeVoort12}) and (to a limited degree) reduce the accretion rates onto haloes (\citealt{vandeVoort11,Faucher,Nelson15a}). In this work we investigate the impact of feedback mechanisms on the hot halo in detail and analyze whether reasonable changes to the feedback implementation result in a change to the mass scale of hot halo formation. Increasing photoionizing flux (higher star formation rate or an active nucleus) from local sources can decrease the net cooling rate of gas in the proximity of the galaxy, potentially suppressing cold gas accretion in low-mass halos ($<5\times 10^{11}\Msun$) and decrease the mass-scale for hot halo formation (\citealt{Cantalupo10}). However, the results are sensitive to the assumed escape fraction and \citet{Vogelsberger12} found only small effects when including local AGN. For simplicity we will assume the gas is only exposed to the metagalactic background radiation.

We use the suite of cosmological hydrodynamical simulations from the EAGLE project (\citealt{Schaye14,Crain15}) to investigate the physical properties of the hot gas in the halo, and their dependence on energy sources like stellar feedback and AGN feedback. Our main goal is to study the thermal properties of gas accreting onto haloes and the gas mass that remains hot in the halo ($M_{\rm{hot}}$). In addition, we develop analytic and semianalytic approaches to calculate the heating rates of gas in the halo and the mass scale of hot halo formation, which we apply in a companion work, Correa et al. (in preparation, hereafter Paper II). In Paper II we derive a physically motivated model for gas accretion onto galaxies that accounts for the hot/cold modes of accretion onto haloes, and for the rate of gas cooling from the hot halo. With this model we aim to provide some insight into the physical mechanisms that drive the gas inflow rates onto galaxies. 

The outline of this paper is as follows. We describe the EAGLE simulations series used in this study and the analysis methodology in Section \ref{simulations_sec}. We present our main results concerning the physical properties of hot and cold gas in the halo in Section \ref{Hothaloformation} and on the modes of gas accretion in Section \ref{sec_hotcold_modes}. In Section ~\ref{Toymodel} we develop an analytic approach to calculate a `critical mass scale', $M_{\rm{crit}}$, for hot halo formation, and compare it with our numerical results and previous works. Finally, in Section \ref{conclusions_sec} we summarize our conclusions.

\section{Simulations}\label{simulations_sec}

\begin{table*}
\centering  
\caption{List of simulations used in this work.  From
  left-to-right the columns show: simulation identifier; comoving box size;
  number of dark matter particles (initially there are equally 
  many baryonic particles); initial baryonic particle mass; dark matter
  particle mass; comoving (Plummer-equivalent) gravitational
  softening; maximum physical softening.} 
\label{Table_sims}
\begin{tabular}{lrrrlrrl}
\hline
  Simulation & L & N & $m_{\rm{b}}$ & $m_{\rm{dm}}$ & $\epsilon_{\rm{com}}$ & $\epsilon_{\rm{prop}}$ \\ 
   & (comoving $\rm{Mpc}$) & & ($\rm{M}_{\sun}$) & ($\rm{M}_{\sun}$) & (comoving $\rm{kpc}$) & (proper $\rm{kpc}$) \\  \hline\hline
   L025N0376 & 25 & $376^{3}$ & $1.81\times 10^{6}$ & $9.70\times 10^{6}$ & 2.66 & 0.70\\
   L025N0752 & 25 & $752^{3}$ & $2.26\times 10^{5}$ & $1.21\times 10^{6}$ & 1.33 & 0.35 \\
   L050N0752 & 50 & $752^{3}$ & $1.81\times 10^{6}$ & $9.70\times 10^{6}$ & 2.66 & 0.70\\ 
   L100N1504 & 100 & $1504^{3}$ & $1.81\times 10^{6}$ & $9.70\times 10^{6}$ & 2.66 & 0.70\\ \hline
\end{tabular}
\end{table*}

\begin{table*}
\centering  
\caption{List of feedback parameters that are varied in the simulations.  From
  left-to-right the columns show: simulation identifier (prefix), asymptotic maximum and minimum values of the efficiency of star formation feedback ($f_{\rm{th}}$), density term denominator ($n_{\rm{H,0}}$) and exponents ($n_{\rm{n}}$ and $n_{\rm{Z}}$) from eq.~(\ref{SNeq}), and temperature increment of stochastic AGN heating ($\Delta T_{\rm{AGN}}$).}
\label{feedback_sims}
\begin{tabular}{lcccl}
\hline
  Simulation & $f_{\rm{th,(max,min)}}$ & $n_{\rm{H,0}}$ & $n_{\rm{n}} (=n_{\rm{Z}})$ & $\Delta T_{\rm{AGN}}$ \\  
  & & (cm$^{-3}$) & (cm$^{-3}$) & (K) \\  \hline\hline
   Ref &  $3.0, 0.30$ & 0.67 & $2/\ln(10)$ & $10^{8.5}$ \\
   Less Energetic FB &  $1.5, 0.15$ & 0.67 & $2/\ln(10)$ & $10^{8.5}$ \\
   More Energetic FB & $6.0, 0.60$ & 0.67 & $2/\ln(10)$ & $10^{8.5}$ \\
   No AGN FB & $3.0, 0.30$ & 0.67 & $2/\ln(10)$ & $-$ \\
   More Explosive AGN FB & $3.0, 0.30$ & 0.67 & $2/\ln(10)$ & $10^{9.5}$ \\
   Recal &  $3.0, 0.30$ & 0.25 & $1/\ln(10)$ & $10^{9}$ \\\hline
\end{tabular}
\end{table*}

To investigate the formation and evolution of hot haloes surrounding galaxies, we use cosmological, hydrodynamical simulations from the Evolution and Assembly of GaLaxies and their Environments project (EAGLE; \citealt{Schaye14,Crain15}). The EAGLE simulations were run using a  modified version of GADGET 3 (\citealt{Springelb}), a $N$-Body Tree-PM smoothed particle hydrodynamics (SPH) code. The EAGLE version contains a new formulation of SPH, new time stepping and new subgrid physics. Below we present a summary of the EAGLE models. For a more complete description see \citet{Schaye14}.

The EAGLE simulations assume a $\Lambda$CDM cosmology with the parameters derived from {\it{Planck-1}} data (\citealt{Planck}), $\Omega_{\rm{m}}=1-\Omega_{\Lambda}=0.307$, $\Omega_{\rm{b}}=0.04825$, $h=0.6777$, $\sigma_{8}=0.8288$, $n_{s}=0.9611$. The primordial mass fractions of hydrogen and helium are $X=0.752$ and $Y=0.248$, respectively. 

Table \ref{Table_sims} lists the box sizes and resolutions of the simulations used in this work. We use the notation $L$xxx$N$yyyy, where xxx indicates box size (ranging from 25 to 100 comoving $\Mpc$) and yyyy indicates the cube root of the number of particles per species (ranging from $376^3$ to $1504^3$, with the number of baryonic particles initially equal to the number of dark mater particles). The gravitational softening was kept fixed in comoving units down to $z=2.8$ and in proper units thereafter. We will refer to simulations with the mass and spatial resolution of L025N0376 as intermediate-resolution runs, and to simulations with the resolution of L025N0752 as high-resolution runs.

\subsection{Baryonic physics}\label{feedback_sec}

Radiative cooling and photo-heating are included as in \citet{Wiersma09a}. The element-by-element radiative rates are computed in the presence of the cosmic microwave background (CMB), and the \citet{Haardt} model for UV and X-ray background radiation from quasars and galaxies. 

Star formation is modelled following the recipe of \citet{Schaye08}. Star formation is stochastic above a density threshold, $n_{\rm{H},0}$, that depends on metallicity (in the model of \citealt{Schaye04}, $n_{\rm{H},0}$ is the density of the warm, atomic phase just before it becomes multiphase with a cold, molecular component), with the probability of forming stars depending on the gas pressure. The implementation of stellar evolution and mass loss follows the work of \citet{Wiersma09b}. Star particles are treated as simple stellar populations with a \citet{Chabrier} initial mass function, spanning the range $0.1-100 \Msun$. Feedback from star formation and supernovae events follows the stochastic thermal feedback scheme of \citet{DallaVecchia12}. Rather than heating all neighboring gas particles within the SPH kernel, they are selected stochastically based on the available energy, and then heated by a fixed temperature increment of $\Delta T = 10^{7.5}$K. The probability that a neighboring SPH particle is heated is determined by the fraction of the energy budget that is available for feedback, $f_{\rm{th}}$. If $\Delta T$ is sufficiently high to ensure that radiative losses are initially small, the physical efficiency of feedback can be controlled by adjusting $f_{\rm{th}}$. The value $f_{\rm{th}} = 1$ corresponds to the expected value of energy injected by core collapse supernovae ($E_{\rm{SN}}=1.736\times 10^{49}$ erg $\rm{M}_{\odot}^{-1}$ per solar mass of stars formed). EAGLE takes $f_{\rm{th}}$ to be a function of the local physical conditions,

\begin{equation}\label{SNeq}
f_{\rm{th}}=f_{\rm{th,min}}+\frac{f_{\rm{th,max}}-f_{\rm{th,min}}}{1+\left(\frac{Z}{0.1Z_{\odot}}\right)^{n_{Z}}\left(\frac{n_{\rm{H,birth}}}{n_{\rm{H},0}}\right)^{-n_{n}}},
\end{equation}

\noindent which depends on maximum and minimum threshold values ($f_{\rm{th,max}}$ and $f_{\rm{th,min}}$, respectively), on density ($n_{\rm{H}}$ refers to hydrogen number density and $n_{\rm{H,birth}}$ to the density inherited by the star particle) and metallicity ($Z$) of the gas particle. The reference simulations (hereafter Ref) use $f_{\rm{th,max}} = 3$, $f_{\rm{th,min}} = 0.3$ and $n_{\rm{H},0}=0.67$ $\rm{cm}^{-3}$. These values were chosen to obtain good agreement with the observed present-day galaxy stellar mass function and disc galaxy sizes (as described by \citealt{Crain15}).

Black hole seeds (of mass $\approx 1.4\times 10^{5}\Msun$) are included in the gas particle with the highest density in haloes of mass greater than $\approx 1.4\times 10^{10}\Msun$ that do not contain black holes (\citealt{Springel05}). Black holes can grow through mergers and gas accretion. The accretion events follow a modified Bondi-Hoyle formula that accounts for the angular momentum of the accreting gas (\citealt{Rosas13,Schaye14}), and a free parameter that is related to the disc viscosity. AGN feedback follows the accretion of mass onto the black hole, where a fraction (0.015) of the accreted rest mass energy is released as thermal energy into the surrounding gas, and is implemented stochastically, as per the stellar feedback scheme, with a fixed free parameter heating temperature, $\Delta T_{\rm{AGN}}$, which is set to $10^{8.5}$ K in the reference simulations.

When the resolution is increased, the parameters may need to be (re-)calibrated to retain the agreement with observations. The high-resolution simulation with recalibrated parameters is called Recal. In addition to Ref and Recal, we also use simulations with different feedback implementations to test the impact of feedback on the formation of the hot halo. Table \ref{feedback_sims} lists the values of the feedback parameters adopted in each simulation. In the table, the simulation identifier describes the differences in the feedback with respect to Ref. In the stellar feedback case, `Less/More Energetic FB' means that in these simulations, the energy injected per mass of stars formed is lower/higher with respect to Ref. In the AGN case, `More Explosive AGN FB' means that AGN feedback is more explosive and intermittent, but the energy injected per unit mass accreted by the BH does not change with respect to Ref. Additional information regarding the performance of the EAGLE simulations, including an analysis of subgrid parameter variations, a study of the evolution of galaxy masses, star formation rates and sizes can be found in \citet{Crain15}, \citet{Furlong15a,Furlong15b} and \citet{Schaye14}.%

\subsection{Hydrodynamics}\label{hydrodynamics_discussion}

There has been much debate regarding the systematic differences between SPH, grid codes and moving mesh grid codes when modelling fluid mixing and gas heating and cooling (e.g. \citealt{Agertz07,Vogelsberger12,Nelson13}). It has been shown by \citet{Hutchings} and \citet{Creasey} that SPH simulations may not adequately resolve shocks of accreted gas. Since shocks are generally spread over several SPH kernel lengths, the heating rate is smoothed over time, potentially making it easier for radiative cooling to become important. In addition, if radiative cooling is able to limit the maximum temperature reached by the gas particle, numerical radiative losses may be enhanced.

In contrast, numerical simulations using grids do not smooth out the shocks, and are thus better at identifying shock temperatures spikes. Numerical simulations using moving mesh codes can also capture shocks accurately. However, in common with grid codes, they may suffer from numerical mixing of hot and cold gas as the fluid moves across cells. Recently, \citet{Nelson13} compared the moving mesh code AREPO (\citealt{AREPO}) with the standard SPH version of GADGET, and calculated the rates of cold mode of gas accretion onto haloes and galaxies. They found that while the rates of gas accreted cold onto haloes are in very good agreement between the simulations run with GADGET and AREPO, the rates of gas accreted cold onto galaxies differ significantly, with galaxies in AREPO having a $20\%$ lower cold fraction in $10^{11}\Msun$ haloes. \citet{Nelson13} concluded that most of the cold gas from filaments experiences significant heating after crossing the virial radius, implying that the numerical deficiencies inherent in different simulation codes may modify the relative contributions of hot and cold modes of accretion onto galaxies. 

Some differences in the contributions of hot and cold modes of accretion onto galaxies and haloes may, however, be due to the method employed to select shock-heated gas. Previous works (e.g. \citealt{Keres05,Keres09,Faucher,vandeVoort11,Nelson13}, among others) followed the thermal history of the gas and applied a fixed temperature cut to the distribution of the maximum past temperature ($T_{\rm{max}}$), to separate hot from cold mode accretion. However, $T_{\rm{max}}$ is not suitable for identifying cold flows if the gas experiences a shock but cools immediately afterwards, as may happen for accretion onto galaxies. In this case, a filament that is mostly cold except at a point near the galaxy would be labeled as hot mode accretion by numerical studies using $T_{\rm{max}}$, but observers would identify it as a cold flow. This practical problem may not be important for SPH simulations that suffer from `in shock cooling' because they do not resolve the accretion shocks onto the galaxy, or, as in the case of \citet{vandeVoort11} and EAGLE, that impose a temperature-density relation onto high-density gas, but it may affect the conclusions inferred from moving mesh codes using the $T_{\rm{max}}$ statistic. To avoid this issue, we use an alternative method to identify shock-heated particles in Section~\ref{sec_hotcold_modes}, based on post-shock temperature values.

Hydrodynamics solvers may also produce differences in the hot/cold modes of accretion. The EAGLE version of GADGET uses the hydrodynamics solver ``Anarchy'', which greatly improves the performance on standard hydrodynamical tests, when compared to the original SPH implementation in GADGET (\citealt{Schaller15b}, see \citealt{Hu14} for similar results). Anarchy makes use of the pressure-entropy formulation derived in \citet{Hopkins13}, alleviating spurious jumps at contact discontinuities. It also uses an artificial viscosity switch advocated by \citet{Cullen10}, that allows the viscosity limiter to be stronger when shocks and shear flows are present. In addition, Anarchy includes an artificial conduction switch (similar to that of \citealt{Price08}), the $C^{2}$ \citet{Wendland95} kernel and the time step limiters of \citet{Durier12}. These changes ensure that ambient particles do not remain inactive when a shock is approaching. 

Recently, \citet{Sembolini16a} compared cosmological simulations of clusters using SPH as well as mesh-based codes. They found that the modern SPH schemes (such as Anarchy) that allow entropy mixing produce gas entropy profiles that are indistinguishable from those obtained with grid-based schemes. In addition, \citet{Schaller15b} compared the EAGLE simulations with simulations run with the same subgrid physics, but using the standard GADGET rather than the Anarchy hydrodynamics solver. They found that while simulations with standard SPH contain haloes with a large number of dense clumps of gas at all radii, Anarchy's ability to mix phases allows dense clumps to dissolve into the hot halo. These substantial improvements of the SPH formulation in the EAGLE simulations motivate a detailed description of the resulting predictions for hot halo formation and of hot/cold mode accretion.

\subsection{Identifying haloes and galaxies}

Throughout this work we select the largest subhalo in each Friends-of-Friends (FoF) group, and use the SUBFIND algorithm (\citealt{Springel,Dolag}) to identify the substructures (subhaloes) within it. The FoF algorithm adopts a dimensionless linking length of 0.2, and the SUBFIND algorithm calculates halo virial masses and radii via a spherical overdensity routine that centers the main subhalo from the FoF group on the minimum of the gravitational potential. We define halo masses, $M_{200}$, as the mass of all matter within the radius, $R_{200}$, for which the mean internal density is 200 times the critical density of the Universe.

To select the gas associated with the central galaxies embedded in each resolved halo, we identify the gravitationally bound cold and dense gas within $R_{200}$ that is star-forming and/or has a hydrogen number density, $n_{\rm{H}}>0.01\rm{cm}^{-3}$, and temperature $T<10^{5}$K. We also require all particles to be contained within a sphere of radius $0.15\times R_{200}$, in order to avoid labelling infalling cold flows (that would be included by the $T-n_{\rm{H}}$ cuts but are mostly at large radii) as part of the galaxy.

\begin{figure*}
  \centering
  \includegraphics[angle=0,width=\textwidth]{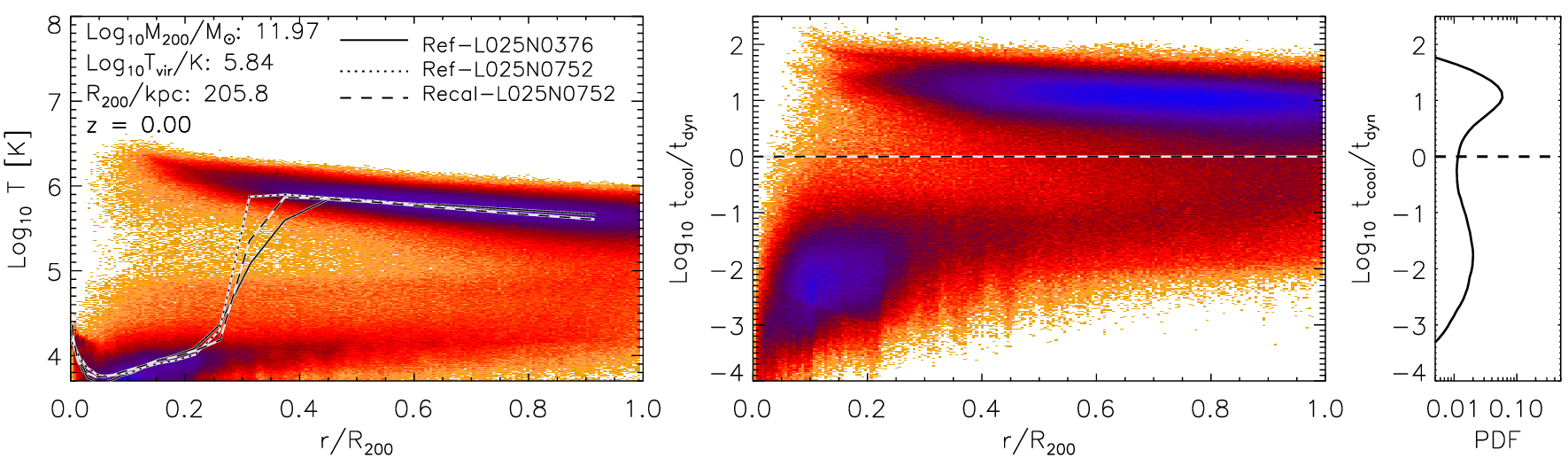}\\
  \includegraphics[angle=0,width=\textwidth]{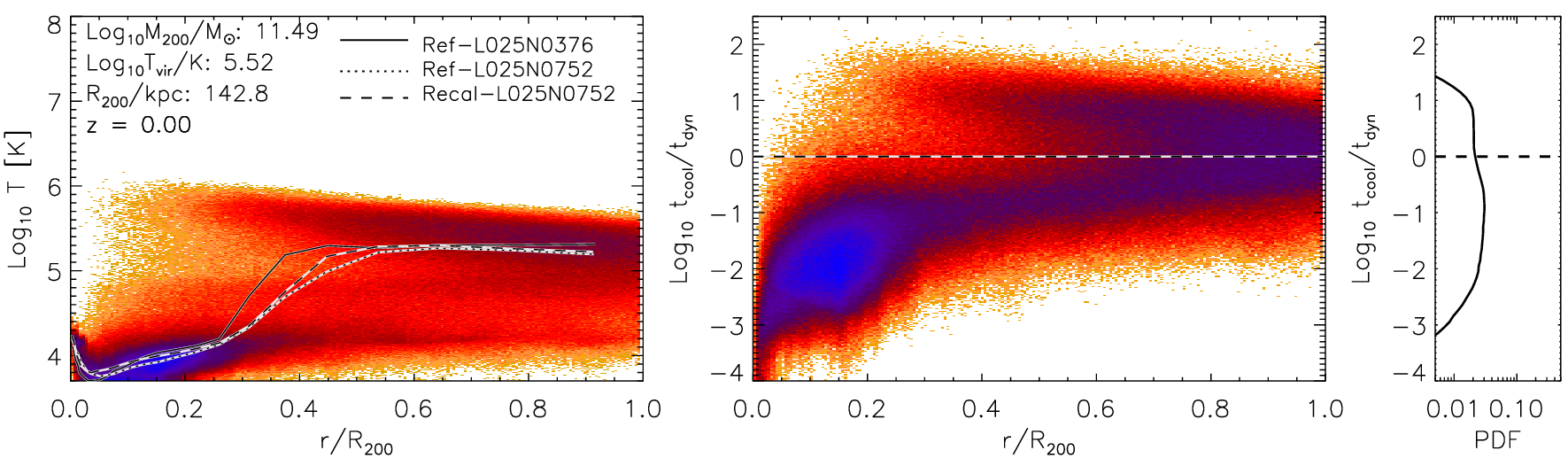}\\
  \includegraphics[angle=0,width=\textwidth]{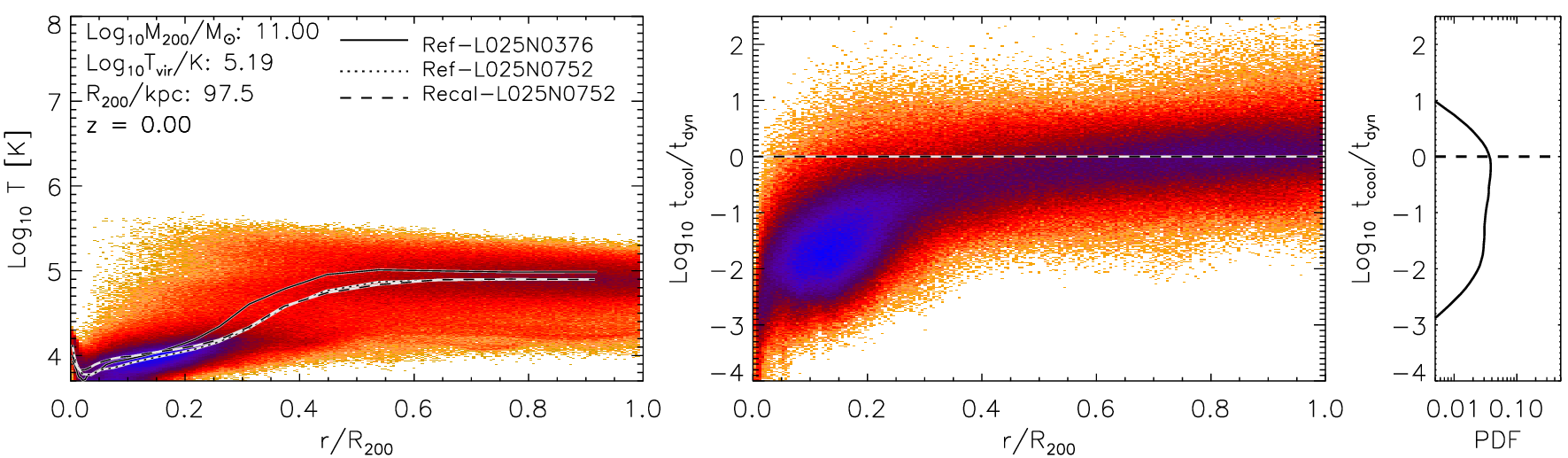}
\caption{Temperature profile (left column), logarithm of the ratio between cooling times and local dynamical times (middle column) and the mass-weighted probability density function (PDF, right column) of $\log_{10}t_{\rm{cool}}/t_{\rm{dyn}}$ of gas from haloes in the mass range $10^{11.9}-10^{12.1}\Msun$ (top row), $10^{11.4}-10^{11.6}\Msun$ (middle row) and $10^{10.9}-10^{11.1}\Msun$ (bottom row) at $z=0$ taken from the Ref-L025N0752 simulation. The number of particles in a pixel is used for colour coding. The solid, dotted and dashed lines in the left panels correspond to the median temperature per radial bin for different simulations.}
\label{Profiles_plot1}
\end{figure*}

\begin{figure*}
  \centering
  \includegraphics[angle=0,width=\textwidth]{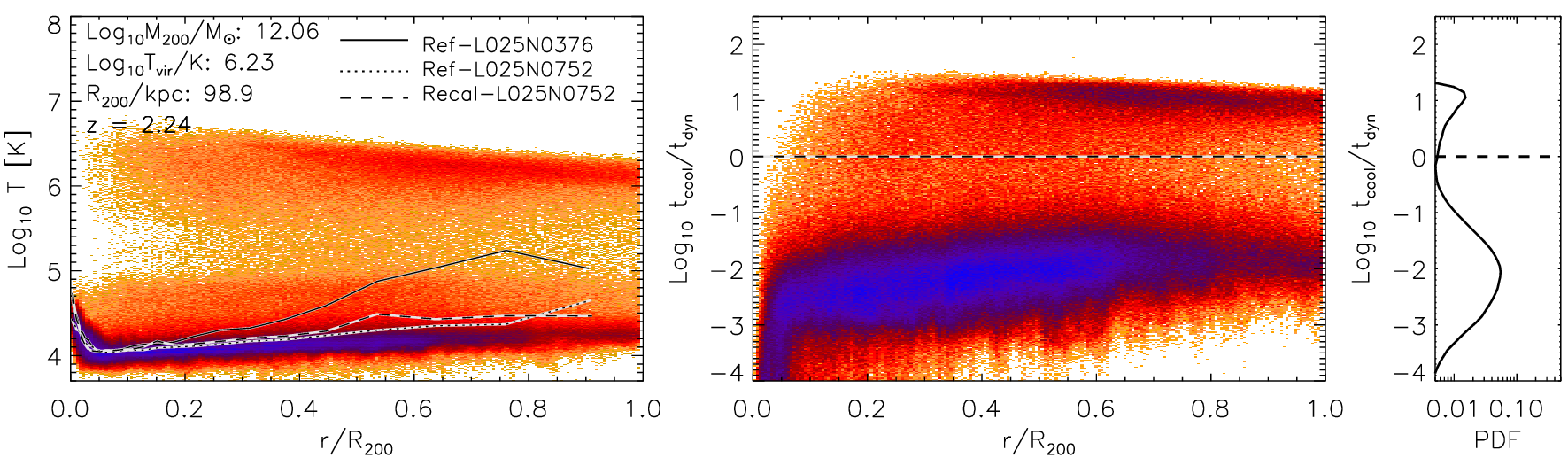}\\
  \includegraphics[angle=0,width=\textwidth]{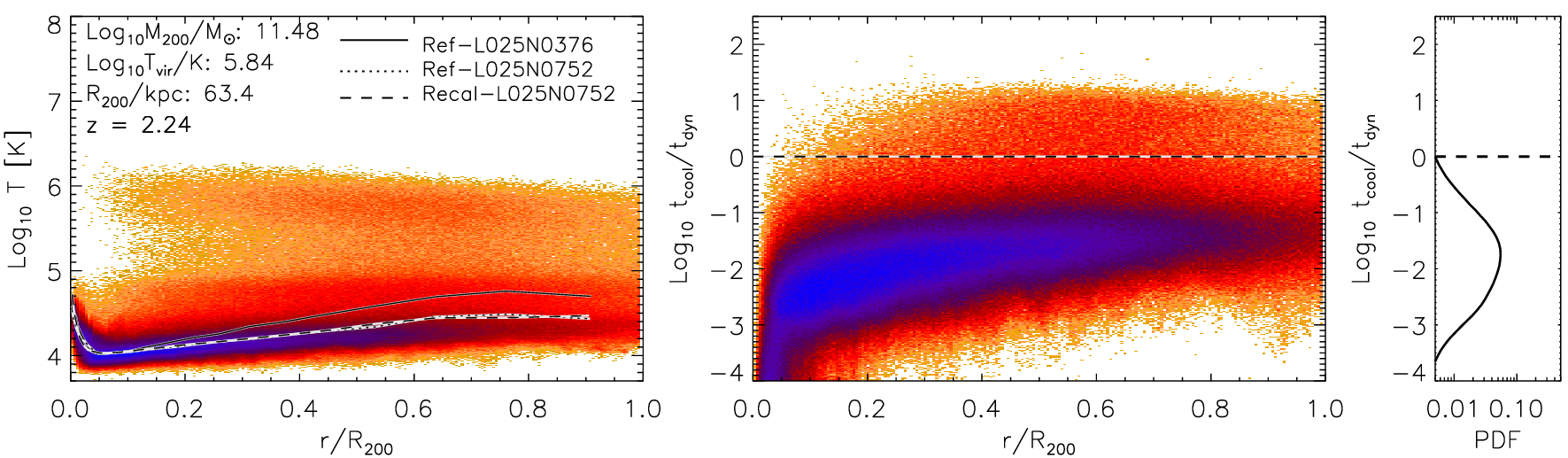}\\
\caption{Same as Fig. 1 but for haloes at $z=2.24$.}
\label{Profiles_plot2}
\end{figure*}

\subsection{Measuring gas accretion}\label{method_accretion}

Once we have identified the haloes, we build merger trees across the simulation snapshots\footnote{The simulation data is saved in 10 discrete output redshifts between redshift 0 to 1, in 4 output redshifts between redshift 1 and 3, and in 8 output redshifts between redshift 3 and 8.}. The standard procedure to build a halo merger tree is to link each progenitor halo with a unique ‘descendant’ halo in the subsequent output (see e.g. \citealt{Fakhouri}). To do so, we identify the main branches of the halo merger trees and compute the halo (and central galaxy) accretion rate between two consecutive snapshots. At each output redshift (snapshot) we select the most massive haloes within each FoF group and consider them to be `resolved' if they contain more than 1000 dark matter particles, which corresponds to a minimum halo mass of $M_{200}=10^{9.8}\Msun$ ($10^{8.8}\Msun$) in the intermediate- (high-) resolution simulations. This limit on the number of dark matter particles results from a convergence analysis that we present in Paper II, where we find that in smaller haloes the accretion onto galaxies does not converge, indicating that the inner galaxies are not well resolved. We refer to the resolved haloes as `descendants', and then link each descendant with a unique `progenitor' at the previous output redshift. This is nontrivial due to halo fragmentation, in which subhaloes of a progenitor halo may have descendants that reside in more than one halo. Such fragmentation can be either spurious or due to a physical unbinding event. To account for this, we link the descendant to the progenitor that contains the majority of the descendant's 25 most bound dark matter particles (see \citealt{Correa15b} for an analysis of halo mass history convergence using these criteria to connect haloes between snapshots). 

We distinguish between gas accreted onto a halo and gas accreted onto a galaxy. For each descendant halo at $z_{i}$ and its linked progenitor at $z_{j}$ ($z_{j}>z_{i}$), we identify the particles that are in the descendant but not in its progenitor by performing particle ID matching. We then select particles that are new in the halo and reside within the virial radius, as particles accreted onto the halo in the redshift range $z_{i}\le z<z_{j}$. The accretion rate onto galaxies is further explored in Paper II, where we follow the methodology described above for calculating accretion rates onto haloes, and we select the new particles within the radius $0.15\times R_{200}$ as particles accreted onto the galaxies in the redshift range $z_{i}\le z<z_{j}$ (see Paper II, Section 2.1, for a discussion on methods for calculating gas accretion onto galaxies).

\section{Hot halo formation}\label{Hothaloformation}

The simple models of galaxy formation (e.g. \citealt{Rees} and \citealt{White}) assume that as long as the cooling time, $t_{\rm{cool}}$, is shorter than the dynamical time, $t_{\rm{dyn}}$, the infalling gas cools (inside a `cooling radius', \citealt{White91}) and settles into the galaxy. Otherwise, the gas is unable to efficiently radiate its thermal energy and forms a hot hydrostatic atmosphere, which is pressure supported against gravitational collapse. More recent semi-analytic models of galaxy formation assume that the cooling radius expands outwards as a function of time, therefore the comparison is done between gas cooling time and a different time scale representing the time available for cooling, like the time since the last major event or the time of halo formation (see e.g \citealt{Lacey16}). In this section we investigate when the hot hydrostatic halo forms in the EAGLE simulations, by analyzing the interplay between the cooling and dynamical times of the gas particles in the halo. Throughout this work we define hot halo gas as all gas particles that have $t_{\rm{cool}}>t_{\rm{dyn}}$, and that do not form part of the galaxy, i.e. $r>0.15R_{200}$. 

We calculate  $t_{\rm{dyn}}$ of the gas particle as

\begin{equation}\label{tdyn}
t_{\rm{dyn}}=r/V_{\rm{c}}(r),
\end{equation}

\noindent where $V_{\rm{c}}(r)=[GM(<r)/r]^{1/2}$ is the circular velocity and $M(<r)$ is the mass enclosed within $r$. We calculate $t_{\rm{cool}}$ as 

\begin{equation}\label{tcool}
t_{\rm{cool}} = \frac{3nk_{\rm{B}}T}{2\Lambda},
\end{equation}

\noindent where $n$ is the number density of the gas particle ($n=\rho_{\rm{gas}}/\mu m_{\rm{p}}$, $\mu$ is the molecular mass weight calculated from the cooling tables of \citet{Wiersma09b}, and $m_{\rm{p}}$ is the proton mass), $k_{\rm{B}}$ is the Boltzmann constant, $T$ is the gas temperature and $\Lambda$ is the cooling rate per unit volume with units of erg cm$^{-3}$s$^{-1}$. To calculate $\Lambda$, we use the tabulated cooling function for gas exposed to the evolving UV/X-ray background from \citet{Haardt} given by \citet{Wiersma09b}, which was also used by the EAGLE simulations. Note that the ``standard" definition for dynamical time of gas within a virialized system depends on $R_{200}$ and $V_{\rm{c}}(R_{200})$, and not on the local radius and local circular velocity as defined here. We use local values rather than to investigate if shorter dynamical times in the inner dense regions give rise to a cooling flow. However, we find that changing eq. (2) to $t_{\rm{dyn}} = R_{200}/V_{\rm{c}}(R_{\rm{vir}})$ does not change our conclusions.

Fig. \ref{Profiles_plot1} shows temperature profiles (left column), the logarithm of the ratio between cooling times and dynamical times (middle column) and the respective mass-weighted probability density function (PDF) of $\log_{10}t_{\rm{cool}}/t_{\rm{dyn}}$ (right column) for gas from haloes in the mass range $10^{11.9}-10^{12.1}\Msun$ (top row), $10^{11.4}-10^{11.6}\Msun$ (middle row) and $10^{10.9}-10^{11.1}\Msun$ (bottom row) at $z=0$, taken from the Ref-L025N0752 simulation. In the right panels and throughout this work, the PDFs are calculated by stacking haloes in the selected mass range and distributing the gas particles in logarithmic bins of size 0.1 dex. We then sum the mass of the gas particles in each bin and normalize the distribution by the total gas mass. In the left panels, the legend lists the median values of the mass, virial temperature (defined as $T_{\rm{vir}}=\frac{\mu m_{\rm{p}}}{2k_{\rm{B}}}V^{2}_{200}$, with $V^{2}_{200}=GM_{200}/R_{200}$) and virial radius of haloes selected in each mass bin. The left panels also show in solid, dotted and dashed lines the median temperature per radial bin of gas from haloes taken from the simulations Ref-L025N0376, Ref-L025N0752 and Recal-L025N0752, respectively. 

The left panels of Fig.~\ref{Profiles_plot1} show that while there is very good agreement in the median temperature profiles at small ($r/R_{200}< 0.2$) and large ($r/R_{200}> 0.4$) radii, at intermediate radii the median temperatures from the intermediate-resolution run (Ref-L025N0376) are larger by up to 0.3 dex than those from the high-resolution runs (L025N0752). This is in agreement with the convergence analysis of \citet{Nelson15b}, who concluded that the physics (different models of stellar winds or AGN feedback) has a greater impact on $T(r)$ than resolution. We also find, that in the radial range $r=[0.2-0.4]R_{200}$, where the convergence with resolution is poorest, the median temperatures drop from $T_{\rm{gas}}\sim T_{\rm{vir}}$ to $T_{\rm{gas}}\sim 10^{4}$ K (in agreement with \citealt{Nelson15b}, $r_{\rm{drop}}\approx 0.25R_{200}$ and \citealt{vandeVoort12}, $r_{\rm{drop}}\approx 0.2-0.4R_{200}$), because of the high densities, that rapidly decrease the gas cooling times, enabling it to radiate away its thermal energy and join the ISM.

The top and middle left panels of Fig. \ref{Profiles_plot1} show that there is relatively little gas with $T\sim 10^{5}$ K at small and intermediate radii, reflecting the short cooling times at these temperatures. The cooling flow in the hot halo is formed by $T=10^{6}$ K gas that slowly decreases in temperature as it loses hydrostatic support due to cooling. The ISM consists of $T=10^{4}$ K gas at $r/R_{200}<0.15$. For a better understanding of the hot halo forming as a function of halo mass and its effect on the infall rate of gas onto the galaxy, we next analyze the middle and right columns. The bottom middle panel shows that in haloes with masses between $10^{10.9}-10^{11.1}\Msun$, most of the gas has low temperature ($T_{\rm{gas}}<T_{\rm{vir}}$), short cooling times ($t_{\rm{cool}}<t_{\rm{dyn}}$) and infalls towards the central galaxy, although a substantial fraction of gas has $t_{\rm{cool}}>t_{\rm{dyn}}$ at $r\sim R_{200}$. At larger halo masses, a larger fraction of the gas is unable to cool and therefore forms a hot halo. The middle panel shows that haloes between $10^{11.4}-10^{11.6}\Msun$ are in the intermediate stage between developing a hot stable atmosphere (gas with $T_{\rm{gas}}\sim T_{\rm{vir}}$ and $t_{\rm{cool}}>t_{\rm{dyn}}$) and continuing to fuel the galaxy. The top middle panel clearly shows a stable hot halo and a reduced amount of gas infalling towards the galaxies (gas at $r>0.3R_{200}$ and $t_{\rm{cool}}<t_{\rm{dyn}}$).

Fig. \ref{Profiles_plot2} is similar to Fig. \ref{Profiles_plot1}, but shows haloes in the mass range $10^{11.9}-10^{12.1}\Msun$ (top panels), $10^{11.4}-10^{11.6}\Msun$ (bottom panels) at $z=2.24$. The top middle panel shows that $10^{12}\Msun$ haloes develop a hot atmosphere, despite the significant fraction of cold gas at large radii that is accreted onto the halo. This cold gas forms part of the cold filamentary flows, that cross the virial radius, and are directly accreted onto the central galaxy. The cold accretion mode is best seen as the gas at $T=10^{4}$ K and $r/R_{200}>0.4$ in the bottom panel of Fig. \ref{Profiles_plot2}, which remains cool (with $T=10^{4}$ K), as it is accreted onto the galaxy.

The presence of cold flows produces gaseous haloes with an isothermal temperature profile of $T_{\rm{gas}}\sim 10^{4}$ K at all radii. Besides analyzing the cooling time profiles, \citet{Nelson15b} and \citet{vandeVoort12} analyzed the entropy profiles of haloes at $z=2$, and concluded the that while the entropy of the cold-mode gas decreases smoothly and strongly towards the centre, the entropy of the hot-mode gas decreases slightly down to $0.2R_{200}$, after which it drops steeply. We find that the cooling time profiles of the hot ($t_{\rm{cool}}>t_{\rm{dyn}}$) and cold ($t_{\rm{cool}}<t_{\rm{dyn}}$) modes follow the entropy profiles.

Figs. \ref{Profiles_plot1} and \ref{Profiles_plot2} show that as the halo mass increases, so does the hot gas mass. In the case of $10^{11}\Msun$ haloes, the bottom left panel of Fig. \ref{Profiles_plot1} shows that there is a large fraction of gas at $r>0.4R_{200}$ with temperatures equal to or larger than the halo virial temperature. This seems to indicate that gas is shock-heating to the halo virial temperature when crossing $R_{200}$ and forming a hot atmosphere. However, gas with $t_{\rm{cool}}\sim t_{\rm{dyn}}$ in the outer parts of $\sim 10^{11}\Msun$ haloes does not imply that the halo formed a stable hot atmosphere via gravitational accretion shocks, since the gas is affected by the extragalactic UV/X-ray background as we show in the next section.

The figures also show that as haloes are growing a hot atmosphere, the $t_{\rm{cool}}/t_{\rm{dyn}}$ PDF begins to present a strong bimodal shape, with a local maximum at $t_{\rm{cool}}>t_{\rm{dyn}}$ followed by a local minimum at $t_{\rm{cool}}<t_{\rm{dyn}}$ (see top right panel). We then conclude that the bimodality in the cooling time PDF provides a clear signature of the formation of a hot halo, and the right panels indicate that the hot hydrostatic halo forms in the halo mass range $10^{11.5}\Msun$ to $10^{12}\Msun$ with a weak dependence on redshift. 

We also analyse the radial velocity distributions of gas and find that in haloes of mass $10^{12}\Msun$ at $z=0$, $92.2\%$ ($61.3\%$) of hot gas has an absolute radial velocity lower than 100 km/s (50 km/s), indicating that most of it is in hydrostatic equilibrium and not accreting onto the galaxy.

\subsection{The impact of photoheating in low-mass haloes}\label{sec_photo_heating}

\begin{figure} 
  \centering
  \subfloat{\includegraphics[angle=0,width=0.42\textwidth]{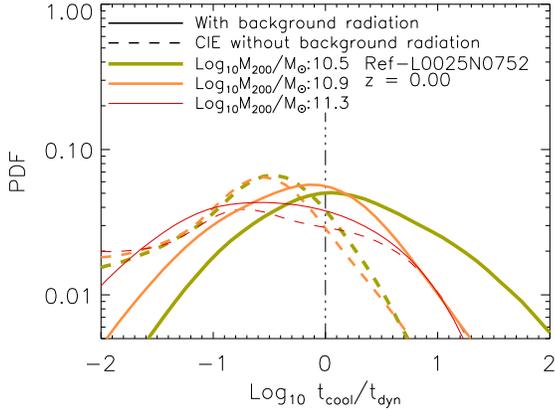}}
  \vspace{-0.15cm}
\caption{Mass-weighted probability density function of the logarithm of the ratio between the net radiative cooling time and the local dynamical time for gas from haloes in the mass range $10^{10.4}-10^{10.6}\Msun$ (green lines),  $10^{10.8}-10^{11}\Msun$ (orange lines),  $10^{11.2}-10^{11.4}\Msun$ (red lines) at $z=0$. The solid lines correspond to the case where the cooling rates are calculated for gas exposed to the evolving UV/X-ray background from \citet{Haardt}, while the dashed lines correspond to the case where the cooling rates are calculated for gas in collisional ionization equilibrium (CIE).}
\label{photo_heating}
\end{figure}

In the previous section we analyzed the dependence of the gas cooling rates on halo mass and concluded that a halo with a hot hydrostatic atmosphere should present a strongly bimodal $t_{\rm{cool}}/t_{\rm{dyn}}$ PDF with a local maximum at $t_{\rm{cool}}>t_{\rm{dyn}}$. As the virial temperature decreases from $10^{5.5}$ K to $10^{5.2}$ K, we would naively expect the peak in the $t_{\rm{cool}}/t_{\rm{dyn}}$ PDF to shift towards shorter cooling times. This is however not the case: we find that at $z=0$ the gas in haloes less massive than $10^{11}\Msun$ ($T_{\rm{vir}}\approx 10^{5.2}$ K) is affected by the extragalactic UV/X-ray background radiation, which strongly suppresses the net cooling rate of gas in the temperature range $T\sim 10^{4}-10^{5}$ K (\citealt{Efstathiou92,Wiersma09b}). As a result, the peak in the $t_{\rm{cool}}/t_{\rm{dyn}}$ PDF remains at $t_{\rm{cool}}\sim t_{\rm{dyn}}$. This can be seen in Fig.~\ref{photo_heating}, where we show the $t_{\rm{cool}}/t_{\rm{dyn}}$ PDF of gas from haloes in the mass range $10^{10.4}-10^{10.6}\Msun$ (olive lines), $10^{10.8}-10^{11.0}\Msun$ (orange lines) and $10^{11.2}-10^{11.4}\Msun$ (red lines). The solid lines correspond to the case where the cooling rates are calculated for gas exposed to the evolving UV/X-ray background from \citet{Haardt}, while the dashed lines correspond to the case where the cooling rates are calculated for gas in collisional ionization equilibrium (CIE) and not exposed to the background radiation. Note, however, that we apply these CIE cooling rates to simulations that were run using cooling rates that did account for photoheating, limiting the gas temperature to $\sim 10^{4}$ K.

Fig.~\ref{photo_heating} shows that there is no large difference in the gas $t_{\rm{cool}}/t_{\rm{dyn}}$ PDF for haloes more massive than $10^{11.3}\Msun$, indicating that there is no strong impact of the background radiation on the cooling rates of gas from haloes with virial temperatures larger than $10^{5}$ K. In smaller haloes, the peak in the $t_{\rm{cool}}/t_{\rm{dyn}}$ PDF curve is shifted to $t_{\rm{cool}}\sim 0.3t_{\rm{dyn}}$ in the case of no background radiation.

\begin{figure*} 
  \centering
  \vspace{-0.3cm}
  \subfloat{\includegraphics[angle=0,width=0.42\textwidth]{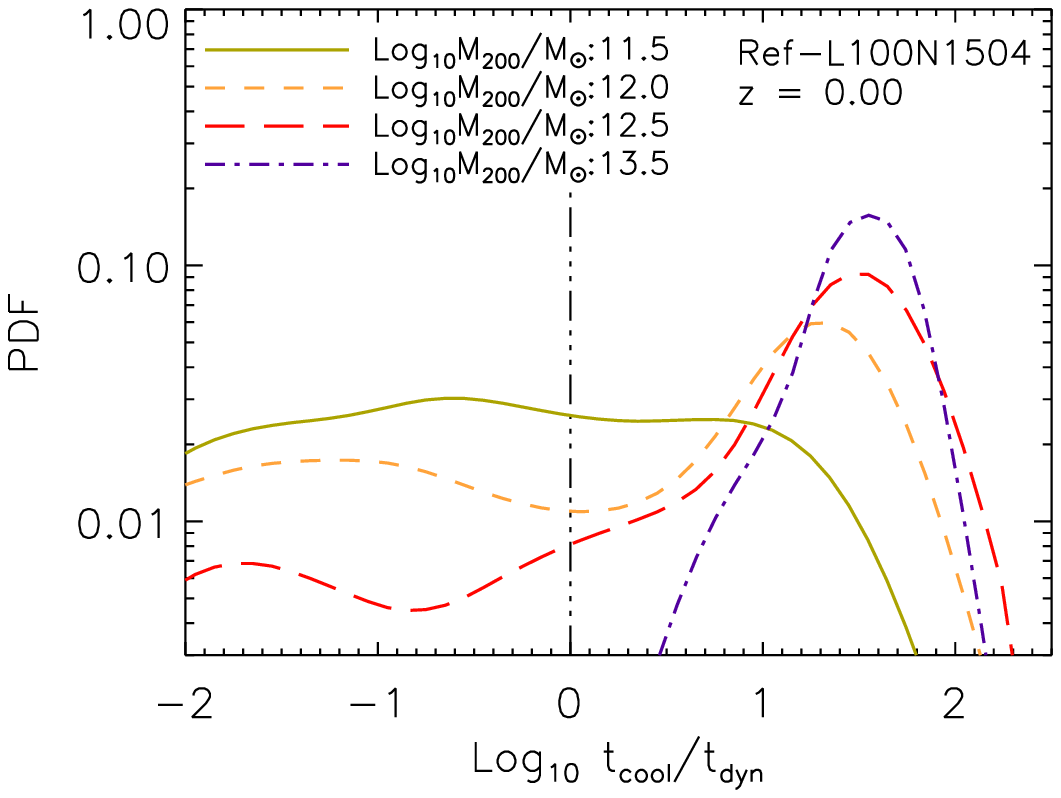}}
  \hspace{0.2cm}
  \subfloat{\includegraphics[angle=0,width=0.42\textwidth]{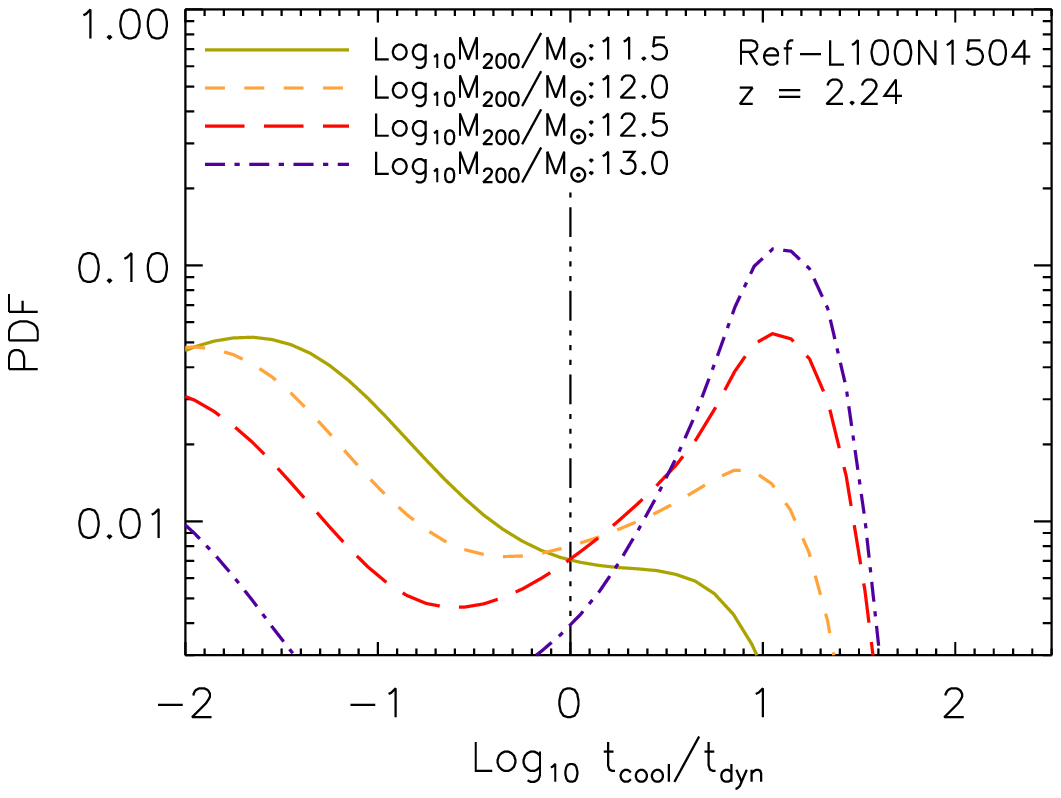}}\\
  \vspace{-0.15cm}
\caption{Mass-weighted probability density function of the logarithm of the ratio between cooling times and local dynamical times of gas from haloes in the mass range $10^{11.4}-10^{11.6}\Msun$,  $10^{11.9}-10^{12.1}\Msun$,  $10^{12.4}-10^{12.6}\Msun$ and $10^{13.4}-10^{13.6}\Msun$ at $z=0$ (left panel) and $z=2.24$ (right panel).}
\label{Distribution_plot1}
\end{figure*}

\begin{figure*} 
  \centering
  \vspace{-0.3cm}
  \subfloat{\includegraphics[angle=0,width=0.42\textwidth]{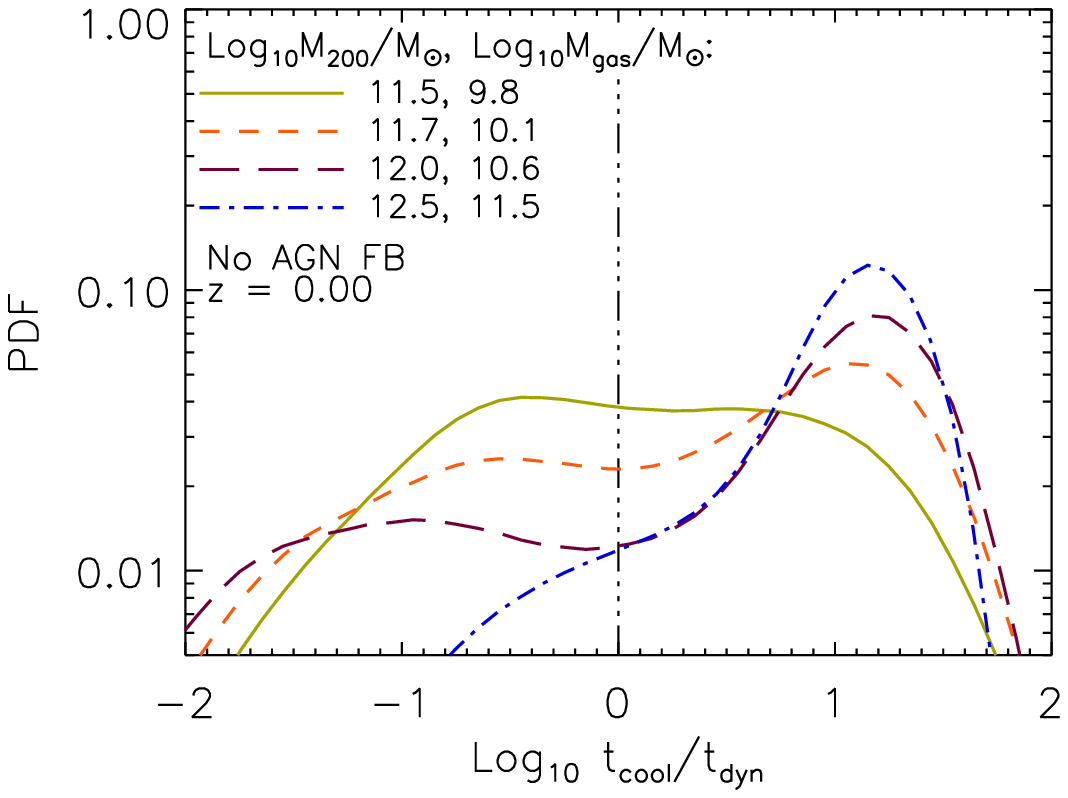}}
  \hspace{0.2cm}
  \subfloat{\includegraphics[angle=0,width=0.42\textwidth]{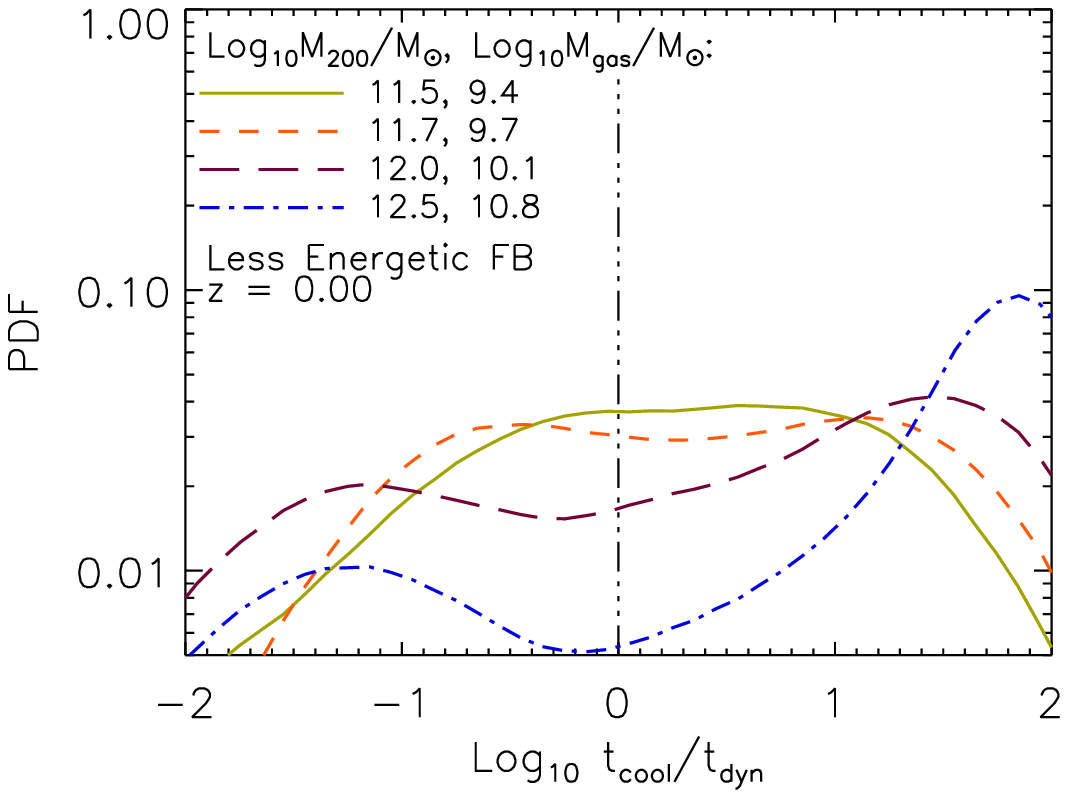}}\\
  \vspace{-0.8cm}
  \subfloat{\includegraphics[angle=0,width=0.42\textwidth]
{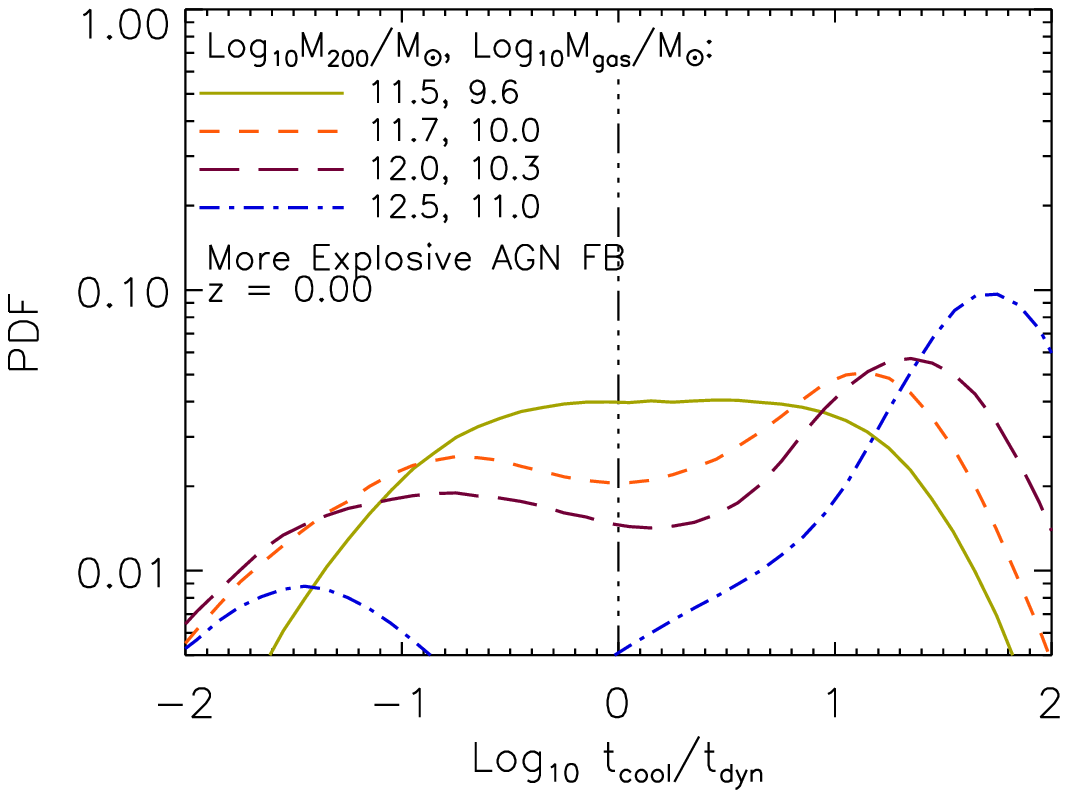}}
  \hspace{0.2cm}
  \subfloat{\includegraphics[angle=0,width=0.42\textwidth]{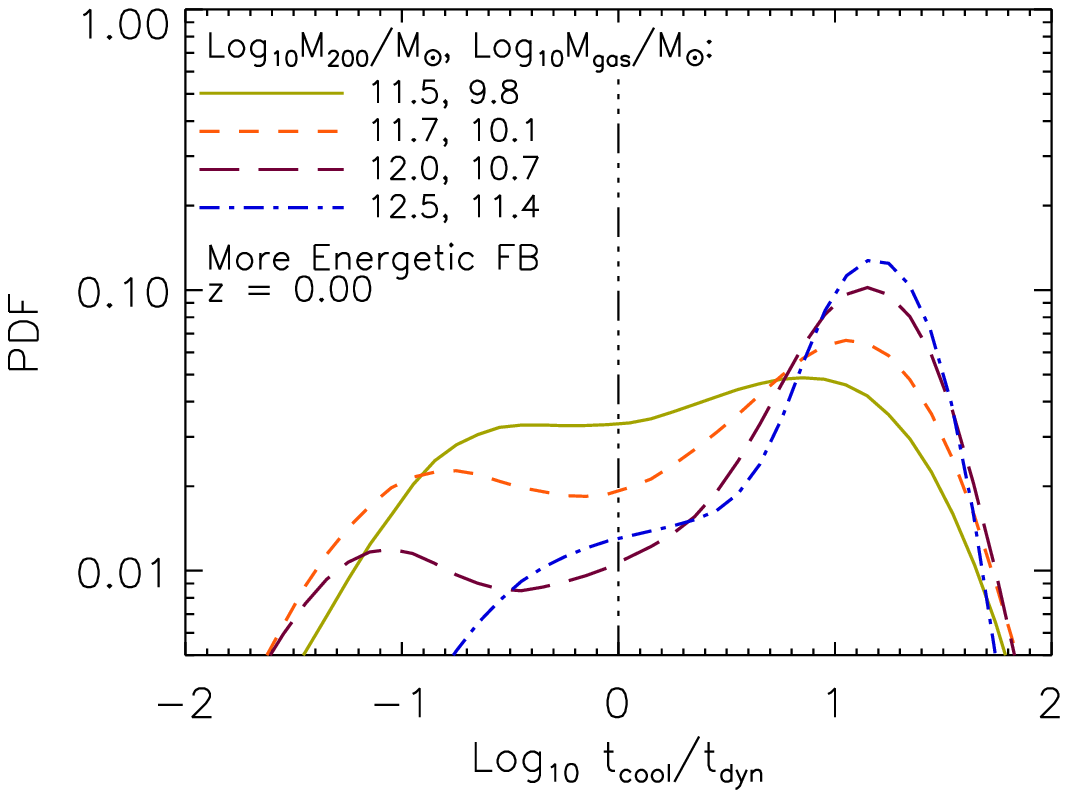}}\\
  \vspace{-0.15cm}
\caption{Mass-weighted probability density function of the logarithm of the ratio between cooling times and local dynamical times of gas from haloes in the mass range $10^{11.4}-10^{11.6}\Msun$, $10^{11.6}-10^{11.8}\Msun$, $10^{11.9}-10^{12.1}\Msun$ and $10^{12.4}-10^{12.6}\Msun$ at $z=0$. The legends show the median mass of the haloes in each mass bin. The different panels show L025N0376 simulations with different feedback prescriptions: no AGN (top left), weak stellar feedback (top right), strong AGN (bottom left) and strong stellar feedback (bottom right).}
\label{Distribution_plot2}
\end{figure*}

\subsection{The impact of feedback}\label{feedback}

Feedback can affect the formation of the hot hydrostatic halo around galaxies. For example, very energetic SN activity generates large outflows and strong winds, that shock against the gaseous halo. As a result, the winds can fill the halo with gas expelled from the galaxy, increasing the amount of hot gas at large radii. In this subsection we compare the $t_{\rm{cool}}/t_{\rm{dyn}}$ mass-weighted PDF at fixed halo mass obtained from simulations with different feedback implementations (see Table \ref{feedback_sims} for reference and section \ref{feedback_sec} for a brief description), and determine, by analysing whether the cooling time PDF is bimodal, the mass range where the hot halo is forming.

Fig.~\ref{Distribution_plot1} shows the mass-weighted PDF of $t_{\rm{cool}}/t_{\rm{dyn}}$ of gas from haloes in the Ref-L100N1504 simulation in the mass range $10^{11.4}-10^{11.6}\Msun$,  $10^{11.9}-10^{12.1}\Msun$,  $10^{12.4}-10^{12.6}\Msun$ and $10^{13.4}-10^{13.6}\Msun$ at $z=0$ (left panel) and $z=2.24$ (right panel). In the figures, the labels show the median halo mass for each mass bin. The region where $t_{\rm{cool}}>t_{\rm{dyn}}$ corresponds to hot gas in the halo.

As the halo mass increases so does the amount of hot gas (Section 3.3). The $t_{\rm{cool}}/t_{\rm{dyn}}$ PDF of gas in $10^{12}\Msun$ haloes at $z=0$ shows a bimodal shape, that becomes mostly unimodal in higher-mass haloes. At $z=2.24$, the bimodality persists up to the highest mass bin ($10^{13}\Msun$) due to the contribution from cold flows that populate the peak at $t_{\rm{cool}}<t_{\rm{dyn}}$. We find that the presence of the bimodality in the $t_{\rm{cool}}/t_{\rm{dyn}}$ PDF indicates the increasing amount of hot gas at large radii and the eventual formation of the hot halo. Then, from visual inspection, we determine that the hot hydrostatic atmosphere is forming in haloes with masses between $10^{11.5}$ and $10^{12}\Msun$ at $z=0$ and $z=2.24$. 

The panels in Fig.~\ref{Distribution_plot2} repeat the analysis shown in the left panel of Fig.~\ref{Distribution_plot1}, but instead show $t_{\rm{cool}}/t_{\rm{dyn}}$ mass-weighted PDFs for the L025N0376 simulations with different feedback prescriptions. In this case, the PDFs correspond to gas from haloes in the mass range $10^{11.4}-10^{11.6}\Msun$, $10^{11.6}-10^{11.8}\Msun$, $10^{11.9}-10^{12.1}\Msun$ and $10^{12.4}-10^{12.6}\Msun$. In the panels, the top left legends indicate the total gas mass in the halo ($M_{\rm{gas}}$).

The simulation shown in the top left panel of Fig.~\ref{Distribution_plot2} does not have AGN feedback while the one in the bottom left panel uses a more explosive AGN feedback. Both include the same feedback from star formation as in Ref. It can be seen that neither the bimodality of the $t_{\rm{cool}}/t_{\rm{dyn}}$ PDF nor the amount of hot gas are strongly affected by AGN feedback in haloes with masses between $10^{11.5}-10^{11.9}\Msun$. The right panels in Fig.~\ref{Distribution_plot2} show the $t_{\rm{cool}}/t_{\rm{dyn}}$ PDFs in the less (top panel) and more energetic stellar feedback (bottom) scenarios (both including the same AGN feedback as in the Ref model). For these halo masses, stellar feedback has a strong impact on the $t_{\rm{cool}}/t_{\rm{dyn}}$ PDFs. While a more energetic stellar feedback increases the fraction of hot gas, at least in the halo mass range probed by these simulations ($<10^{12}\Msun$) and thus limits the build-up of cold-mode gas in the halo centre (in agreement with \citealt{vandeVoort12}), a less energetic stellar feedback maintains the bimodality in the $t_{\rm{cool}}/t_{\rm{dyn}}$ PDF but shifts the peak in the $t_{\rm{cool}}/t_{\rm{dyn}}$ PDF of hot gas towards larger cooling times. We find that the bimodality of the $t_{\rm{cool}}/t_{\rm{dyn}}$ PDF is present in $10^{11.7}\Msun$ haloes with more energetic stellar feedback and in $10^{12}\Msun$ haloes with less energetic stellar feedback. 

In the following section we further analyse the dependence of the total hot gas mass on halo mass, redshift and feedback. In Section \ref{Toymodel} we derive an analytic model for the formation of a stable hot hydrostatic atmosphere. In the model we calculate a halo mass scale for which the gravitational heating rate of the hot halo gas balances the gas cooling rate, thus keeping the gas hot and enabling the formation of a hot atmosphere. With the model we show that the ability of a halo to develop a hot hydrostatic atmosphere depends on the amount of hot gas that the halo already contains, which we calculate in the next subsection, and on the fraction of accreted gas that shock-heats to the halo virial temperature (Section \ref{sec_hotcold_modes}). 


\begin{figure}
  \centering
  \includegraphics[angle=0,width=0.45\textwidth]{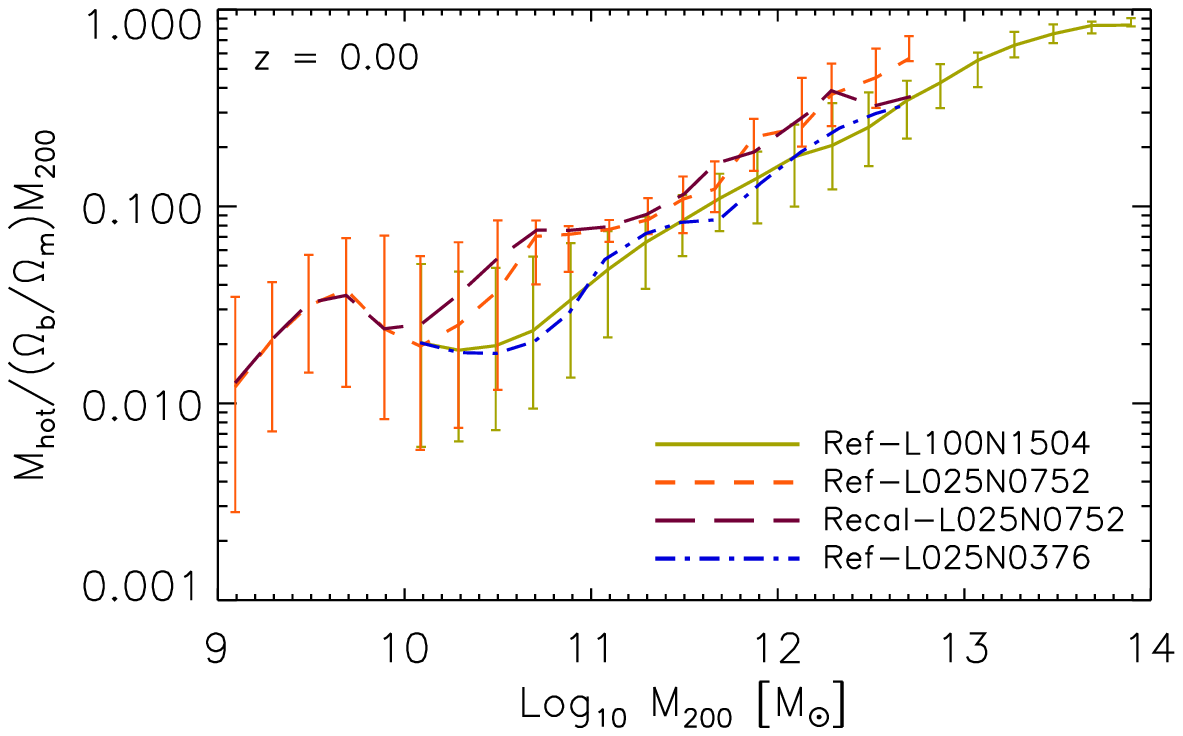}\\
  \vspace{-0.4cm}
  \includegraphics[angle=0,width=0.45\textwidth]{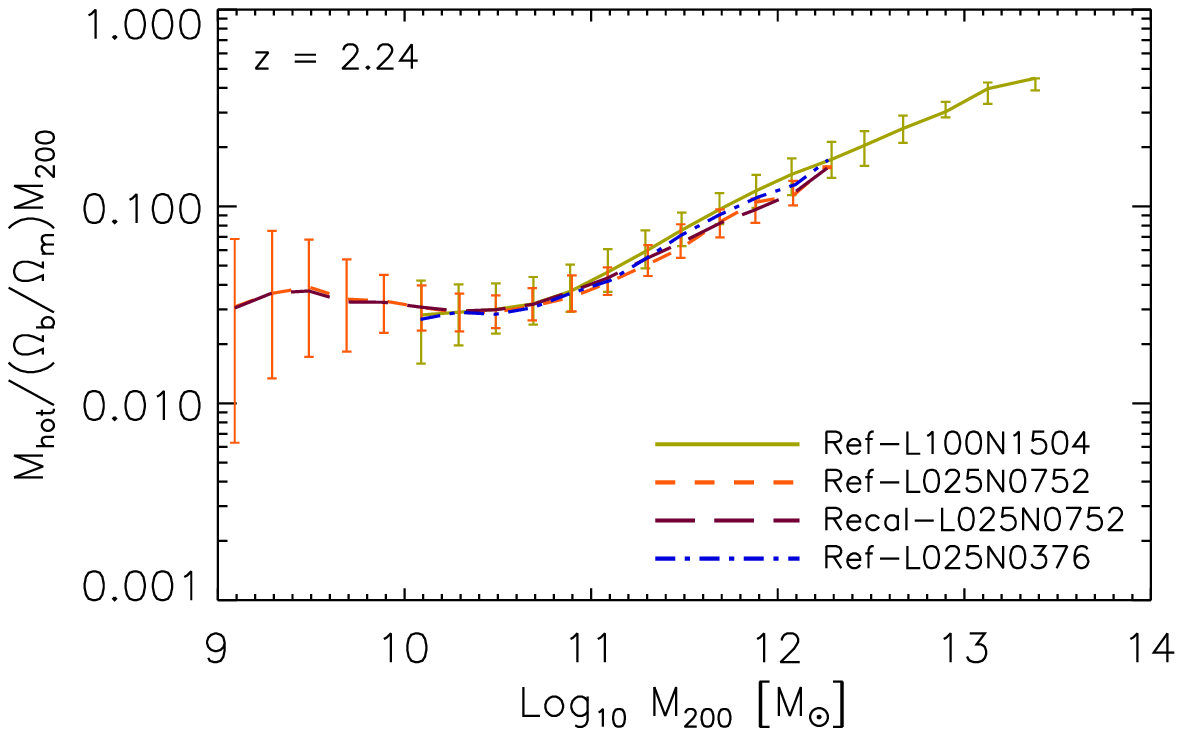}\\
  \vspace{-0.4cm}
  \includegraphics[angle=0,width=0.45\textwidth]{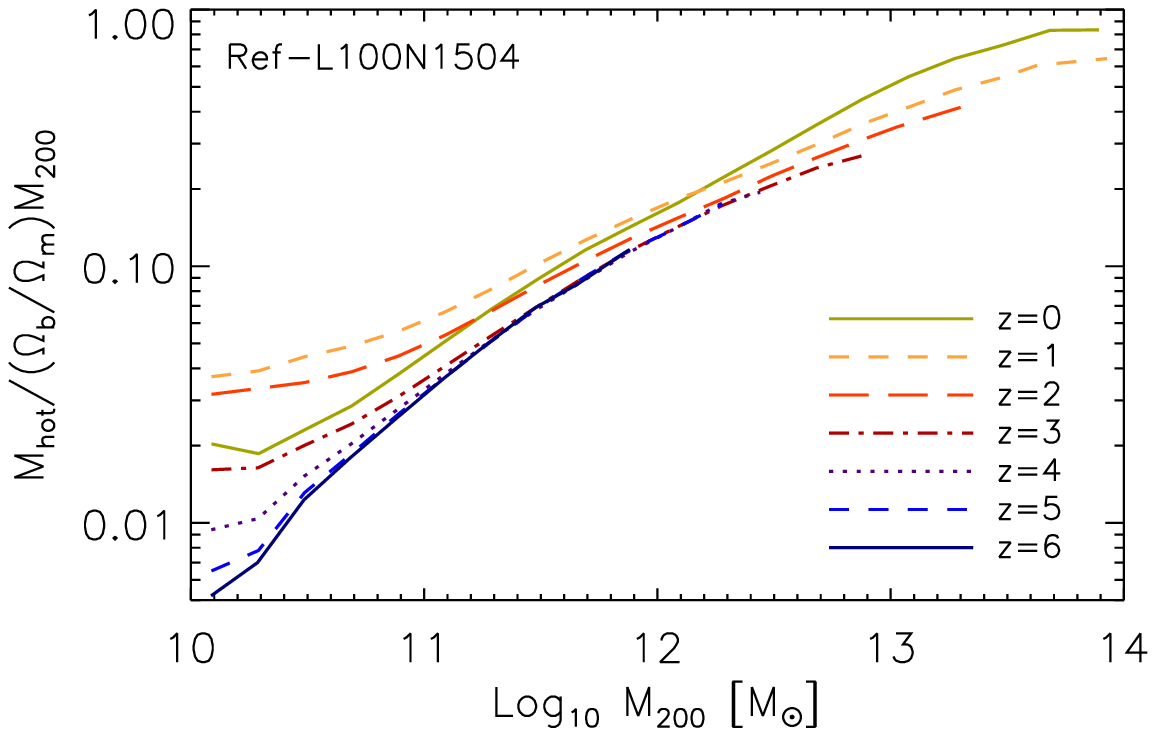}\\
    \includegraphics[angle=0,width=0.45\textwidth]{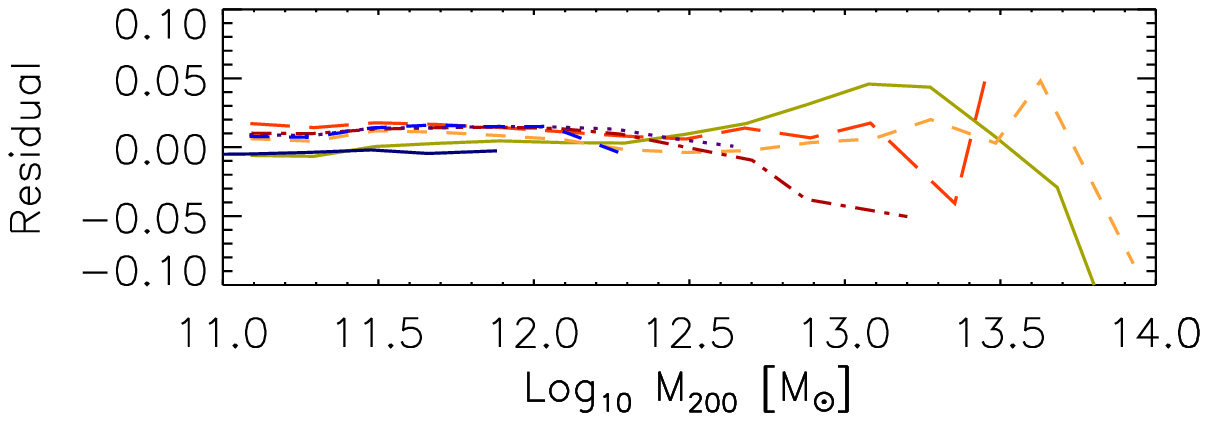}
  \vspace{0.3cm}
  \caption{Fraction of hot (with $t_{\rm{cool}}>t_{\rm{dyn}}$) gas mass with respect to the total halo mass, $M_{\rm{200}}$ (normalized by the universal baryon fraction), as a function of $M_{\rm{200}}$ for different simulations at $z=0$ (top panel), at $z=2.24$ (second panel from the top), and for Ref-L100N1504 at various redshifts (from $z=0$ to $z=6$, third panel from the top). The error bars in the top and middle panels show the $1\sigma$ scatter. Each bin contains at least 5 haloes. The bottom panel shows the residual of the data points with respect to the best-fit expression (eqs. 7-10).}
\label{Mhot_plot1}
\end{figure}

\subsection{Hot gas mass}

In order to better understand the build up of the hot gas mass, $M_{\rm{hot}}$ (mass of gas with $t_{\rm{cool}}>t_{\rm{dyn}}$ and $r>0.15R_{200}$) as haloes evolve, in this section we look for a correlation between $M_{\rm{hot}}$ and the total halo mass as a function of redshift. Fig.~\ref{Mhot_plot1} shows the median ratio of $M_{\rm{hot}}/(\Omega_{\rm{b}}/\Omega_{\rm{m}})M_{\rm{200}}$ (with $\Omega_{\rm{b}}/\Omega_{\rm{m}}=0.146$ the universal baryon fraction) taken from a range of simulations (as indicated in the legends) at $z=0$ (top panel) and at $z=2.24$ (second panel from the top). In these panels the error bars show the $1\sigma$ scatter for the Ref-L025N0752 and Ref-L100N1504 simulations. The median ratio of $M_{\rm{hot}}/(\Omega_{\rm{b}}/\Omega_{\rm{m}})M_{200}$ is also shown in the third panel from the top, but in this case the values are taken from the Ref-L100N1504 simulation and at various output redshifts.

The top panel of Fig.~\ref{Mhot_plot1} highlights the relatively poor agreement between the intermediate- and high-resolution simulations, with the latter predicting somewhat higher hot gas fractions. Good agreement is however achieved at $z=2.24$ (middle panel). Although the Ref-L100N1504 simulation is not fully converged with respect to the numerical resolution at $z=0$, the convergence with box size is excellent at all redshifts. The intermediate-resolution runs show that the hot gas represents $<10\%$ of the total halo gas mass for $M_{200}<10^{11.6}\Msun$ at $z=0$. The hot gas mass fraction reaches $80-90\%$ in $10^{13.6}\Msun$ haloes and remains roughly constant for higher masses. In very low-mass haloes ($M_{200}<10^{10.5}\Msun$), the hot gas mass fraction also remains roughly constant ($M_{\rm{hot}}/(\Omega_{\rm{m}}/\Omega_{\rm{b}})M_{200}\approx 0.02-0.03$). In these haloes cold accretion dominates, therefore the heating mechanism that maintains $M_{\rm{hot}}$ is the UV background as discussed in Section~\ref{sec_photo_heating}.

The third from the top panel of Fig.~\ref{Mhot_plot1} shows the evolution of the hot gas fraction. In haloes larger than $10^{11.5}\Msun$, $M_{\rm{hot}}/(\Omega_{\rm{b}}/\Omega_{\rm{m}})M_{\rm{200}}$ remains constant over the redshift range 3 to 6 and at lower redshift it increases somewhat with time. In smaller haloes ($M_{200}<10^{11.5}\Msun$), $M_{\rm{hot}}/(\Omega_{\rm{b}}/\Omega_{\rm{m}})M_{\rm{200}}$ increases with time until $z=1$ but decreases thereafter. We calculate the cooling rate of gas exposed to the UV background, and in the absence of it and compare the hot gas mass. We find that the hot gas mass in low-mass haloes increases due to the heating produced by the background radiation. In the case of gas not being exposed to the UV background, the total hot gas mass decreases with increasing redshift at fixed halo mass. We also find that the differences between $M_{\rm{hot}}$ occurs in haloes lower than $10^{11.3}\Msun$.

We next perform a least-square minimization to determine the best-fit relation $M_{\rm{hot}}-(\Omega_{\rm{b}}/\Omega_{\rm{m}})M_{\rm{200}}$ as a function of redshift. We apply equal weighting for each mass bin from the Ref-L100N1504 simulation (which we use to cover a large halo mass range) and minimize the quantity $\Delta_{j} = \frac{1}{N}\sum_{i=1}^{N}Y_{i}^{2}$, where

\begin{equation}
Y_{i}(z_{j}) =\log_{10}\left[\frac{M_{\rm{hot}}}{(\frac{\Omega_{\rm{b}}}{\Omega_{\rm{m}}})M_{\rm{200}}}\right]_{i}-F[M_{200,i},\alpha(z_{j}),\beta(z_{j}),\gamma(z_{j})],
\end{equation}

\noindent $N$ is the number of bins at each output redshift $z_{j}$, and $F$ is

\begin{eqnarray}
F &=& \alpha(z_{j})+\beta(z_{j})x_{i}+\gamma(z_{j})x_{i}^{2},\\
x_{i} &=& \log_{10}(M_{200,i}/10^{12}\Msun).
\end{eqnarray}

\noindent We obtain the best-fitting values for $\alpha$, $\beta$ and $\gamma$ at each redshift $z_{j}$, and following the same methodology we look for the best-fit expression of these parameters as functions of redshift.

We find that the following expression best reproduces the relation in the halo mass range $M_{200}=10^{11}-10^{14}\Msun$,

\begin{eqnarray}\label{Mhot1}
\log_{10}\left(\frac{M_{\rm{hot}}}{(\frac{\Omega_{\rm{b}}}{\Omega_{\rm{m}}})M_{200}}\right)&=& \alpha(z)+\beta(z)x+\gamma(z)x^{2},\\\label{Mhot2}
x &=& \log_{10}(M_{200}/10^{12}\Msun),
\end{eqnarray}

\noindent where $\alpha$, $\beta$ and $\gamma$ are functions of $z$ given by 

\begin{equation}\label{Mhot3}
{\rm{if }}\hspace{0.1cm}z \le 2 \left\{
\begin{array}{ll}
\alpha(z) = -0.79+0.31\tilde{z}-0.96\tilde{z}^{2},\\
\beta(z) = 0.52-0.57\tilde{z}+0.85\tilde{z}^{2},\\
\gamma(z) = -0.05,
\end{array} \right.
\end{equation}
\begin{equation}\label{Mhot4}
{\rm{if }}\hspace{0.1cm}z > 2 \left\{
\begin{array}{ll}
\alpha(z)=-0.38-1.56\tilde{z}+1.17\tilde{z}^2,\\
\beta(z)=0.12+0.94\tilde{z}-0.55\tilde{z}^2,\\
\gamma(z)=-0.05,
\end{array} \right.
\end{equation}

\noindent where $\tilde{z}=\log_{10}(1+z)$. The bottom panel of Fig.~\ref{Mhot_plot1} shows the residual of the data points with respect to the best-fit expression.

\begin{figure}
  \centering
  \includegraphics[angle=0,width=0.45\textwidth]{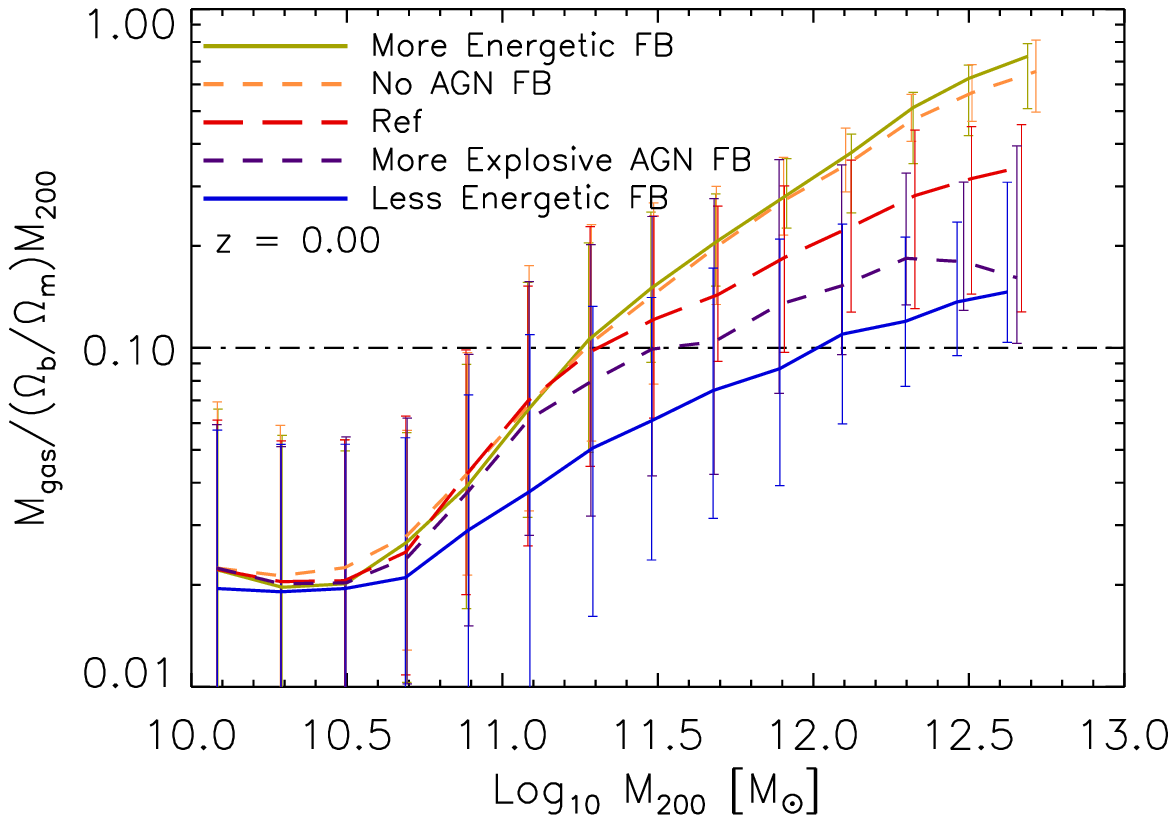}\\
  \includegraphics[angle=0,width=0.45\textwidth]{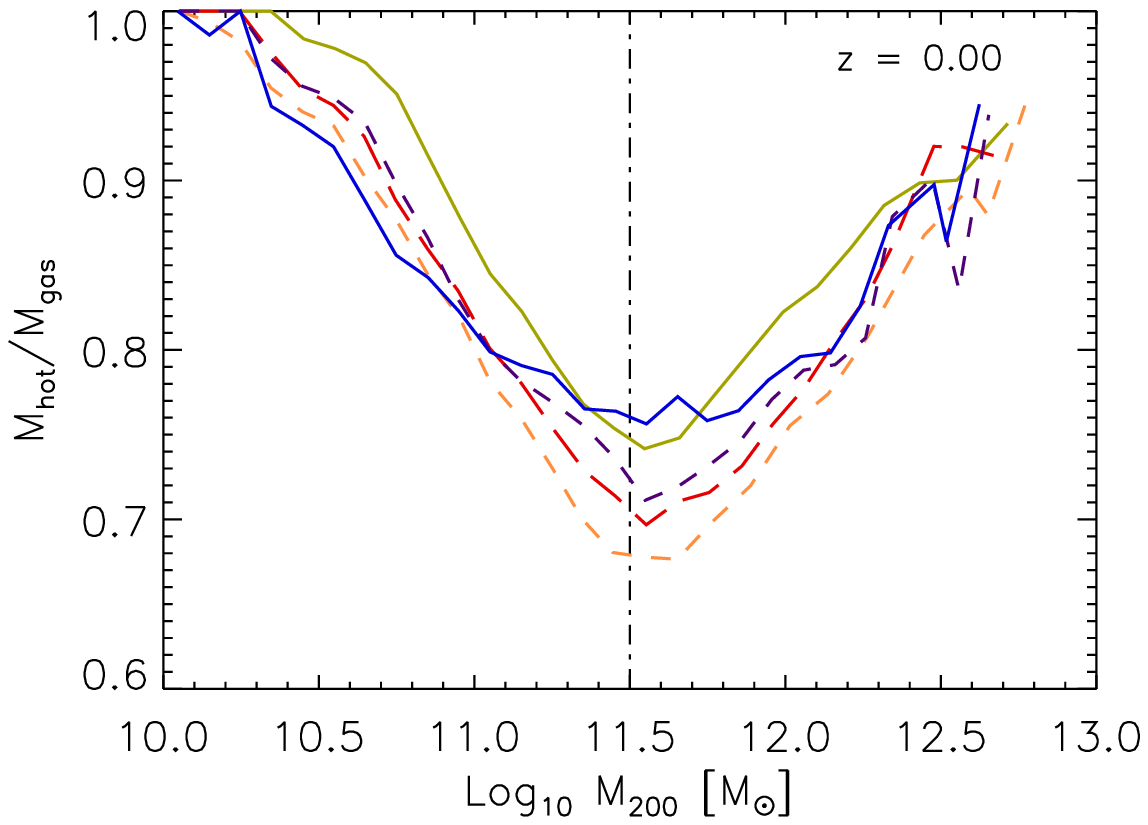}\\
  \caption{Top panel: Fraction of gas mass ($M_{\rm{gas}}(0.15R_{200}<r<R_{200})$ with respect to the total halo mass, $M_{\rm{200}}$ (normalized by the universal baryon fraction), as a function of $M_{\rm{200}}$ at $z=0$. The different lines correspond to L025N0376 simulations with different feedback prescriptions (see Table \ref{feedback_sims} and/or Section \ref{feedback_sims}). Bottom panel: Fraction of hot gas (gas with $t_{\rm{cool}}>t_{\rm{dyn}}$) with respect to $M_{\rm{gas}}$ as a function of $M_{200}$.}
\label{Mhot_plot2}
\end{figure}

Next, we investigate how the presence of different feedback mechanisms affect the hot gas as well as the total gas mass ($M_{\rm{gas}}$) in the halo (all gas contained between $0.15-1R_{200}$). The top panel of Fig. \ref{Mhot_plot2} shows the $M_{\rm{gas}}-(\Omega_{\rm{b}}/\Omega_{\rm{m}})M_{200}$ relation for haloes in the mass range $10^{10}-10^{13}\Msun$ at $z = 0$ for the L025N0376 simulations. The different coloured lines correspond to simulations with different feedback prescriptions. 

This panel shows that the impact of feedback increases with halo mass and that stellar feedback has a larger impact on the amount of gas in the halo than AGN feedback. We find that doubling the efficiency of the stellar feedback increases the gas mass fraction by a factor of 1.3 in $10^{12}\Msun$ haloes relative to the Ref model, whereas halving the efficiency decreases the gas mass fraction by a factor of 2.5. No (Explosive) AGN feedback results in an increase (decrease) by a factor of 1.5 in the gas mass fraction. 

While efficient stellar feedback increases the gas mass in the halo, more explosive AGN feedback decreases it. Overall, it seems that in haloes more massive than $10^{12}\Msun$ there is a greater difference in the gas mass fraction between Ref and More Energetic FB than between Ref and More Explosive AGN FB. This is due to two different reasons. Physically, AGN feedback mainly ejects gas mass from the halo, or prevents it from falling into the halo, whereas stellar feedback ejects gas out of the galaxy into the inner halo. Numerically, although stellar and AGN feedback use a similar thermal implementation (\citealt{DallaVecchia12}), there is a difference in the actual energetics of the processes. The energy injected per mass of stars formed changes between Ref and More Energetic FB, whereas the energy injected per unit mass accreted by the BH does not change between Ref and More Explosive AGN FB. In the latter, it is only the intermittency and the explosiveness that changes as a consequence of the change in the temperature of the AGN. In the case of Less Energetic FB, we find that the gas mass in the halo decreases because more gas is accreted by the galaxy and locked up in stars.

The bottom panel of Fig. \ref{Mhot_plot2} shows the variation of the ratio $M_{\rm{hot}}/M_{\rm{gas}}$ with feedback. It can be seen that in haloes less massive than $10^{11.5}\Msun$, the $M_{\rm{hot}}/M_{\rm{gas}}$ ratio increases with decreasing halo mass, indicating that most of the halo gas is heated by the X-ray/UV background (see Section 3.1 for a discussion). In haloes more massive than $10^{11.5}\Msun$, $M_{\rm{hot}}/M_{\rm{gas}}$ increases with halo mass. While a more energetic stellar feedback increases the hot mass fraction by 10$\%$, no AGN feedback decreases it by $8\%$ in $10^{12}\Msun$ haloes. In the case of Less Energetic FB and More Explosive AGN, $M_{\rm{hot}}/M_{\rm{gas}}$ increases by $10\%$ and $3\%$ (on average) with respect to Ref, respectively, in the halo mass range $10^{11.5}-10^{12}\Msun$.

\begin{figure*} 
\centering
\subfloat{\includegraphics[angle=0,width=0.7\textwidth]{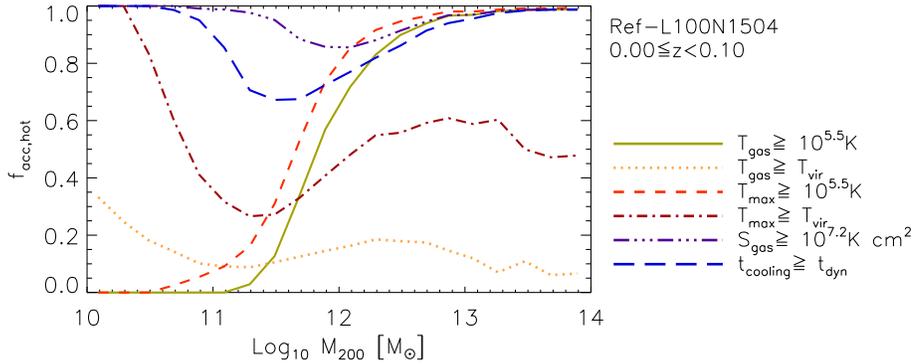}}
\caption{Fraction of gas accreted hot during the redshift range $0\le z<0.1$ as a function of halo mass. The various curves correspond to the hot fraction calculated using different methods as indicated in the legends.}
\label{Def_hot_frac}
\end{figure*}

In this section, gas particles with long cooling times ($t_{\rm{cool}}>t_{\rm{dyn}}$) are considered hot and counted in the calculation of $M_{\rm{hot}}$. Different from this work, \citet{vandeVoort12} separated hot and cold gas by performing a $T_{\rm{max}}$ cut, and found that the hot fraction as a function of radius decreases not only when AGN feedback is switched on, but also when stellar winds are enhanced. The reason for this is the way stellar feedback is implemented. In the more energetic stellar feedback simulation used by \citet{vandeVoort12}, the wind velocity scales with the local sound speed, so it largely overcomes the pressure of the ISM, blowing the gas out of the galaxy and halo, thus decreasing the amount of hot gas. In our work, the efficiency of stellar feedback is regulated by the fraction of the energy budget available ($f_{\rm{th}}$), which makes it more/less energetic and controls the frequency of feedback events, but the temperature increase is kept fixed.

So far we have analyzed the behavior of the hot gas mass in the halo. In the next section we investigate the fraction of gas that is accreted via hot and cold modes as a function of halo mass and redshift.

\section{Hot and cold modes of accretion}\label{sec_hotcold_modes}

Over the last decade, numerical simulations have shown that gas accretion onto haloes occurs in two different modes, gas either shock-heats to the halo virial temperature near the virial radius (the hot accretion mode), or crosses the virial radius unperturbed (the cold accretion mode). Several works have found that cold accretion dominates in low-mass haloes ($M_{200}< 10^{12}\Msun$) and that the transition mass increases weakly with increasing redshift (e.g. \citealt{Katz03,Keres05,Keres09,Ocvirk,vandeVoort11,vandeVoort12}). The two modes coexist at high redshift in massive haloes, which develop a hot hydrostatic atmosphere despite experiencing significant cold accretion through filaments, generally referred to as `cold flows' (\citealt{Keres05,Dekel09}). Cold flows are important for galaxy formation, because even if they experience significant heating when crossing the hot atmosphere (\citealt{Nelson13}), they are responsible for delivering cold, star-forming, gas deep within the halo (e.g. \citealt{Dekel09}). In the following sections we investigate different definitions that can be used to calculate the modes of accretion in the EAGLE simulations, analyse the impact of the hydrodynamics scheme and obtain best-fitting relations.

\subsection{Definition of hot accretion}\label{Shock_definition_sec}

The contributions from the two different modes of accretion have generally been calculated from the temperature history of the accreted gas (e.g. \citealt{Keres05,vandeVoort11,Nelson13}, among others), by following the maximum temperature, $T_{\rm{max}}$, each gas particle has ever reached. \citet{Keres05} found a clear bimodality in the distribution of $T_{\rm{max}}$ of accreted particles, and they proposed a threshold value, of $T_{\rm{max}}=2.5\times 10^{5}$ K, given by the minimum in the distribution of $T_{\rm{max}}$ values, to determine whether gas is accreted hot ($T_{\rm{max}}\ge 2.5\times 10^{5}$ K) or cold ($T_{\rm{max}}< 2.5\times 10^{5}$ K). This is an often used method but other approaches have also been taken. For example, \citet{Brooks09} identified hot gas accretion based on an entropy jump criterion, and concluded that their method led to a distinction between hot/cold modes in very good agreement with the selection of hot/cold gas based on the use of a constant temperature threshold. In this section, we compare $T_{\rm{max}}$ with other variables that can also give us some insight into whether a particle experienced a shock when crossing the virial radius.

We follow the method described in Section~\ref{method_accretion}, and look for gas particles that crossed the virial radius between two consecutive snapshots ($z_{i}\le z< z_{j}$). Then, we calculate the mass-weighted PDFs of the gas particles $T_{\rm{max}}$, temperature ($T_{\rm{gas}}$) and entropy ($S_{\rm{gas}}$) at redshift $z_{i}$ (see Fig.~\ref{Max_temp_plot}). We find that the PDFs for the redshift interval $0.0<z\le 0.1$ and $2.0<z\le 2.2$ are bimodal, but only for haloes larger than $10^{12}\Msun$, with the location of the local minimum changing with $M_{200}$. A detailed analysis of the PDFs can be found in Appendix~\ref{Analysis_PDF}. We next use the minimum of the PDFs from the $10^{12}\Msun$ haloes as a threshold value to separate particles accreted hot and cold. For $T_{\rm{gas}}$ and $T_{\rm{max}}$ PDFs we select the threshold to be $T_{\rm{min}}=10^{5.5}$ K, and for $S_{\rm{gas}}$ we use $S_{\rm{min}}=10^{7.2}$ K cm$^{2}$.

Fig.~\ref{Def_hot_frac} shows the fraction of gas particles accreted hot, $f_{\rm{acc,hot}}$, in the redshift range 0 to 0.1 as a function of halo mass. The different lines correspond to $f_{\rm{acc,hot}}$ calculated using different definitions. We first calculated $f_{\rm{acc,hot}}$ requiring that the gas particles at redshift 0 have temperatures higher than $10^{5.5}$ K (olive solid line), or that the gas particles have temperatures higher than the host halo virial temperature (orange dotted line). Then, we calculated $f_{\rm{acc,hot}}$ requiring that the gas particles maximum past temperature is higher than $10^{5.5}$ K (red dashed line) or higher than the host halo virial temperature (dark red dot-dashed line). Finally, we calculated $f_{\rm{acc,hot}}$ requiring that the entropy of the gas particles is larger than $10^{7.2}$ K cm$^{2}$ (purple dot-dot-dot-dashed line) or that the gas particles cooling time is larger than the local dynamical time (blue dashed line).  

As previous works have shown (e.g. \citealt{vandeVoort11,Nelson13}), the hot fraction depends very much on the definition. The fixed temperature cut, $T_{\rm{gas}}\ge 10^{5.5}$ K ($T_{\rm{max}}\ge 10^{5.5}$ K), gives a hot fraction that increases from 0.2 (0.5) in $10^{11.5}\Msun$ haloes to 0.7 (0.9) in $10^{12}\Msun$ haloes, thus showing a smooth transition from cold to hot accretion in the halo mass range $10^{11}$ to $10^{13}\Msun$. On the other hand, the criterion $T_{\rm{gas}}\ge T_{\rm{vir}}$ ($T_{\rm{max}}\ge T_{\rm{vir}}$) indicates that for the most massive haloes the hot mode accounts for only $20\%$ ($60\%$) of the accreted gas particles. For the low-mass haloes, $T_{\rm{gas}}\ge T_{\rm{vir}}$ ($T_{\rm{max}}\ge T_{\rm{vir}}$) gives a hot fraction that increases from 0.1 (0.35) in $10^{11}\Msun$ haloes to 0.35 (1.0) in $10^{10}\Msun$ haloes. This seems to indicate that there is another mechanism, such as shocks driven by winds or heating by the extragalactic UV/X-ray background radiation, that makes gas reach temperatures higher than the halo's $T_{\rm{vir}}$. In the case of $T_{\rm{max}}$, stellar feedback events certainly increase the hot fraction, since we obtain a peak in the $T_{\rm{max}}$ PDFs at $10^{7.5}$ K for all halo mass, which disappears when stellar feedback is switched off. 

Fig.~\ref{Def_hot_frac} also shows that $f_{\rm{acc,hot}}$ calculated using $S_{\rm{gas}}\ge 10^{7.2}$ K cm$^{2}$ and $t_{\rm{cooling}}\ge t_{\rm{dyn}}$, decreases in the halo mass range $10^{10}-10^{12}\Msun$ and increases for higher halo masses, being in agreement with $f_{\rm{acc,hot}}$ from $T_{\rm{gas}}\ge 10^{5.5}$ K in haloes larger than $10^{12}\Msun$. From this upturn we conclude that for haloes with $T_{\rm{vir}}\gg 10^{5}$ K ($M_{200}>10^{12}\Msun$), a large fraction of the accreted gas goes through a virial shock, and thus we can safely separate hot and cold accretion using $T_{\rm{gas}}\ge 10^{5.5}$ K or $T_{\rm{max}}\ge 10^{5.5}$ K. In lower-mass haloes ($M_{200}<10^{12}\Msun$), separating hot and cold accretion is not so easy. Although the hot halo is not expected to form (\citealt{Dekel}) and the $T_{\rm{gas}}$, $T_{\rm{max}}$ and $S_{\rm{gas}}$ PDFs are unimodal, UV background radiation (as discussed in Section 3.1) can significantly increase the cooling time and entropy of accreted gas.

Throughout this work we have investigated how gas heated by either stellar or AGN feedback, UV/X-ray background radiation or accretion shocks, evolves with halo mass and redshift. We next aim to identify gas that is mostly heated by accretion shocks when crossing the virial radius. While stellar or AGN feedback does not strongly impact gas falling onto the halo (\citealt{vandeVoort11}, it does strongly affect gas falling onto the galaxy, \citealt{vandeVoort11}; Paper II), UV radiation does, for low-mass haloes ($<10^{11.5}\Msun$). This can be seen in Fig.~\ref{Def_hot_frac} through the unexpected increase in $f_{\rm{acc,hot}}$ towards very low halo masses using $S_{\rm{gas}}\ge 10^{7.2}$ K cm$^{2}$, $t_{\rm{cool}}\ge t_{\rm{dyn}}$, $T_{\rm{gas}}\ge T_{\rm{vir}}$ and $T_{\rm{max}}\ge T_{\rm{vir}}$. These methods clearly indicate that gas is hot after falling onto halo, but not necessarily due to shock-heating. For that reason we decide to use a fixed temperature cut to calculate the hot mode of accretion. Note that $T_{\rm{max}}$ is updated whenever the gas particle reaches a higher temperature. However, if the gas particle is star-forming, $T_{\rm{max}}$ is not updated, because we impose a lower limit on the temperature of such gas. As in \citet{vandeVoort11}, ignoring shocks in the ISM is appropriate because we are interested in the $T_{\rm{max}}$ the gas reached before accreting onto the galaxy.

Because the gas temperature after shock-heating can be slightly higher or lower than $T_{\rm{vir}}$, we analyse the dependence of $f_{\rm{acc,hot}}$ on the fraction of $T_{\rm{vir}}$ used in the definition of hot accretion. As expected, at fixed halo mass $f_{\rm{acc,hot}}$ increases with decreasing fraction of $T_{\rm{vir}}$, reaching $f_{\rm{acc,hot}}=0.3$ and 0.6 in $10^{12}\Msun$ halos for $T_{\rm{gas}}\ge 0.8T_{\rm{vir}}$ and $T_{\rm{gas}}\ge 0.5T_{\rm{vir}}$, respectively. In addition, the virial shock can be located slightly inwards or outwards of $R_{200}$. We therefore calculated $f_{\rm{acc,hot}}$ (using $T_{\rm{gas}}\ge 10^{5.5}$ K) for gas crossing $0.8R_{200}$ and $1.2R_{200}$. We found that for halos more massive than $10^{12}\Msun$, $f_{\rm{acc,hot}}$ is insensitive to the precise value of the radius. In lower mass halos, the difference between $f_{\rm{acc,hot}}$ for gas crossing $0.8R_{200}$ and $1.2R_{200}$ can be as high as 0.2dex in $10^{11.5}\Msun$ halos.

\begin{figure} 
  \centering
  \subfloat{\includegraphics[angle=0,width=0.45\textwidth]{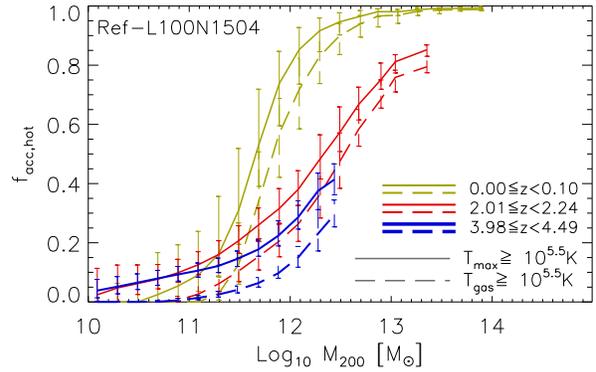}}
  \vspace{-0.15cm}
  \caption{Fraction of gas accreted hot during the redshift ranges $0\le z<0.1$ (green lines), $2.0\le z<2.2$ (red lines) and $4.0\le z<4.49$ (thick blue lines), against halo mass. The solid curves correspond to the hot fraction calculated using $T_{\rm{max}}\ge 10^{5.5}$ K, whereas the dashed curves correspond to $T_{\rm{gas}}\ge 10^{5.5}$ K. The error bars show the $1\sigma$ scatter of the fractions.}
\label{HotFraction_plot_1}
\end{figure}

Fig.~\ref{HotFraction_plot_1} shows $f_{\rm{acc,hot}}$ calculated using $T_{\rm{max}}\ge 10^{5.5}$ K (solid lines) and $T_{\rm{gas}}\ge 10^{5.5}$ K (dashed lines) at $z=0$ (green lines), $z=2$ (red lines) and $z=4$ (thick blue lines). For $10^{12}\Msun$ haloes the fraction of shock-heated gas particles with $T_{\rm{gas}}\ge 10^{5.5}$ K is a factor 1.25 lower than $f_{\rm{acc,hot}}$ given by $T_{\rm{max}}\ge 10^{5.5}$ K at $z=0$, a factor of 1.6 at $z=2$ and a factor of 2 at $z=4$. Although changing the threshold value can bring the $f_{\rm{acc,hot}}$ curves into better agreement, some disagreement is expected because gas that was heated once may cool later. 

$T_{\rm{gas}}$ and $T_{\rm{max}}$ are able to identify the gas particles that are shock-heated when crossing the virial radius, however since we have found that $T_{\rm{max}}$ is affected by stellar feedback, we decide to use the gas particle temperature, $T_{\rm{gas}}$, after accretion. We find that a lower limit on $T_{\rm{gas}}$ is the most suitable method to calculate $f_{\rm{acc,hot}}$, since it also does not include gas that goes through a shock but cools immediately afterwards and therefore does not contribute to the hot halo formation process.

\begin{figure}   \centering
  \includegraphics[angle=0,width=0.45\textwidth]{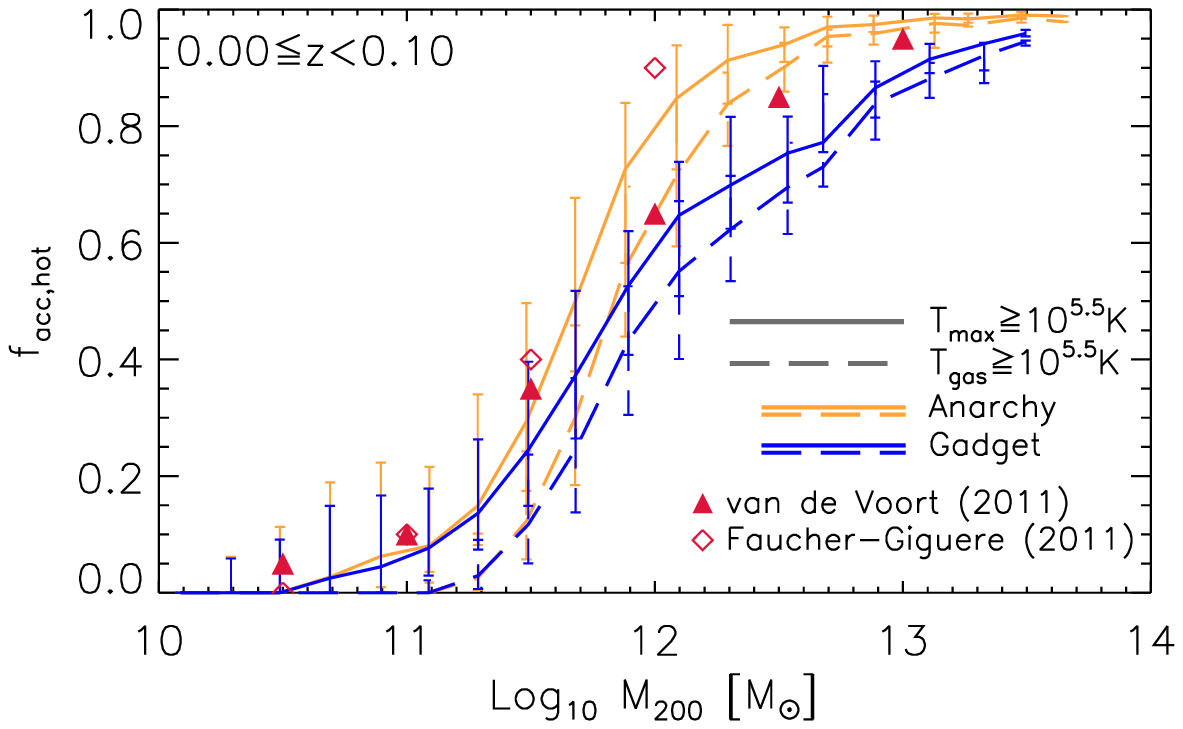}\\
  \vspace{-0.43cm}
  \includegraphics[angle=0,width=0.45\textwidth]{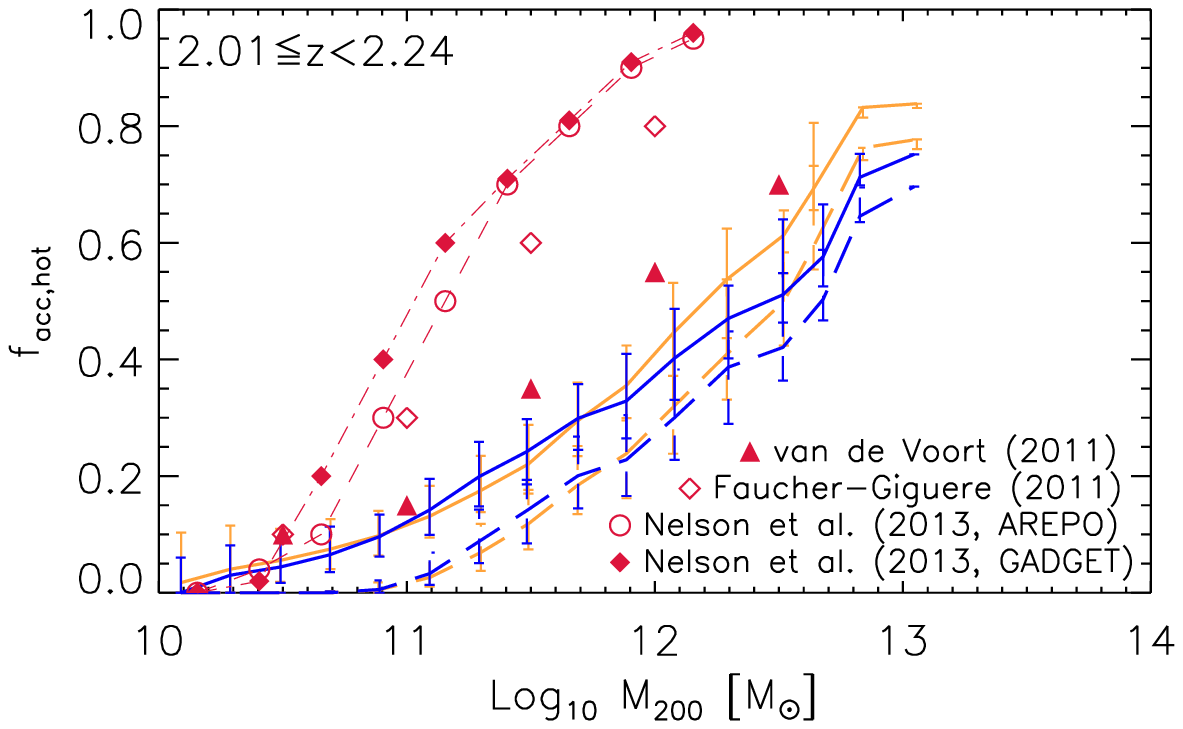}
  \vspace{-0.15cm}
  \caption{Fraction of gas accreted hot during the redshift ranges $0\le z<0.1$ (top) and $2.0\le z<2.2$ (bottom) against halo mass. The panels show the same as Fig.~\ref{HotFraction_plot_1}, but in this case the curves correspond to the hot fraction calculated from the Ref-L050N0752 Anarchy (orange lines) and GADGET (blue lines) simulations. The error bars show the $1\sigma$ scatter of the fractions. The symbols correspond to the hot fraction estimates of van de Voort et al (2011, blue triangles), Faucher-Giguere et al (2011, open diamonds) and Nelson et al. (2013, open circles).}
\label{HotFraction_plot_3}
\end{figure}

\subsubsection{Gadget and Anarchy}

In this section we extend the discussion presented in Section~\ref{hydrodynamics_discussion}, and analyse the differences in the hot/cold modes of accretion onto haloes when the formulation of the hydrodynamics scheme is changed. We compare two L050N0752 simulations that use the same subgrid models, one employs the standard SPH code GADGET, while the other employs the Anarchy hydrodynamics solver used in the fiducial EAGLE runs. Fig.~\ref{HotFraction_plot_3} shows the same as Fig.~\ref{HotFraction_plot_1} for $z=0$ (top panel) and $z=2$ (bottom panel), but the lines correspond to the standard GADGET (blue lines) and Anarchy (orange lines) simulations. 

The top and bottom panels show excellent agreement between GADGET and Anarchy in haloes less massive than $10^{11.5}\Msun$ and $10^{12}\Msun$, respectively, and modest differences in larger haloes, irrespective of whether we use the $T_{\rm{max}}$ hot gas accretion fractions (solid lines) or the $T_{\rm{gas}}$ hot gas accretion fractions (dashed lines). The Anarchy simulation exhibits a somewhat larger fraction of hot accretion onto massive haloes than its GADGET counterpart. This is expected, since the spurious surface tension appearing in the GADGET formulation of SPH prevents the cold dense clumps of gas from being disrupted, mixed and heated when crossing the virial shock (e.g. \citealt{Schaller15b}). 

The top panel of Fig.~\ref{HotFraction_plot_3} shows that the $T_{\rm{max}}$ hot gas accretion fractions taken from the GADGET simulation are in agreement with \citet{vandeVoort11} analysis of the OWLS simulations (\citealt{Schaye}). The bottom panel also shows broad agreement with \citet{vandeVoort11}, but substantial differences with \citet{Nelson13}. While \citet{vandeVoort11} used the standard GADGET hydrodynamic solver in their simulations, \citet{Nelson13} analysed two simulation series that employed either the moving mesh code AREPO (\citealt{AREPO}) or standard GADGET, both without stellar, AGN feedback or metal cooling. \citet{Nelson13} traced the evolution of the gas properties using a Monte Carlo tracer particle technique that enable them to compute $T_{\rm{max}}$. They did not find large differences between GADGET and AREPO in the cold mode of accretion onto haloes, and concluded that the cold fraction onto haloes mainly depends on the manner (either with $T_{\rm{max}}$ or other cut-off temperature) in which it is measured. 

\begin{figure*} 
  \centering
  \includegraphics[angle=0,width=0.45\textwidth]{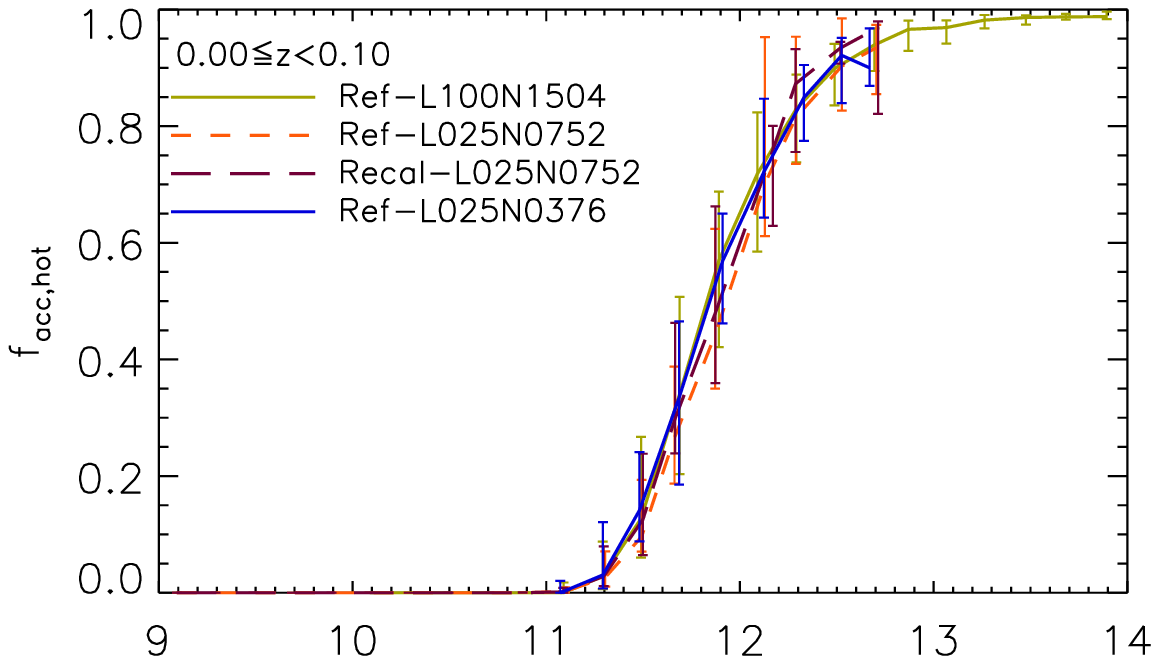}
  \includegraphics[angle=0,width=0.45\textwidth]{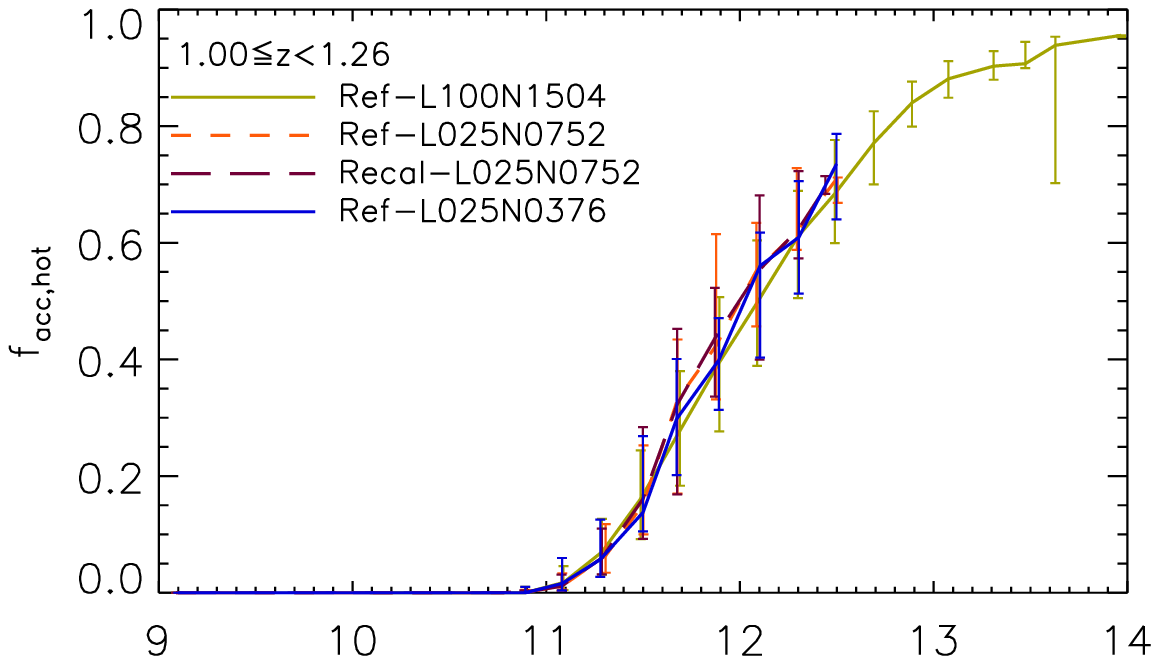}\\
  \vspace{-0.4cm}
  \includegraphics[angle=0,width=0.45\textwidth]{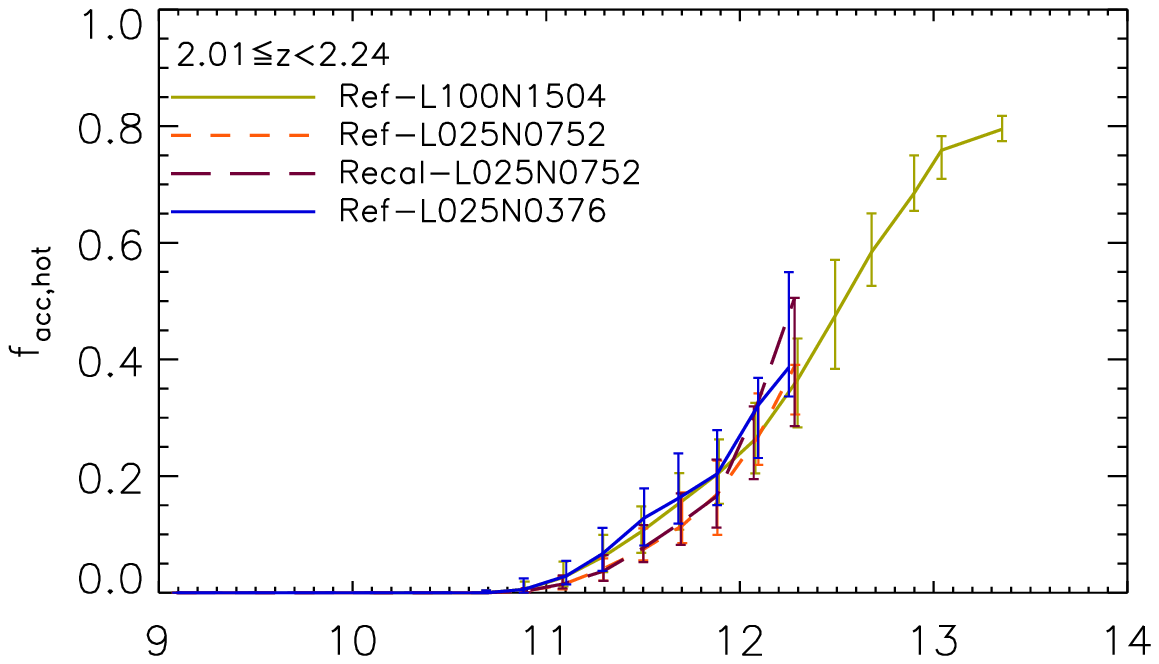}
  \includegraphics[angle=0,width=0.45\textwidth]{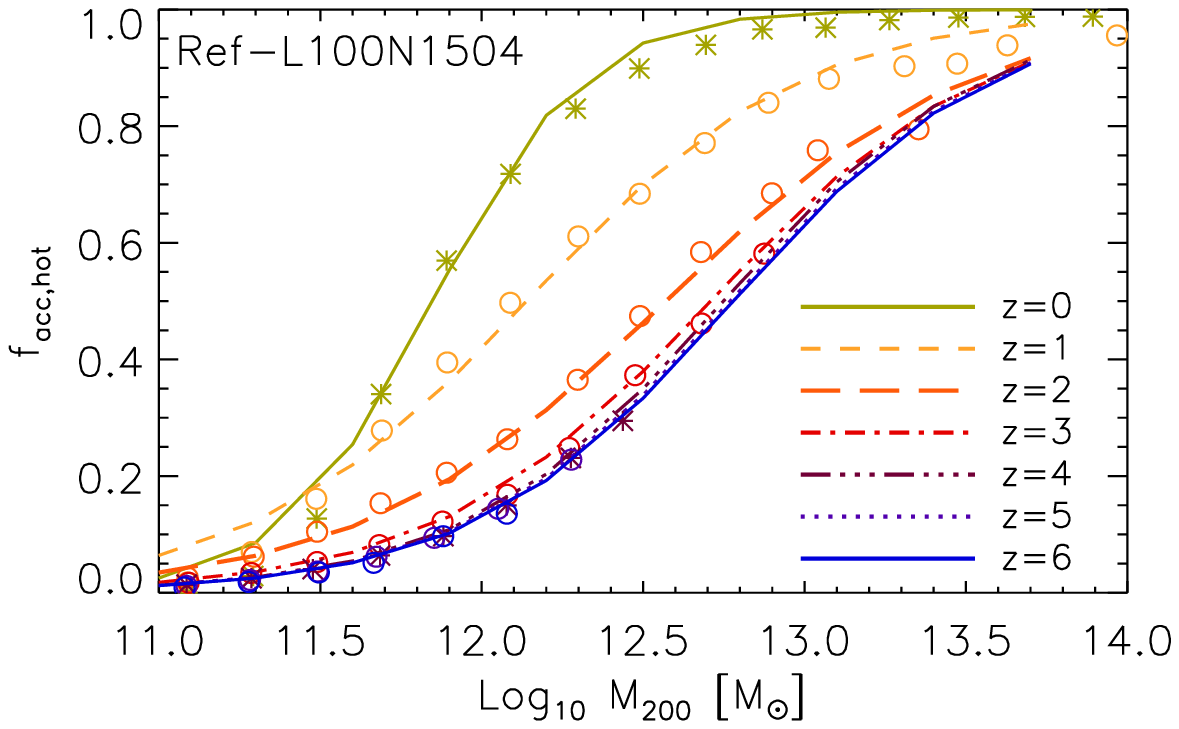}\\
  \vspace{-0.15cm}
  \caption{Fraction of hot mode gas accretion during $0\le z<0.1$ (top left panel), $1.0\le z<1.26$ (top right panel), $2.0\le z<2.24$ (bottom left panel) against halo mass. In these panels, the different lines correspond to simulations with different resolutions and box sizes and the error bars show the $1\sigma$ scatter. The bottom right panel compares different redshifts (symbols) to the best-fit expression (lines).}
\label{HotFraction_plot_2}
\end{figure*}

The large differences between our work and that of \citet{Nelson13} is intriguing. \citet{Nelson13} calculated the accretion rates and so the hot and cold fractions considering only smooth accretion (without including the merger contribution) and over an accretion time window of 1 Gyr. In this work we did not separate gas accreted smoothly or through mergers and used smaller time windows, however we find that considering only smooth accretion and/or larger time window does not significantly change the fraction of gas accreted hot (but increases the total rate of gas accretion). In addition, \citet{Nelson13} used simulations without metal cooling, stellar feedback or AGN feedback. We compared the fraction of hot gas accretion between our reference model and a model without feedback and found that in the simulation without feedback the hot fraction increases from $10\%$ to $20\%$ in the mass range $10^{12}\Msun$ to $10^{12.5}\Msun$, and agree in $10^{13}\Msun$. However this increase is not enough to explain the large differences we find with \citet{Nelson13}. Unfortunately, we do not have a model without metal cooling, to test whether this plausible explanation is sufficient to account for the remaining differences.

We also compare our results with the work of \citet{Faucher}, who used a series of cosmological simulations run with the standard SPH code GADGET and including stellar feedback and metal cooling but no AGN feedback. They calculated the rates of gas crossing a virial shell and, similar to this work, they used an instantaneous temperature ($T_{\rm{gas}}>2.5\times 10^{5}$ K rather than $T_{\rm{max}}$) to separate hot from cold accretion. We find good agreement at $z=0$, but at $z=2$ we find large differences. The \citet{Faucher} transition mass (i.e. the halo mass where the hot mode of accretion equals the cold mode) at $z=2$ is between $10^{11.3}\Msun$ and $10^{11.5}\Msun$ (depending on the stellar feedback).

\subsection{Hot/Cold fraction}

The final ingredient for our model of hot halo formation, which we will present in Section \ref{Toymodel}, is the fraction of gas accreted onto haloes in the hot and cold modes. Throughout this work we calculate the fraction of hot mode gas accretion, $f_{\rm{acc,hot}}(M,z)$, using $T_{\rm{gas}}\ge 10^{5.5}$ K. $f_{\rm{acc,hot}}$ can be considered as an indirect measure of the presence of hot gas in the halo, since large values of $f_{\rm{acc,hot}}$ imply large values of $M_{\rm{hot}}/M_{200}$. Fig.~\ref{HotFraction_plot_2} shows $f_{\rm{acc,hot}}(M,z)$ at $z=0-0.1$ (top left panel), $z=1.0-1.26$ (top right panel) and $z=2.0-2.24$ (bottom left panel). For each redshift range we find excellent agreement between the $f_{\rm{acc,hot}}(M,z)$ curves taken from simulations with different resolution and box size. We find that $f_{\rm{acc,hot}}(M,z)$ increases smoothly with halo mass and decreasing redshift.

We look for the best-fit expression for $f_{\rm{acc,hot}}(M_{\rm{halo}},z)$ by performing a least-square minimization. We follow the method described in Section 3.2, we apply equal weighting for each mass bin from the Ref-L100N1504 simulation and minimize the quantity $\Delta_{j}=\frac{1}{N}\sum_{i}^{N}[f_{\rm{acc,hot}}(M_{i},z_{j})-F(M_{200,i},a(z_{j}),M_{1/2}(z_{j}))]^{2}$, where $N$ is the number of mass bins at each output redshift $z_{j}$, and $F$ is

\begin{eqnarray}
F &=& 1/(1+[M_{200,i}/M_{1/2}(z_{j})]^{a(z_{j})}).
\end{eqnarray}

\noindent We calculate the best-fitting values of $a$ and $M_{1/2}$ at each redshift $0\le z_{j}<6$ and for the halo mass range $10^{10}\le M_{200}<10^{14}\Msun$. We then look for the best-fitting expressions of $a$ and $M_{1/2}$ as a function of redshift. We find that the relations

\begin{eqnarray}\label{fhot}
f_{\rm{acc,hot}}(M_{200},z) &=& 1/(1+[M_{200}/M_{1/2}(z)]^{a(z)}).
\end{eqnarray}

\begin{eqnarray}\label{az}
a(z) &=& \left\{
\begin{array}{cl}
-1.86\times 10^{-1.26\tilde{z}+1.29\tilde{z}^{2}} & {\rm{if }}\hspace{1mm} 0\le z < 2,\\
-0.46\times10^{0.81\tilde{z}-0.42\tilde{z}^{2}} & {\rm{if }}\hspace{1mm} 2\le z < 4,\\
-1.07 & {\rm{if }}\hspace{1mm} z \ge 4,\\
\end{array} \right.\\
\tilde{z} &=& \log_{10}(1+z),
\end{eqnarray}

\begin{eqnarray}\nonumber
M_{1/2}(z) &=& 10^{12}\Msun\times\\\label{Mhalf}
& & \left\{
\begin{array}{cl}
-0.15+0.22z+0.07z^{2} & {\rm{if }}\hspace{1mm} 0\le z < 2,\\
-0.25+0.53z-0.07z^{2} & {\rm{if }}\hspace{1mm} 2\le z < 4,\\
0.72+0.01z & {\rm{if }}\hspace{1mm} z \ge 4,
\end{array} \right.
\end{eqnarray}

\noindent best reproduce the fraction of hot mode accretion as a function of halo mass and redshift. The bottom right panel of Fig.~\ref{HotFraction_plot_2} compares the fraction of hot accretion in various redshift ranges ($0\le z<6$, symbols) with the best-fit expression (lines). We find that $f_{\rm{acc,hot}}$ evolves similarly to $M_{\rm{hot}}/(\Omega_{\rm{b}}/\Omega_{\rm{m}})M_{200}$ (shown in the bottom panel Fig.~\ref{Mhot_plot1}). For all halo masses, $f_{\rm{acc,hot}}$ increases with time until $z=1$. At $z<1$ $f_{\rm{acc,hot}}$ increases further in high-mass haloes ($M_{200}>10^{11.5}\Msun$) but decreases in low-mass haloes ($M_{200}<10^{11.5}\Msun$). This is in agreement with \citet{vandeVoort11}, who calculated $f_{\rm{acc,hot}}$ using the $T_{\rm{max}}$ criterion applied to the OWLS simulations. 

We have also investigated how $f_{\rm{acc,hot}}(M,z)$ is affected when we vary the feedback mechanisms. We find that although the total gas accretion rate onto the halo remains nearly unchanged under varying feedback mechanisms (in agreement with \citealt{vandeVoort11}), $f_{\rm{acc,hot}}(M,z)$ increases somewhat at fixed halo mass for the strong stellar feedback and no AGN feedback scenarios. The impact of stellar feedback is largest. For example, strong stellar feedback increases $f_{\rm{acc,hot}}(M,z)$ by a factor of 1.2 in $10^{12}\Msun$ haloes, whereas weak stellar feedback decreases it by a factor of 1.26. Strong AGN feedback decreases $f_{\rm{acc,hot}}(M,z)$ but only in high-mass haloes and by up to a factor of 1.1. As in \citet{vandeVoort11}, we find that the impact of feedback mechanisms on the fraction of hot mode gas accretion is small.

In the next section we derive an analytic model for hot halo formation. In the model we assume that the halo develops a hot atmosphere depending on the fraction of hot mode gas accretion, and on the amount of hot gas already in the halo.    						

\section{Toy model}\label{Toymodel}

In Section 3 we investigated the formation of the hot halo in the EAGLE simulations. We found that the development of a strong bimodality in the cooling time PDF of the halo gas provides a clear signature of hot halo formation, which occurs in the halo mass range $10^{11.5}-10^{12}\Msun$. We noticed however that, even when a stable hot atmosphere has not yet been formed, there is already some hot gas in haloes less massive than $10^{11.5}\Msun$. This is because gas can be heated by the extragalactic UV/X-ray background or by shocks with stellar or AGN outflows. In this section we aim to determine the heating rate of gas produced by accretion shocks only, and the halo mass at which this heating overcomes the cooling. To do so, we present an analytic model for the shock-heating rate that takes into account the hot gas mass already in the halo, and the fraction of gas accretion occurring in the hot mode. With the model we aim to assess the impact of feedback mechanisms (that change the hot gas mass), as well as filamentary cold accretion (that decreases the hot mode fraction of gas accretion), on the formation of a stable hot halo. The model assumes that the hot halo forms when the heating rate produced by accretion shocks is able to balance the radiative cooling rate. 




We calculate the variation of the post-shock gas internal energy, ${\mathscr{E}}$ (in units of erg), due to the transformation of kinetic energy into thermal energy through accretion shocks and due to radiative losses as

\begin{equation}\label{energy}
\dot{\mathscr{E}} = \Gamma_{\rm{heat}}-\Gamma_{\rm{cool}},
\end{equation}

\noindent where $\Gamma_{\rm{heat}}=\frac{{\rm{d}}}{{\rm{d}}t}(\frac{3}{2}k_{\rm{B}}TN_{\rm{hot}})$ is the gas heating rate (in units of erg s$^{-1}$), defined as the variation in time of the thermal energy of the hot gas in the halo, and $\Gamma_{\rm{cool}}$ is the net radiative cooling rate (in units of erg s$^{-1}$). In the definition of $\Gamma_{\rm{heat}}$, $N_{\rm{hot}}$ is number of hot gas particles in the halo and $T$ the mean gas temperature, which we assume to be $T=T_{\rm{vir}}$. Also, we assume the gas to be monatomic (i.e. the ratio of specific heat is 5/3). Therefore, when

\begin{equation}
\Gamma_{\rm{heat}}>\Gamma_{\rm{cool}},
\end{equation}

\noindent the accumulated shock-heated gas at the virial radius gains the necessary pressure through external shock heating to overcome the energy loss from radiative cooling. We follow \citet{Dekel} and define a critical mass, $M_{\rm{crit}}$, above which haloes develop a hot atmosphere. We define $M_{\rm{crit}}$ as the halo mass at which the cooling rate, $\Gamma_{\rm{cool}}$, of the hot gas in the halo equals the heating rate, $\Gamma_{\rm{heat}}$, produced by the accretion shocks. In the following subsections we present the calculations for the critical halo mass.

\subsection{Virial heating rate and accretion history}\label{Heating}

In a $\Lambda$CDM cosmology, haloes grow through mergers and smooth accretion. Rapid mass accretion and mergers dynamically heat the gas when haloes form, transforming gravitational potential energy into kinetic energy of baryons and dark matter. For the gaseous component, kinetic energy associated with bulk and turbulent motions is transformed into thermal energy through shocks and viscous dissipation (e.g. \citealt{Wang}). As a result, the heating rate defined above is driven by the transformation of the gravitational potential energy of baryons and dark matter into thermal energy through $\dot{T}_{\rm{vir}}$, and by the accretion rate of gas undergoing shocks through $\dot{N}_{\rm{hot}}$, as follows

\begin{equation}\label{Gamma_heat1}
\Gamma_{\rm{heat}} = \frac{3}{2}k_{\rm{B}}\dot{T}_{\rm{vir}}N_{\rm{hot}}+\frac{3}{2}k_{\rm{B}}T_{\rm{vir}}\dot{N}_{\rm{hot}}.
\end{equation}

\noindent We next rewrite eq. (\ref{Gamma_heat1}) assuming that ${N_{\rm{hot}}=M_{\rm{hot}}/\mu m_{\rm{p}}}$ (with $\mu=0.59$ and invariant), that ${f_{\rm{hot}} =M_{\rm{hot}}/[(\Omega_{\rm{b}}/\Omega_{\rm{m}})M_{200}]}$, that ${\dot{M}_{\rm{hot}}=f_{\rm{acc,hot}}(\Omega_{\rm{b}}/\Omega_{\rm{m}})\dot{M}_{200}}$, and that $\dot{T}_{\rm{vir}}=\frac{2}{3}\frac{\dot{M}_{200}}{M_{200}}T_{\rm{vir}}$. Eq. (\ref{Gamma_heat1}) then yields 

\begin{eqnarray}\label{gamma_heat}
\Gamma_{\rm{heat}} &=& \frac{3}{2}\frac{k_{\rm{B}}T_{\rm{vir}}}{\mu m_{\rm{p}}}\frac{\Omega_{\rm{b}}}{\Omega_{\rm{m}}}\dot{M}_{200}\left[\frac{2}{3}f_{\rm{hot}}+f_{\rm{acc,hot}}\right].
\end{eqnarray}

The virial temperature of a halo formed at redshift $z$ is related to the total mass $M_{200}$ as

\begin{equation}\label{Tvir}
T_{\rmn{vir}}=10^{5.3}\,{\rm{K}}\,\left(\frac{M_{200}}{10^{12}\Msun}\right)^{2/3}(1+z),
\end{equation}

\noindent where we assumed that the halo encloses a characteristic virial overdensity $\Delta_{c}=200$ relative to the critical density at redshift $z$, $\rho_{\rm{crit}}(z)=\left(\frac{3 H_{0}^{2}}{8\pi G}\right)[\Omega_{\rm{m}}(1+z)^{3}+\Omega_{\rm{\Lambda}}]$. 

To calculate the halo accretion rate, we use the analytic derivation based on Press-Schechter theory from \citet{Correa15c},

\begin{eqnarray}\nonumber
\dot{M}_{200}(z) &=& 71.6\,{\rm{M}}_{\sun}{\rm{yr}}^{-1}\,\left(\frac{M_{200}(z)}{10^{12}\rm{M}_{\sun}}\right)\left(\frac{h}{0.7}\right)[-\tilde{\alpha}-\tilde{\beta}]\\\label{dMdt}
&&\times (1+z)[\Omega_{\rm{m}}(1+z)^{3}+\Omega_{\Lambda}]^{1/2},
\end{eqnarray}

\noindent where $\tilde{\alpha}$ and $\tilde{\beta}$ depend on halo mass and the linear power spectrum. This formula gives the accretion rate at redshift $z$. See \citet{Correa15a,Correa15b,Correa15c} for more details on the accretion rate model.

\subsection{Cooling rate}\label{Cooling}

To find haloes for which the infalling gas is shock-heated and prevented from cooling onto the inner halo, we compare the mechanical heating rate in equation (\ref{gamma_heat}) to the net radiative cooling rate, $\Gamma_{\rmn{cool}}$ $[\rmn{erg}$ $\rmn{s}^{-1}]$,

\begin{eqnarray}\label{gamma_cool}
\Gamma_{\rmn{cool}} &=& M_{\rm{hot}}\frac{\Lambda(T_{\rm{hot}},Z_{\rm{hot}},\rho_{\rm{hot}})}{\rho_{\rm{hot}}},\\\nonumber
&=&f_{\rm{hot}}\frac{\Omega_{\rm{b}}}{\Omega_{\rm{m}}}M_{200}\frac{\Lambda(T_{\rm{hot}},Z_{\rm{hot}},\rho_{\rm{hot}})}{\rho_{\rm{hot}}},
\end{eqnarray}

\noindent where $\Lambda(T_{\rm{hot}},Z_{\rm{hot}},\rho_{\rm{hot}})$ $[\rmn{erg}$ $\rmn{cm}^{-3}\rmn{s}^{-1}]$ is the net cooling rate per unit volume and $M_{\rm{hot}}/\rho_{\rm{hot}}$ is the volume the hot gas occupies. In the calculation of $\Gamma_{\rm{cool}}$ we assume that $T_{\rm{hot}}=T_{\rm{vir}}$, that the density of the hot gas is $\rho_{\rm{hot}}=10^{0.6}\rho_{\rm{crit}}$ and that the metallicity, $Z_{\rm{hot}}$, is $Z_{\rm{hot}}= 0.1Z_{\odot}$ (both constant with redshift). 
The $\rho_{\rm{hot}}$ and $Z_{\rm{hot}}$ values were chosen after the analyses of the hot gas density and metallicity, as well as the dependence on the halo mass, which is included in Appendices \ref{Metallicity_Sec} and \ref{Density_Sec}, respectively. In particular, Appendix \ref{Density_Sec} shows that at $z=0$, the mean density of the hot gas in the halo is around $10^{0.6}\rho_{\rm{crit}}$ in both $10^{12}\Msun$ and $10^{11.5}\Msun$ haloes, is slightly higher at $z=2.2$ than at $z=0$ and does not change significantly with the host halo mass.

\subsection{Critical halo mass for the formation of a hot halo}

\subsubsection{Analytic estimate}

In this subsection we use the energy condition of post-shock gas given by eq.~(\ref{energy}) and calculate the critical halo mass, $M_{\rm{crit}}$, for which the mechanical heating rate, $\Gamma_{\rm{heat}}$ (eq.~\ref{gamma_heat}), equals the gas cooling rate, $\Gamma_{\rm{cool}}$ (eq.~\ref{gamma_cool}). $M_{\rm{crit}}$ is the halo mass above which the heating rate exceeds the cooling rate, and as a result the halo develops a stable hot hydrostatic atmosphere. To calculate $M_{\rm{crit}}$, we assume values for the fraction of hot mode gas accretion ($f_{\rm{acc,hot}}$), as well as the fraction of hot gas mass in the halo ($f_{\rm{hot}}$).

\begin{figure} 
\centering
\subfloat{\includegraphics[angle=0,width=0.45\textwidth]{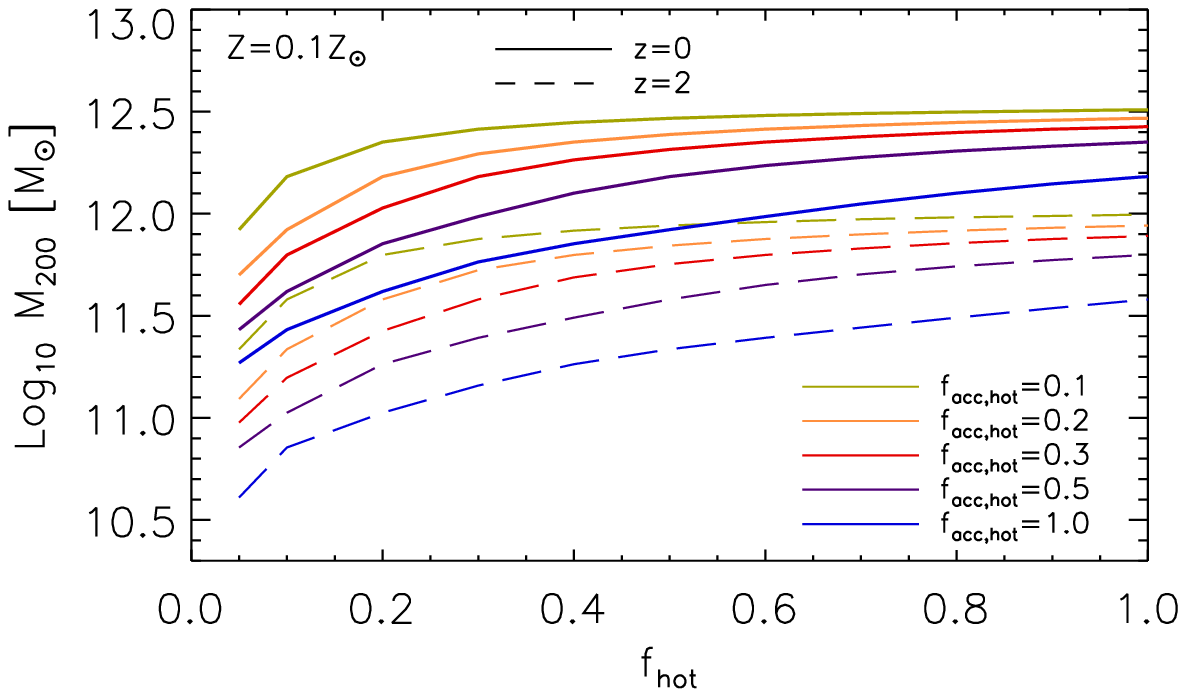}}\\
\vspace{-0.2cm}
\subfloat{\includegraphics[angle=0,width=0.45\textwidth]{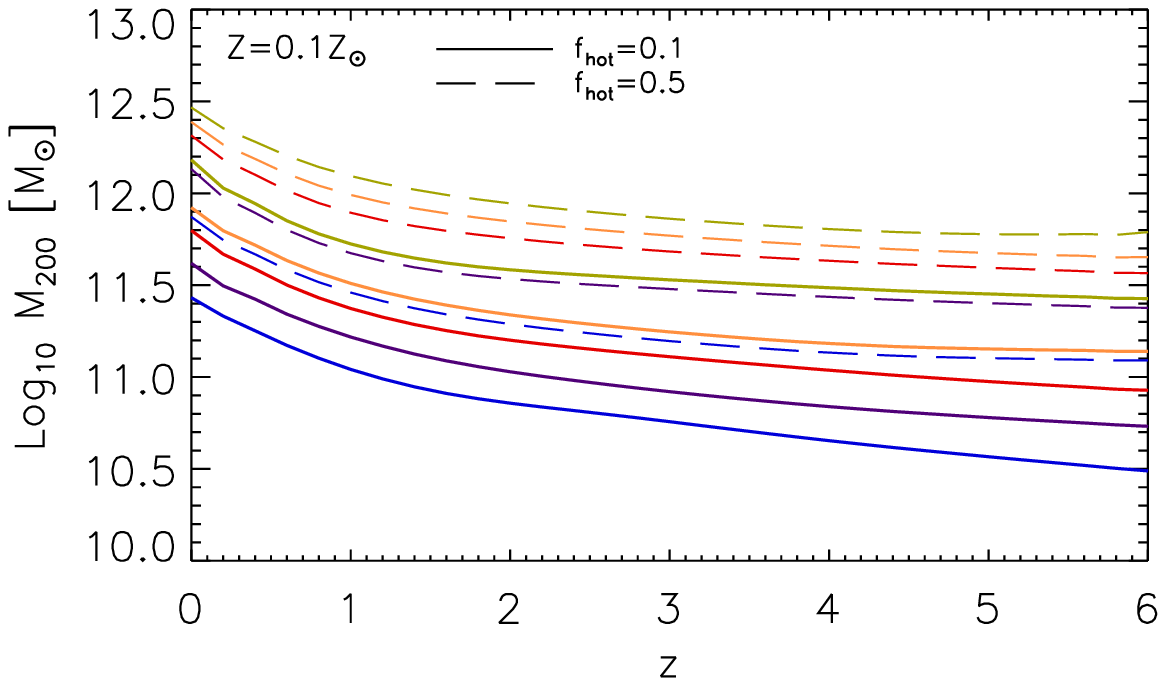}}
\vspace{-0.1cm}
\caption{Top panel: Halo mass obtained by equating $\Gamma_{\rm{heat}}$ and $\Gamma_{\rm{cool}}$ for redshift $z=0$ (solid lines) and $z=2$ (dashed lines) as a function of $f_{\rm{hot}}=M_{\rm{hot}}/(\Omega_{\rm{b}}/\Omega_{\rm{m}})M_{200}$. The different color lines correspond to $M_{\rm{crit}}$ calculated assuming fixed values for the fraction of the hot mode gas accretion ($f_{\rm{acc,hot}}$, as indicated in the legends). Bottom panel: Same as the top panel, but in this case $M_{\rm{crit}}$ is calculated assuming $f_{\rm{hot}}=0.1$ (solid lines) and $f_{\rm{hot}}=0.5$ (dashed lines) as a function of redshift.}
\label{mcrit_limits}
\end{figure}

The top panel of Fig.~\ref{mcrit_limits} shows $M_{\rm{crit}}$ as a function of $f_{\rm{hot}}$ and $f_{\rm{acc,hot}}$ for $z=0$ (solid lines) and $z=2$ (dashed lines). From the panel it can be seen that for fixed $f_{\rm{acc,hot}}$, $M_{\rm{crit}}$ increases with increasing $f_{\rm{hot}}$. This is because as $f_{\rm{hot}}$ increases, so does $\Gamma_{\rm{cool}}$, and therefore for $\Gamma_{\rm{heat}}$($\propto M_{200}^{2/3}\dot{M}_{200}$) to be able to balance $\Gamma_{\rm{cool}}$, $M_{200}$ needs to be larger. 

For fixed $f_{\rm{hot}}$, $M_{\rm{crit}}$ increases with decreasing $f_{\rm{acc,hot}}$. In this case, when $f_{\rm{acc,hot}}$ decreases, the heating rate is able to balance the cooling rate only if the accretion rate is large ($\Gamma_{\rm{heat}}\propto \dot{M}_{200}T_{\rm{vir}}$), and since the accretion rate increases with halo mass, the halo needs to grow in mass in order to develop a heating rate large enough to keep the gas hot. The top panel of Fig.~\ref{mcrit_limits} also shows that for fixed $f_{\rm{acc,hot}}$ and $f_{\rm{hot}}$, $M_{\rm{crit}}$ at $z=2$ is lower than $M_{\rm{crit}}$ at $z=0$. This can also be explained in terms of the halo accretion rate. If $f_{\rm{acc,hot}}$ and $f_{\rm{hot}}$ do not change, the heating rate for fixed halo mass still increases because $\dot{M}_{200}$ and $T_{\rm{vir}}$ increase with increasing redshift. As a result, lower mass haloes are able to produce a heating rate that balances the gas cooling rate.


It is challenging to calculate analytically the hot gas mass in the halo, and the fraction of gas accreted hot, as a function of halo mass and redshift. For that reason, we follow our analysis from Section 3 and make the ansatz that a halo develops a hot atmosphere when the hot gas mass is $10\%$ (which is roughly the hot fraction in haloes with masses between $10^{11}-10^{12}\Msun$ in the redshift range 0-6, see bottom panel from Fig.~\ref{Mhot_plot2}). In the case of the fraction of hot gas accretion, it is known that for fixed halo mass, $f_{\rm{acc,hot}}$ is large at low redshift, and it decreases with increasing redshift due to the presence of cold gas accretion from filaments (as shown in Section 4.2). We assume that at $z=0$ $f_{\rm{acc,hot}}\sim 0.5-1$, and obtain that the mass-scale of hot halo formation is between $10^{11.4}-10^{11.7}\Msun$. This is in agreement with the analysis from Section 3 where, by visual inspection of the gas cooling time PDF, we concluded that the hot halo forms between $10^{11.5}-10^{12}\Msun$. 

We next analyse how $M_{\rm{crit}}$ changes with redshift. The bottom panel of Fig.~\ref{mcrit_limits} shows $M_{\rm{crit}}$ as a function of redshift for $f_{\rm{hot}}=0.1$ (solid lines) and $f_{\rm{hot}}=0.5$ (dashed lines). The different color lines correspond to $M_{\rm{crit}}$ calculated assuming fixed values for $f_{\rm{acc,hot}}$ (as indicated in the legend in the top panel). It can be seen from the panel that, for any redshift, the higher $f_{\rm{hot}}$ and lower $f_{\rm{acc,hot}}$, the higher $M_{\rm{crit}}$. It can also be seen that for any $f_{\rm{hot}}$ and $f_{\rm{acc,hot}}$ values, $M_{\rm{crit}}$ remains roughly constant in the redshift range 6-2, and then increases. This is possibly due to the rapid drop of the accretion rate ($\dot{M}_{200}$, hence $\Gamma_{\rm{heat}}$) in the redshift range 0-1 caused by the accelerated expansion of the Universe.


\citet{Ocvirk}, along with DB06, argued that chemical enrichment has a crucial impact on shock stability, since metallicity, as well as gas density, determines the cooling rate. DB06 and \citet{Ocvirk} found that increasing the metallicity increases the critical halo mass for shock stability. We analyse the impact of metallicity on $M_{\rm{crit}}$ in Appendix A, where we show that increasing metallicity increases $M_{\rm{crit}}$, but if the hot gas metallicity is lower than $0.1Z_{\sun}$, it does not strongly impact the normalization of $M_{\rm{crit}}$. This is expected, since metal cooling only becomes important for $Z\gtrsim 0.1Z_{\sun}$ (e.g.\citealt{Wiersma09b}).

\begin{figure} 
\centering
\subfloat{\includegraphics[angle=0,width=0.45\textwidth]{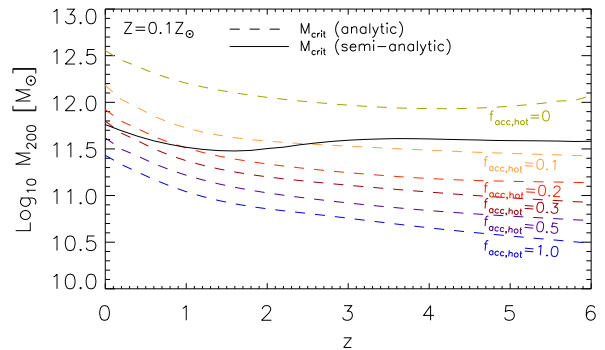}}
\vspace{-0.1cm}
\caption{Halo mass obtained by equating $\Gamma_{\rm{heat}}$ and $\Gamma_{\rm{cool}}$, for constant values of $f_{\rm{hot}}$ and $f_{\rm{acc,hot}}$ (analytic $M_{\rm{crit}}$, coloured dashed lines) and for $f_{\rm{hot}}$ and $f_{\rm{acc,hot}}$ as a function of halo mass and redshift (semi-analytic $M_{\rm{crit}}$ using eqs.~\ref{Mhot1}-\ref{Mhot4} and \ref{fhot}-\ref{Mhalf}, black solid line).}
\label{mcrit_comparison_2}
\end{figure}

\subsubsection{Semi-analytic estimate}

In the previous subsection we used the analytic model to analyse how the mass scale of hot halo formation changes with the fraction of hot gas mass in the halo, $f_{\rm{hot}}$, and the fraction of gas that shock-heats when crossing the virial radius, $f_{\rm{acc,hot}}$. Since it is challenging to derive analytically $f_{\rm{hot}}$ and $f_{\rm{acc,hot}}$ as a function of halo mass and redshift, we assumed typical values and concluded that the hot halo forms in the halo mass range $10^{11.4}-10^{11.7}\Msun$ at $z=0$, and remains roughly constant with redshift. In this section we make a `semi-analytic' estimate of $M_{\rm{crit}}$ (hereafter $M_{\rm{crit,sa}}$), by using the best-fitting relations from our simulations for $f_{\rm{hot}}(M_{200},z)$ (eqs.~\ref{Mhot1}-\ref{Mhot4}), $f_{\rm{acc,hot}}(M_{200},z)$ (eqs.~\ref{fhot}-\ref{Mhalf}). Note that $f_{\rm{acc,hot}}(M_{200},z)$ relation may under estimate the fraction of hot gas accretion for halos with virial temperatures lower than $10^{5.5}$K, see Section 4.1 for a discussion. In this section however, we do not intend to predict a mass scale for hot halo formation, but compare the result with the analysis done in the previous subsection.

Fig.~\ref{mcrit_comparison_2} shows the semi-analytic estimate of $M_{\rm{crit}}$ in black solid lines and the analytic estimates calculated in the previous subsection in coloured dashed lines. By using the best-fitting relations, we obtain a critical mass scale of $10^{11.75}\Msun$ at $z=0$ that remains roughly constant with redshift, in agreement with the analysis done in the previous section. 

We find that $f_{\rm{acc,hot}}(M_{\rm{crit,sa}})\approx 0.3$ at $z=0$ and $f_{\rm{acc,hot}}(M_{\rm{crit,sa}})< 0.3$ at $z>0$, from which we conclude that $M_{\rm{crit}}$ does not necessarily correspond to the mass scale where the hot and cold modes of accretion contribute equally ($f_{\rm{acc,hot}}=0.5$), because equal hot/cold modes of accretion do not imply the existence of a stable hot atmosphere or lack thereof. In fact, $f_{\rm{acc,hot}}(M_{\rm{crit,sa}})< 0.3$ at $z>0$ implies that massive haloes are able to develop a hot atmosphere (and hence virial shocks), even when they are accreting the majority of gas in the cold mode. 

We also find that $M_{\rm{crit,sa}}$ is somewhat lower than the analytic halo mass calculated by DB06 at $r=R_{200}$. In their work, DB06 found that the critical halo mass of shock-heating occurring in the inner halo at $r=0.1R_{200}$ is $6\times 10^{11}\Msun$ and $2\times 10^{12}\Msun$ at $r=R_{200}$, in both cases the critical halo mass was nearly constant with redshift. In our case we focus on shocks occurring at $R_{200}$ and find that $M_{\rm{crit,sa}}$ is a factor 3.5 lower and changes slightly with redshift. The small change of $M_{\rm{crit,sa}}$ with redshift is driven by the interplay between the accretion rate, $\dot{M}_{200}$, which increases with redshift, and the fraction of hot gas accretion, $f_{\rm{acc,hot}}$, which decreases with redshift, but it is also due to the fact that we assume a fixed value of $10^{0.6}\rho_{\rm{crit}}$ for the hot gas density. In Appendix \ref{comparison_Dekel} we do a more detailed comparison with the work of DB06 and in Appendix \ref{Density_Sec} we discuss how the density of the hot gas in the halo changes with redshift.





\begin{figure}
  \centering
  \includegraphics[angle=0,width=0.45\textwidth]{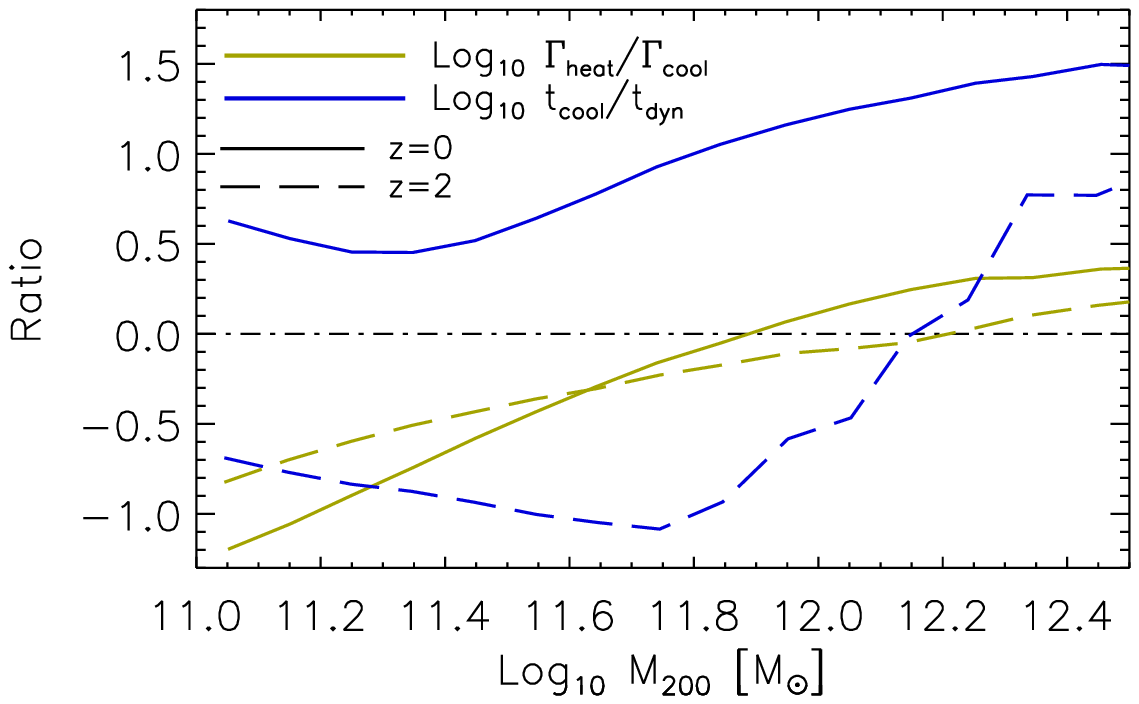}\\
  \vspace{0.1cm}
    \includegraphics[angle=0,width=0.45\textwidth]{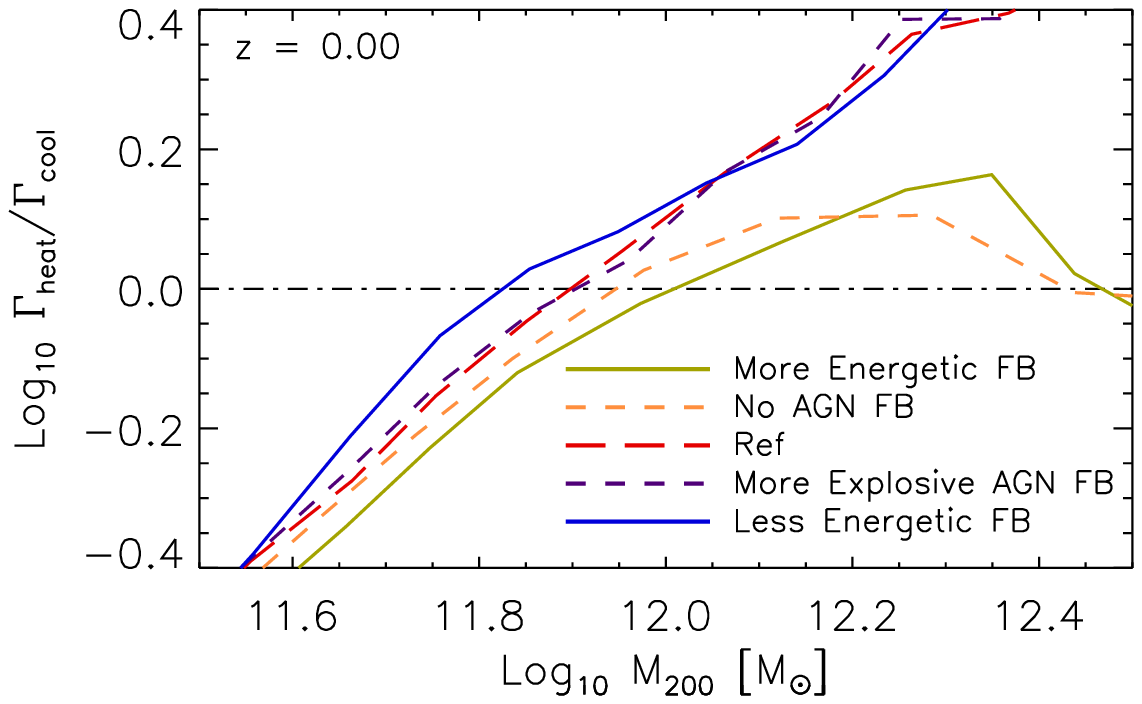}
  \caption{Top panel: median logarithmic ratio between $\Gamma_{\rm{heat}}$ and $\Gamma_{\rm{cool}}$ (green lines), and between $t_{\rm{cool}}$ and $t_{\rm{dyn}}$ (blue lines) as a function of halo mass ($M_{200}$) for gas in haloes at $z=0$ (solid lines) and $z=2$ (dashed lines). Bottom panel: median logarithmic ratio of $\Gamma_{\rm{heat}}$ and $\Gamma_{\rm{cool}}$ calculated from simulations with different feedback prescriptions as indicated in the legend.}
\label{Mcritical_check}
\end{figure}

\begin{figure*}
  \centering
  \includegraphics[angle=0,width=0.95\textwidth]{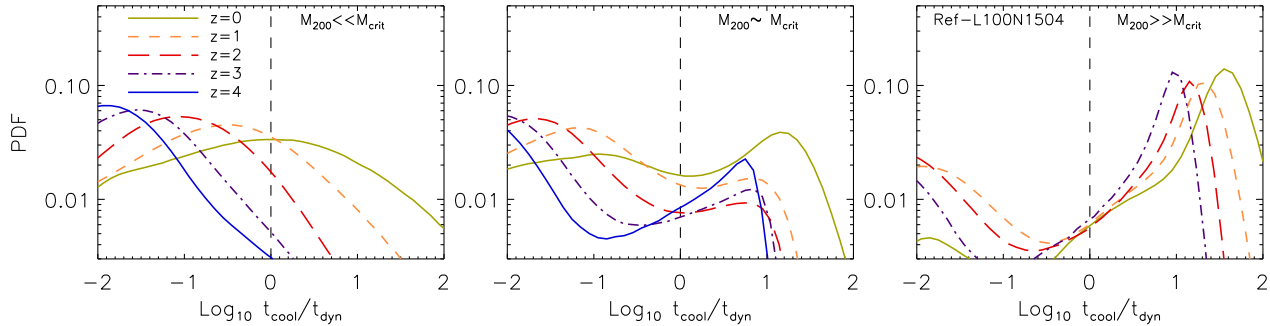}\\
    \vspace{-0.1cm}
  \caption{Mass-weighted PDF of the ratio of the cooling and local dynamical time for gas in haloes with masses $M_{200}\ll M_{\rm{crit}}$ (left panel), $M_{200}\sim M_{\rm{crit}}$ (middle panel) and $M_{200}\gg M_{\rm{crit}}$ (right panel) at $z=0$, 1, 2, 3, 4 and 5, taken from the Ref-L100N1504 simulation.}
\label{cooling_time_Mcritical}
\end{figure*}

\subsection{Comparison between $t_{\rm{cool}}/t_{\rm{dyn}}$ and $\Gamma_{\rm{heat}}/\Gamma_{\rm{cool}}$}\label{comparison2} 

In this subsection we aim to show that the analytical model gives a better prediction for $M_{\rm{crit}}$ compared to the ratio of the halo dynamical and cooling timescales. To do so, we compute $\Gamma_{\rm{heat}}$ and $\Gamma_{\rm{cool}}$, as well as $t_{\rm{cool}}$ and $t_{\rm{dyn}}$, for each halo in the simulations.

In the calculation of $\Gamma_{\rm{heat}}$, we compute the individual gas accretion rates for each halo, as well as the fraction of particles that shock-heat, and define hot gas as all particles with temperatures larger than the host halo's $T_{\rm{vir}}$ (to avoid using $t_{\rm{cool}}/t_{\rm{dyn}}$ as we do in section 3.3). In the case of $t_{\rm{cool}}$ and $t_{\rm{dyn}}$, we assume that the hot halo is formed when the cooling time at the virial radius is larger than the dynamical time, and to calculate them we only use gas between $(0.8-1)\times R_{200}$. 


The top panel of Fig.~\ref{Mcritical_check} shows the ratio between $\Gamma_{\rm{heat}}$ and $\Gamma_{\rm{cool}}$ (olive lines) and $t_{\rm{cool}}$ and $t_{\rm{dyn}}$ (blue lines). We find that the halo mass where $\Gamma_{\rm{heat}}=\Gamma_{\rm{cool}}$ at $z=0$ is in very good agreement with the semi-analytic prediction ($M_{\rm{crit,sa}}=10^{11.8}$). At $z=2$, the halo mass is larger than $M_{\rm{crit,sa}}$ due to the hot gas density being slightly lower than the fiducial value adopted in Section 5.2. We find that at $z=0$ the median gas $t_{\rm{cool}}$ is always larger than $t_{\rm{dyn}}$ in the halo mass range $10^{11}-10^{12.5}\Msun$, indicating that the hot halo should form, in contradiction with the results from Section 3. Such long cooling times at all masses are due to the presence of additional heating mechanisms (like UV/X-ray background). At $z=2$ the gas median cooling times at large radii are shorter, due to the cold, dense filamentary gas, and $t_{\rm{cool}}$ overcomes $t_{\rm{dyn}}$ only in haloes larger than $10^{12.2}\Msun$. 

We believe that $\Gamma_{\rm{heat}}=\Gamma_{\rm{cool}}$ is a better method to determine when the hot halo forms because, unlike $t_{\rm{cool}}=t_{\rm{dyn}}$, $\Gamma_{\rm{heat}}$ by definition only considers the heating due to halo growth and accretion shocks.

\subsubsection{Feedback variations}

The model presented in this section assumes that the formation of the hot halo is only driven by the heating from gravitational accretion shocks. However, in the presence of other energy sources, like stars or AGN, the heating rate should increase, and therefore extra terms (like $\Gamma_{\rm{stellar}}$ or $\Gamma_{\rm{AGN}}$) should be added to eq.~(\ref{energy}). Although we do not include extra heating sources accounting for the presence of feedback, we still find good agreement between the analytical results and the numerical analysis. Also, varying feedback may change the metallicity and the hot gas density (e.g. \citealt{Crain13}) and hence $\Gamma_{\rm{cool}}$. The sign of the effect is however difficult to predict. On the one hand, a more efficient feedback will reduce the stellar and hence the total metal mass. On the other hand, a greater fraction of the metals may reside in the hot halo gas. 

We next investigate how the critical halo mass is affected by different feedback implementations. To do so, we calculate $\Gamma_{\rm{heat}}$ and $\Gamma_{\rm{cool}}$ from simulations with feedback variations and show these in the the bottom panel of Fig.~\ref{Mcritical_check}. We find that at $z=0$, in the less energetic stellar feedback scenario $M^{'}_{\rm{crit}}$ (at which $\Gamma_{\rm{heat}}=\Gamma_{\rm{cool}}$) is $\sim 10^{11.8}\Msun$, in the more explosive AGN FB $M^{'}_{\rm{crit}}\sim 10^{11.9}\Msun$, in the no AGN FB (but moderate stellar feedback) $M^{'}_{\rm{crit}}\sim 10^{11.9}\Msun$, and in the more energetic stellar feedback case $M^{'}_{\rm{crit}}\sim 10^{12}\Msun$. This is in excellent agreement with the analytic model, which predicts that $M_{\rm{crit}}$ increases when $f_{\rm{hot}}$ increases, which occurs when stellar feedback is more energetic and/or there is no AGN feedback. The result is also in good agreement with Section 3.2, where we concluded that the hot halo formation is not as strongly affected by AGN feedback as it is by stellar feedback.

\subsection{Comparison with simulated $t_{\rm{cool}}/t_{\rm{dyn}}$}\label{comparison}

In this subsection we compare the semi-analytic predictions for $M_{\rm{crit,sa}}$ presented in the previous section with results from our simulations. To do so, we investigate the halo mass range for which haloes develop a hot atmosphere. The model predicts that haloes with masses $M_{200}>M_{\rm{crit,sa}}$ form a hot atmosphere. We test this by analysing the PDF of $t_{\rm{cool}}/t_{\rm{dyn}}$ for haloes with masses $M_{200}\ll M_{\rm{crit,sa}}$ (left panel of Fig.~\ref{cooling_time_Mcritical}), $M_{200}\sim M_{\rm{crit,sa}}$ (middle panel) and $M_{200}\gg M_{\rm{crit,sa}}$ (right panel) at $z=0-4$. We select gas particles in the halo (located between $r=[0.15-1]R_{200}$) that are not star-forming, and use the Ref-L100N1504 simulation.

The left panel of Fig. \ref{cooling_time_Mcritical} shows that in haloes with masses $M_{200}\sim 0.1M_{\rm{crit,sa}}$ most of the gas has $t_{\rm{cool}}<t_{\rm{dyn}}$ and is thus able to cool effectively and accrete onto the galactic disk. In larger haloes ($M_{200}\sim M_{\rm{crit,sa}}$, middle panel), the PDF is bimodal with a peak in either side of $t_{\rm{cool}}/t_{\rm{dyn}}=1$, indicating that a hot atmosphere has been formed in these haloes at each redshift, despite the increasing contribution of cold gas from filaments at higher redshifts. In haloes with masses $M_{200}\sim 10M_{\rm{crit,sa}}$ (right panel), the increase in shock-heated hot gas enhances the peak at $t_{\rm{cool}}\gg t_{\rm{dyn}}$ at the expense of the peak at $t_{\rm{cool}}\ll t_{\rm{dyn}}$. We conclude that the semi-analytic model for hot halo formation captures the mass-scale of hot halo formation in the simulations.

\section{Summary and conclusions}\label{conclusions_sec}

We have studied the formation of the hot hydrostatic halo, its dependence on feedback mechanisms, on the hot gas mass that is already in the halo, and on the fraction of gas accreted hot using the EAGLE suite of hydrodynamical simulations, as well as analytic calculations.

We began by analysing the PDF of the ratio of the radiative cooling time and the dynamical time for gas in the halo and found that when the hot halo is formed, it produces a strong bimodality in the PDF (Figs.~\ref{Profiles_plot1} and \ref{Profiles_plot2}, top right panels). By inspection of cooling time PDFs, we found that the mass scale for hot halo formation is $10^{11.5}-10^{12}\Msun$ at $z=0-4$ (Fig. 4). 

We found, however, that the cooling time PDF is strongly affected by the extragalactic UV/X-ray background radiation and stellar feedback. The UV/X-ray background radiation suppresses the net cooling rate of gas in the temperature range $T\sim 10^{4}-10^{5}$ K, therefore the peak of the cooling time PDF shifts towards larger cooling times as the halo's virial temperature decreases to these values (Fig. 3). 

In the case of stellar feedback, galactic winds expel gas from the galaxy into the halo, changing the distribution of the hot gas. As a result, more energetic stellar feedback increases the fraction of hot gas. We also analysed the build up of the total gas mass, $M_{\rm{gas}}$, as well as the hot gas mass, $M_{\rm{hot}}$, as haloes evolve, and concluded that stellar feedback has a large impact on the amount of gas in the halo. For example, doubling the strength of the stellar feedback increases the gas mass fraction by a factor of 1.3 in $10^{12}\Msun$ haloes relative to the Ref model, whereas halving the strength of the stellar feedback decreases the gas mass fraction by a factor of 2.5 (Fig. 7).

In the case of AGN feedback, neither the bimodality of the cooling time PDF nor the hot gas mass are strongly affected in haloes smaller than $10^{12}\Msun$ since they do not form massive black holes. However, the PDFs and hot gas mass do change at higher halo masses and in a manner opposite from that of stellar feedback. While efficient stellar feedback increases the gas mass in the halo, more explosive AGN feedback decreases it. For example, with strong (without) AGN feedback the total gas mass in the halo decreases (increases) by a factor of 1.5 (Fig. 7). The effect on $M_{\rm{hot}}$ is slightly different, efficient stellar feedback increases the hot mass fraction by $10\%$ relative to Ref, but no AGN feedback decreases it by $8\%$ in $10^{12}\Msun$ haloes. In the case of less energetic stellar feedback and more explosive AGN feedback, the ratio $M_{\rm{hot}}/M_{\rm{gas}}$ increases by $10\%$ and $3\%$ (on average) with respect to Ref, respectively, in the halo mass range $10^{11.5}-10^{12}\Msun$.

In addition to the hot gas in the halo, we calculated the fraction of gas accretion occurring in the hot mode (Fig.~\ref{HotFraction_plot_2}), $f_{\rm{acc,hot}}$. Rather than using a lower limit on the maximum past temperature to select shock-heated gas particles, as done in most previous works, we used the gas temperature after accretion, $T_{\rm{gas}}$, which yields lower $f_{\rm{acc,hot}}$ values than the maximum past temperature. We believe that a lower limit on $T_{\rm{gas}}$ is a better method to select hot gas accretion, because it excludes the gas that goes through a shock but cools immediately afterwards or that has not passed through an accretion shock but was heated in the past by stellar feedback and has since cooled, and therefore does not contribute to the formation of a hot halo.

We derived an analytic model for hot halo formation that depends on the fraction of hot gas mass that is already in the halo, $f_{\rm{hot}}$, as well as on the accretion rates and the fraction of gas accretion occurring in the hot mode, $f_{\rm{acc,hot}}$. We assumed that a hot halo develops when the heating rate from accretion shocks balances the radiative cooling rate. We computed $M_{\rm{crit}}$, the critical mass scale above which the hot halo forms, as a function of $f_{\rm{hot}}$ and $f_{\rm{acc,hot}}$. We found that $M_{\rm{crit}}$ increases with increasing $f_{\rm{hot}}$ and decreasing $f_{\rm{acc,hot}}$ (Fig.~\ref{mcrit_limits}). The analytic model yields a mass estimate of $M_{\rm{crit}} \approx 10^{12}-10^{11.5}\Msun$ at $z=0$, which agrees with the simulation results. 

Because estimating $f_{\rm{hot}}$ and $f_{\rm{acc,hot}}$ analytically is very challenging, we combined the analytic model with fits to the hot gas mass, and hot mode accretion rates as a function of mass and redshift. We computed a semi-analytic critical mass, $M_{\rm{crit,sa}}$, and found that $M_{\rm{crit,sa}}=10^{11.75}\Msun$ at $z=0$. At higher redshift $M_{\rm{crit,sa}}$ remains roughly constant (Fig.~\ref{mcrit_comparison_2}). We tested the $M_{\rm{crit,sa}}$ values by inspecting the cooling time PDF of hot gas in haloes with mass $M_{\rm{crit,sa}}$ and confirmed that at all redshifts, the PDF has a clear bimodal shape (Fig.~\ref{cooling_time_Mcritical}). Note that because the semi-analytic model uses our simulation results as input, its prediction for the mass scale for hot halo formation cannot be tested using the same simulations. We can, however, use it to compare with the analytic analysis.

We compared the ratio of the heating due to accretion ($\Gamma_{\rm{heat}}$, derived in the analytic model) and the radiative cooling ($\Gamma_{\rm{cool}}$) rates of hot gas, with the ratio of the cooling ($t_{\rm{cool}}$) and dynamical ($t_{\rm{dyn}}$) times of gas at the virial radius, and found that unlike $\Gamma_{\rm{heat}}/\Gamma_{\rm{cool}}$, the median $t_{\rm{cool}}/t_{\rm{dyn}}$ of gas is always greater than unity in the halo mass range $10^{11}-10^{12.5}\Msun$ (Fig. 14). On the contrary, $\Gamma_{\rm{heat}}/\Gamma_{\rm{cool}}$ is only greater than unity for haloes more massive than $10^{11.8}\Msun$, indicating that this ratio better captures the heating due to halo growth and accretion shocks. We believe that compared with the ratio of cooling and dynamical times, the analytic model of hot halo formation is better at indicating when a hot halo forms.

Finally, we investigated how feedback impacts the hot halo formation mass scale. We calculated $\Gamma_{\rm{heat}}/\Gamma_{\rm{cool}}$ using simulations with different feedback prescriptions, and concluded that while a hot hydrostatic atmosphere forms in more (less) massive haloes in scenarios with more (less) energetic stellar feedback, the mass scale of hot halo formation is not strongly affected by AGN feedback. This result is driven by the dependence of $\Gamma_{\rm{heat}}$ and $\Gamma_{\rm{cool}}$ on the hot gas mass fraction, $f_{\rm{hot}}$. When $f_{\rm{hot}}$ increases (i.e. in  the strong stellar feedback scenario), so does the rate of cooling and therefore the halo needs to grow in mass in order to develop a heating rate that overcomes  the cooling rate.

Cosmological hydrodynamical simulations have shown that the manner in which galaxies accrete gas depends on the complex interaction between the hot halo, AGN and stellar feedback (see e.g. \citealt{vandeVoort11,Faucher,Nelson15a}). In Paper II we will make use of the semi-analytic calculations presented in this work and derive a semi-analytic model for gas accretion onto galaxies that accounts for the hot/cold modes of gas accretion onto haloes and for the rate of gas cooling from the hot halo. By doing so, we aim to provide some insight into the physical mechanisms that drive the gas inflow rates onto galaxies.

\section*{Acknowledgments}
We are grateful to the referee for fruitful comments that substantially improved the original manuscript. This work used the DiRAC Data Centric system at Durham University, operated by the Institute for Computational Cosmology on behalf of the STFC DiRAC HPC Facility (www.dirac.ac.uk). This equipment was funded by BIS National E-infrastructure capital grant ST/K00042X/1, STFC capital grant ST/H008519/1, and STFC DiRAC Operations grant ST/K003267/1 and Durham University. DiRAC is part of the National E-Infrastructure. The EAGLE simulations were performed using the DiRAC-2 facility at Durham, managed by the ICC, and the PRACE facility Curie based in France at TGCC, CEA, Bruyeres-le-Chatel. This work was supported by the European Research Council under the European Union's Seventh Framework Programme (FP7/2007-2013)/ERC Grant agreement 278594-GasAroundGalaxies and by the Netherlands Organisation for Scientific Research (NWO) through VICI grant 639.043.409. RAC is a Royal Society University Research Fellow. We also acknowledge support from STFC (ST/L00075X/1).

\bibliography{biblio}
\bibliographystyle{mn2e}

\appendix

\section{Shock analysis}\label{Analysis_PDF}

\begin{figure*} 
\centering
\subfloat{\includegraphics[angle=0,width=0.42\textwidth]{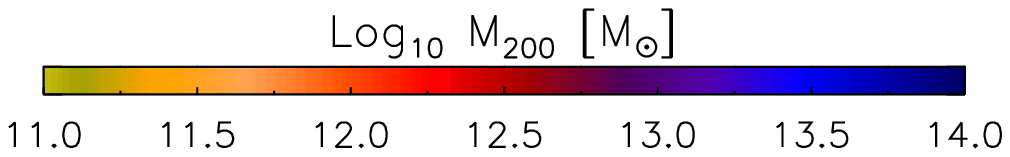}}
\subfloat{\includegraphics[angle=0,width=0.42\textwidth]{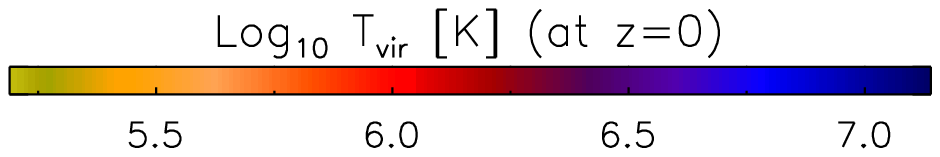}}\\
\vspace{-0.3cm}
\subfloat{\includegraphics[angle=0,width=0.42\textwidth]{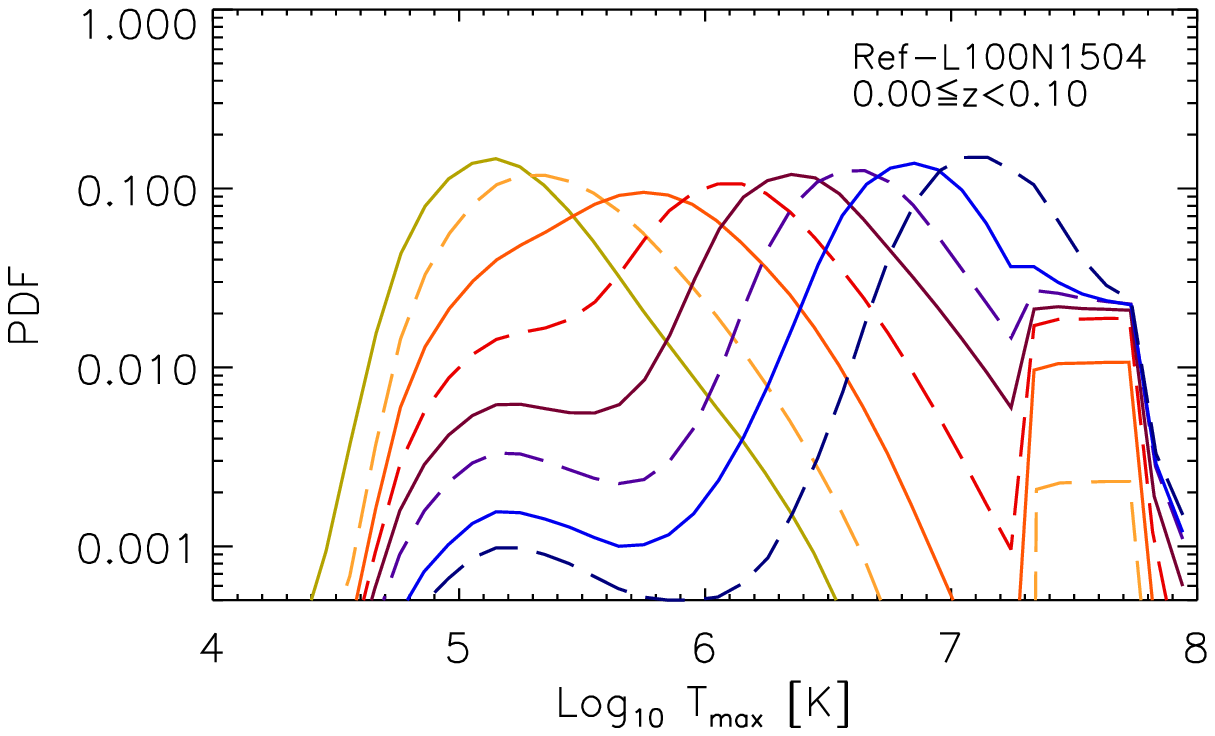}}
\subfloat{\includegraphics[angle=0,width=0.42\textwidth]{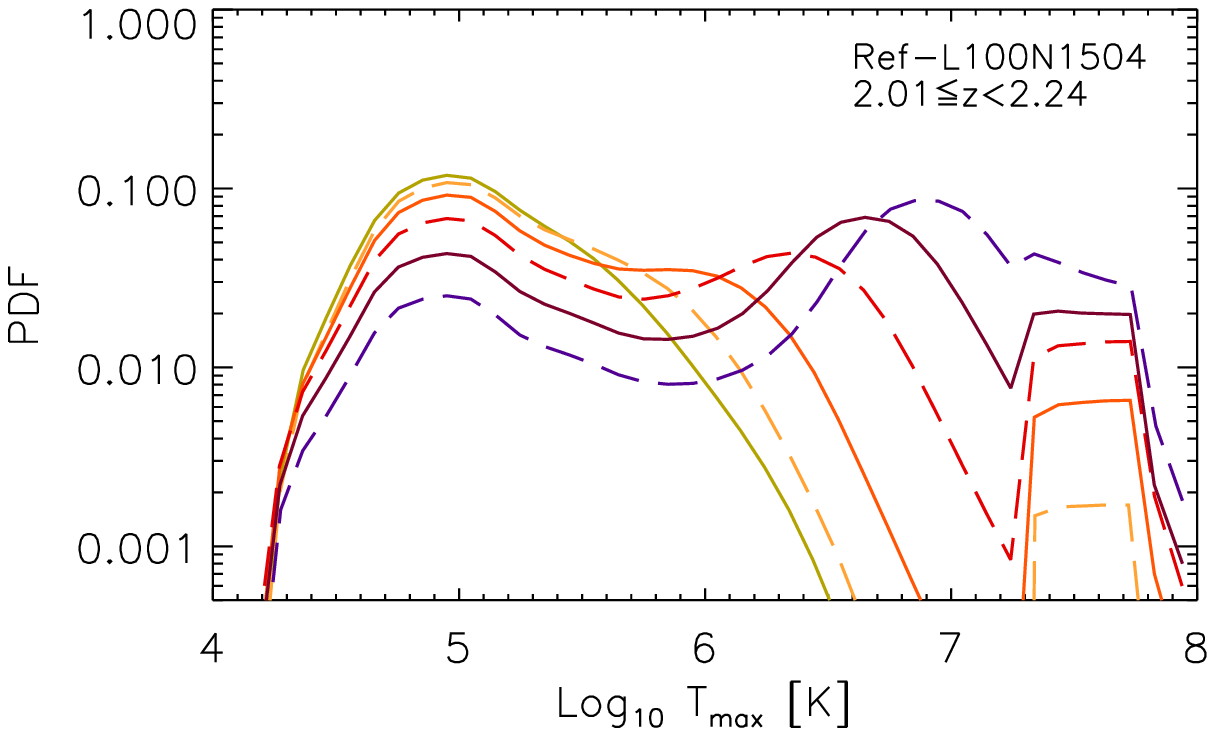}}\\
\vspace{-0.3cm}
\subfloat{\includegraphics[angle=0,width=0.42\textwidth]{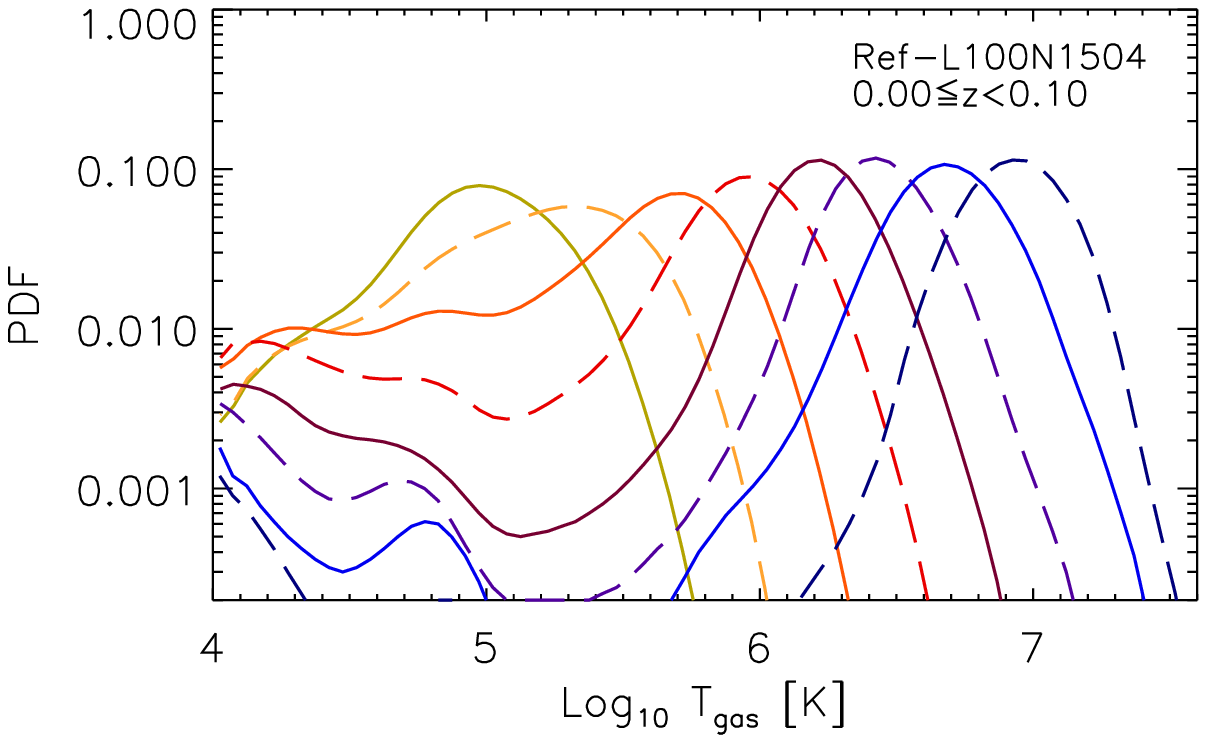}}
\subfloat{\includegraphics[angle=0,width=0.42\textwidth]{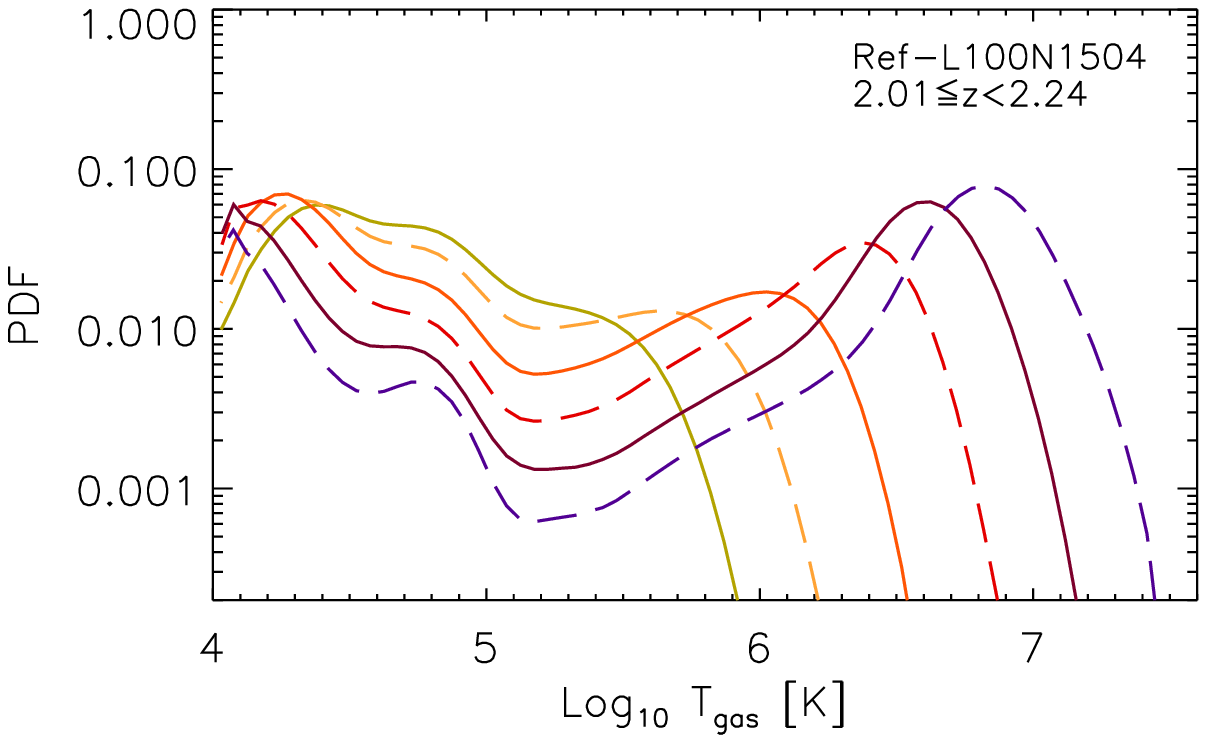}}\\
\vspace{-0.3cm}
\subfloat{\includegraphics[angle=0,width=0.42\textwidth]{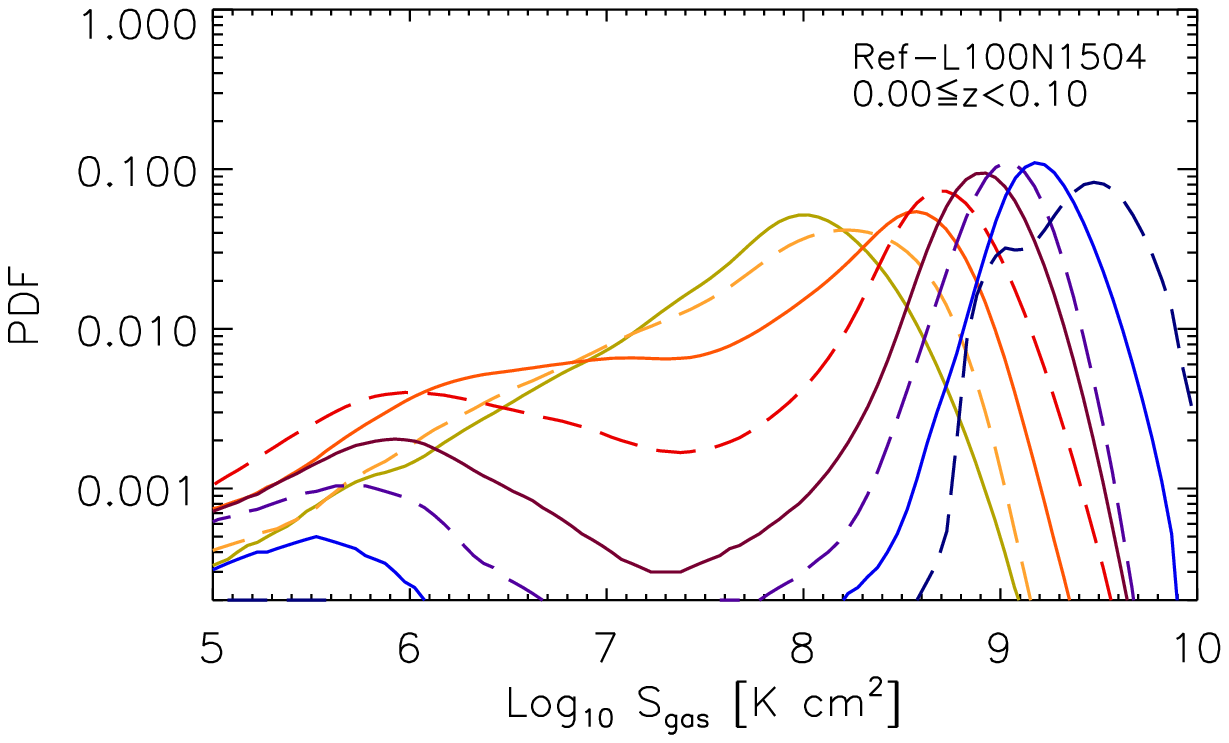}}
\subfloat{\includegraphics[angle=0,width=0.42\textwidth]{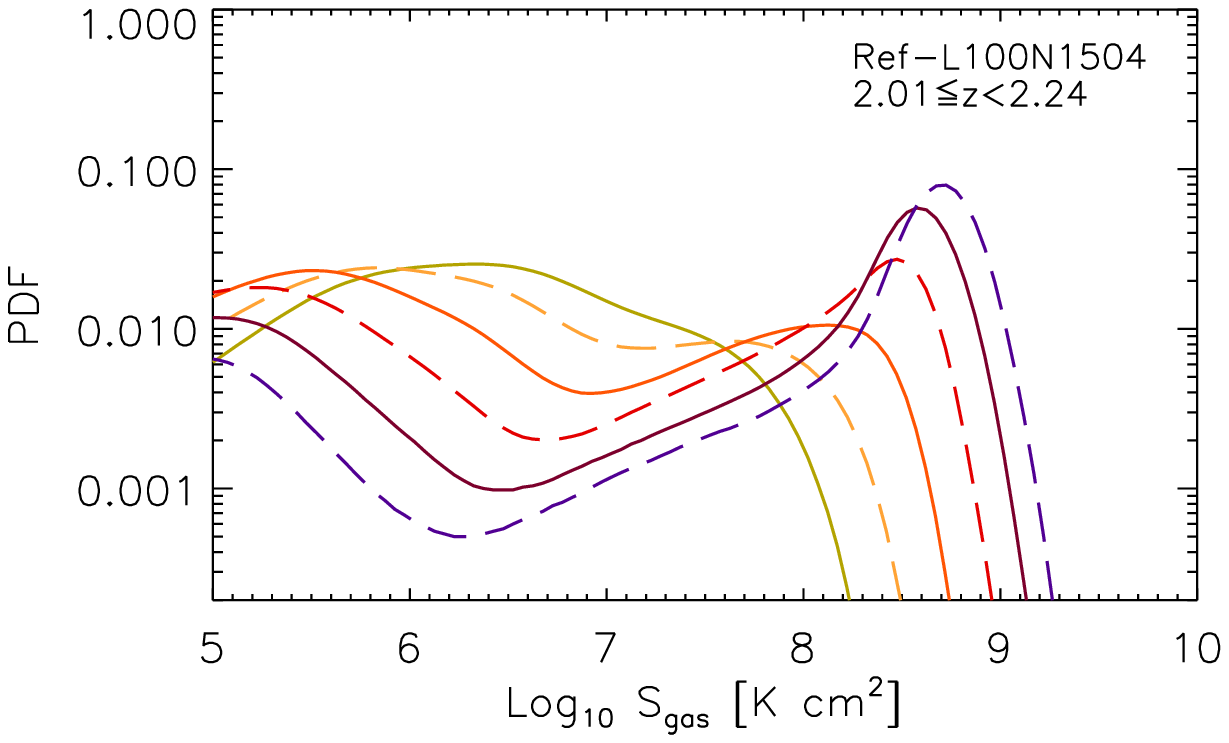}}
\caption{Probability density function of maximum past temperature (top panels), temperature (middle panels) and entropy (bottom panels) of gas accreted on to haloes in the redshift ranges $z=0-0.1$ (left panels) and $z=2.0-2.2$ (right panels). The curves are colored according to the color bars at the top of the figure, which indicate the halo mass and virial temperature.}
\label{Max_temp_plot}
\end{figure*}

In this section we analyse the mass-weighted PDF of $T_{\rm{max}}$, $T_{\rm{gas}}$, $S_{\rm{gas}}$. Fig.~\ref{Max_temp_plot} shows the PDFs for the redshift interval $0.0<z\le 0.1$ (top left panel) and $2.0<z\le 2.2$ (top right panel). The curves are colored according to the color bars at the top of the figure, which indicate the halo mass (and virial temperature) of the halo that gas is accreted onto. It can be seen that the $T_{\rm{max}}$ PDF varies with $M_{200}$, being unimodal in low-mass haloes and bimodal in high-mass haloes. The location of the local minimum of the bimodal distribution also changes with $M_{200}$, going from $T_{\rm{max,min}}\sim 10^{5.5}\hspace{1mm}\rm{K}$ in $10^{12}\Msun$ haloes to $T_{\rm{max,min}}\sim 10^{6}\hspace{1mm}\rm{K}$ in $10^{14}\Msun$ haloes. Besides this local minimum, there is a local maximum at $10^{7.5}\hspace{1mm}\rm{K}$ for all halo masses. This peak is produced by stellar feedback instead of accretion shocks. Some of the gas that is ejected out of the halo due to stellar feedback, is eventually re-accreted. However, if it does not reach a temperature larger than $10^{7.5}\hspace{1mm}\rm{K}$ when crossing $R_{200}$, $T_{\rm{max}}$ is not updated, and the gas will be considered hot mode accretion by the maximum temperature criterion. When applying a $T_{\rm{max}}$ criterion to separate hot from cold accretion, rather than calculating a $T_{\rm{max,min}}$ threshold value that changes with $M_{200}$, we follow previous works from the literature (e.g. \citealt{vandeVoort11,Nelson13}) and use $T_{\rm{max}}=10^{5.5}\hspace{1mm}\rm{K}$.

As per $T_{\rm{max}}$, the middle and bottom panels of Fig.~\ref{Max_temp_plot}  show that the $T_{\rm{gas}}$ and $S_{\rm{gas}}$ PDFs have a bimodal shape. To identify the accreted gas that does not cool immediately after the shock, we analyze the post-shock gas temperature and entropy. We select the gas particles that were accreted during the redshift interval $z_{i}-z_{j}$ ($z_{i}<z_{j}$), and are hot using a temperature and entropy threshold value. We calculate the gas temperature and entropy mass-weighted PDFs at redshift $z_{i}$. We use the local minima of the bimodal distribution as the threshold values ($T_{\rm{min}}=10^{5.5}\hspace{1mm}\rm{K}$ and $S_{\rm{min}}=10^{7.2}\hspace{1mm}\rm{K}\hspace{1mm}\rm{cm}^{2}$) to calculate the fraction of the gas accreted hot. Our motivation for using the gas entropy (besides temperature) to identify shocked gas is based on the fact that gas generally undergoes a large entropy increase when it encounters a shock (\citealt{Brooks09}).

\section{Metallicity}\label{Metallicity_Sec}

\begin{figure*} 
  \centering
  \vspace{-0.4cm}
  \subfloat{\includegraphics[angle=0,width=0.45\textwidth]{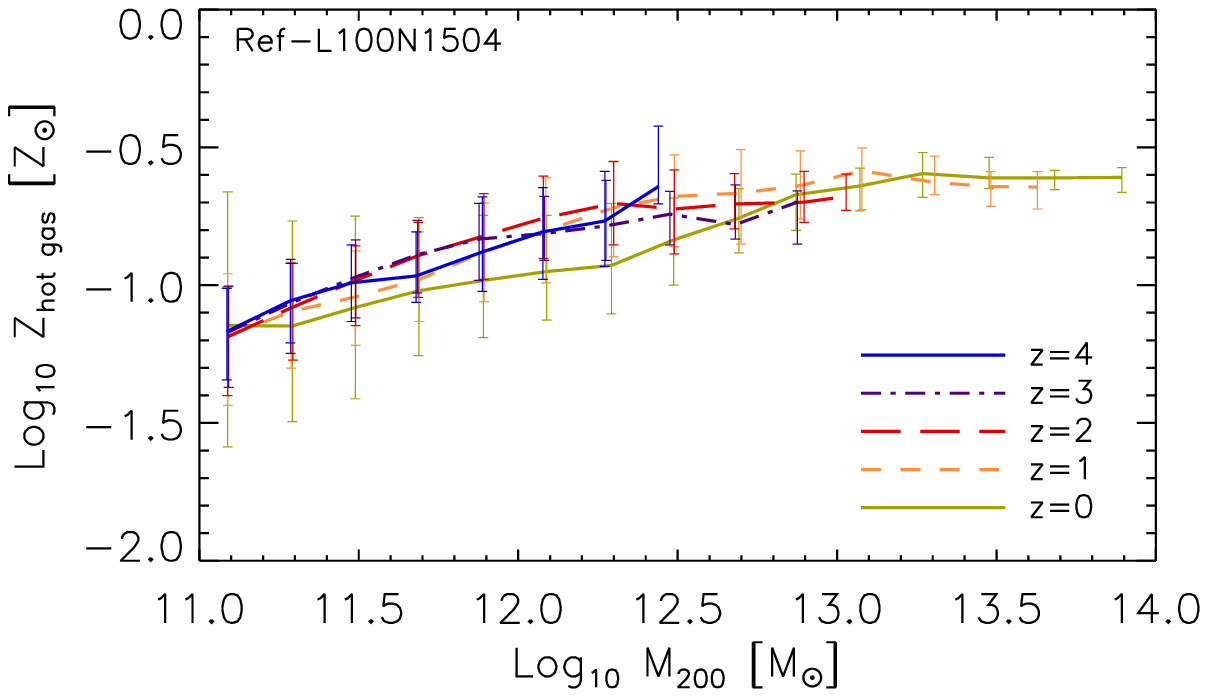}}
  \subfloat{\includegraphics[angle=0,width=0.45\textwidth]{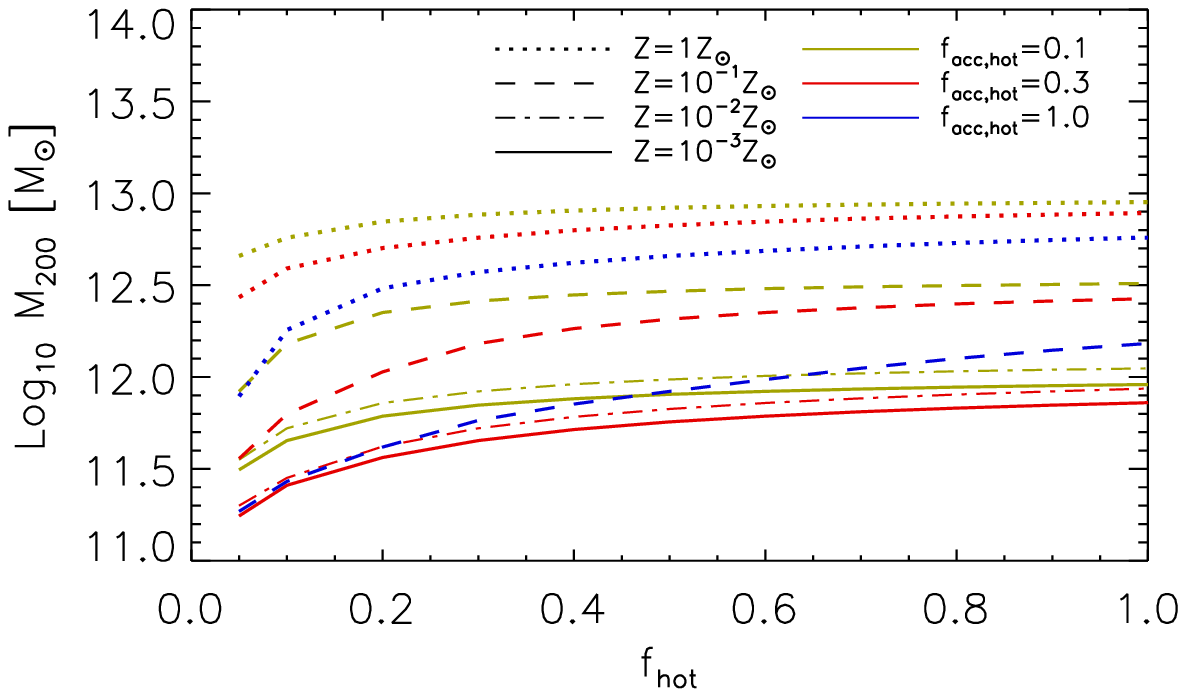}}
  \caption{Left panel: mass weighted median metallicity of the hot gas in the halo as a function of halo mass for various output redshifts, as indicated in the legends. Hot gas is defined as gas within $R_{200}$ that has a cooling time longer than the local dynamical time. The error bars in the panel correspond to the 16-84th percentiles (i.e. $1\sigma$ scatter) and each bin contains at least 5 haloes. Right panel: Halo mass obtained by equating $\Gamma_{\rm{heat}}$ and $\Gamma_{\rm{cool}}$ for redshift $z=0$ as a function of $f_{\rm{hot}}=M_{\rm{hot}}/(\Omega_{\rm{b}}/\Omega_{\rm{m}})M_{200}$. The different color lines and types correspond to $M_{\rm{crit}}$ calculated assuming fixed values for the fraction of the hot mode gas accretion ($f_{\rm{acc,hot}}$) and metallicity, respectively, as indicated in the legends.}
\label{metallicity_plot}
\end{figure*}

In Section \ref{Toymodel} we derived an analytic model for hot halo formation. The model predicts that a hot halo develops when the heating rate from accretion shocks balances the radiative cooling rate. To calculate the gas cooling rate, we assumed a constant hot gas metallicity, $Z_{\rm{hot}\hspace{0.05cm}\rm{gas}}$, of $0.1Z_{\odot}$. However, it has been argued that chemical enrichment has a crucial impact on shock stability (e.g. DB06, \citealt{Ocvirk}). Therefore in this section we investigate how the hot gas metallicity in haloes changes with halo mass and redshift, and analyse how the halo mass scale, $M_{\rm{crit}}$, at which the heating rate balances the gas cooling rate, depends on metallicity. 

Using the Ref-L100N1504 simulation, we define hot gas as all gas that is within $R_{200}$ and that has a cooling time longer than the local dynamical time, and calculate the mass weighted median metallicity per halo mass bin. The left panel of Fig.~\ref{metallicity_plot} shows the $Z_{\rm{hot}\hspace{0.05cm}\rm{gas}}-M_{200}$ relation for various output redshifts (here $Z_{\rm{hot}\hspace{0.05cm}\rm{gas}}$ is normalized by solar metallicity, which we assume to be $Z_{\odot}=0.0129$). The error bars show the 16-84th percentiles and each bin contains at least 5 haloes. Interestingly, at fixed halo mass the mass weighted median metallicity of the hot gas slightly increases with redshift (i.e. by up to a factor of 1.6 in $10^{12}\Msun$ haloes). Many studies of metal abundance have shown that galaxies tend to have lower metallicities at higher redshift (see e.g. \citealt{Prochaska,Nagamine,Savaglio,Kulkarni07,Peroux}). It is important to note that the left panel does not show the median metallicity of the ISM ($Z_{\rm{ISM}}$), but of the hot mostly ionised gas in the halo. In the case of the ISM, we obtain $Z_{\rm{ISM}}-M_{*}$ relations (with $M_{*}$ stellar mass) in agreement with the galaxy mass-metallicity relation from \citet{Andrews13} and \citet{Zahid13}, as recently shown by \citet{Somerville15}. As expected, $Z_{\rm{ISM}}$ at fixed stellar mass decreases with increasing redshift. 

We find that for all halo masses, the median mass weighted metallicity of the gas in the halo increases towards the halo centre (in agreement with \citealt{Ocvirk,vandeVoort12}). Interestingly, we find that the diffuse hot gas in the halo has lower median mass weighted metallicity (by up to 0.5 dex in haloes larger than $10^{11}\Msun$) than the cold gas. However, if we define hot gas as all gas within $R_{200}$ that has a maximum past temperature lower than $10^{5.5}$ K, we obtain that the median mass weighted metallicity of the hot gas is higher (also by to 0.5 dex in haloes larger than $10^{11}\Msun$), as in \citet{vandeVoort12}.

We next analyse how $M_{\rm{crit}}$ changes with metallicity. The right panel of Fig.~\ref{metallicity_plot} shows $M_{\rm{crit}}$ at $z=0$, calculated assuming constant values of the fraction of hot mode gas accretion, $f_{\rm{acc,hot}}$, and metallicity (as indicated in the legends), as a function of the fraction of hot gas mass in the halo, $f_{\rm{hot}}$. The panel shows that increasing metallicity increases $M_{\rm{crit}}$, but if the hot gas metallicity is lower than $10^{-1}Z_{\sun}$, it does not strongly impact on the normalization of $M_{\rm{crit}}$. This is expected, since metal cooling only becomes important for $Z\gtrsim 0.1Z_{\sun}$ (e.g.\citealt{Wiersma09b}). The left panel of Fig.~\ref{metallicity_plot} shows that the typical metallicity reached by hot gas in the redshift range 0-4 is $\sim 0.1Z_{\odot}$ in $\sim 10^{12}\Msun$ haloes, we then find that assuming $Z_{\rm{hot}\hspace{0.05cm}\rm{gas}}\sim 0.1Z_{\odot}$ in the analysis of Section 5.3 is a good approximation.

\section{Density}\label{Density_Sec}

\begin{figure*} 
  \centering
  \vspace{-0.4cm}
  \subfloat{\includegraphics[angle=0,width=0.35\textwidth]{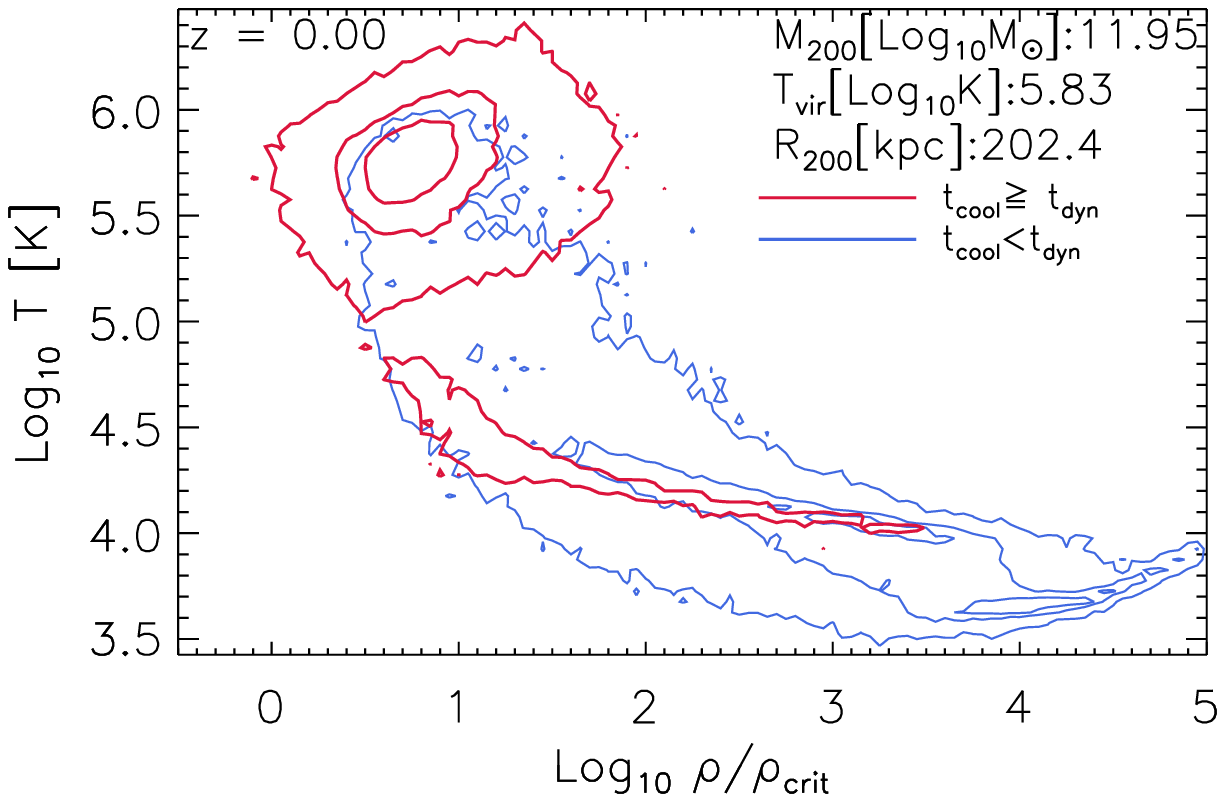}}
  \hspace{1cm}
  \subfloat{\includegraphics[angle=0,width=0.35\textwidth]{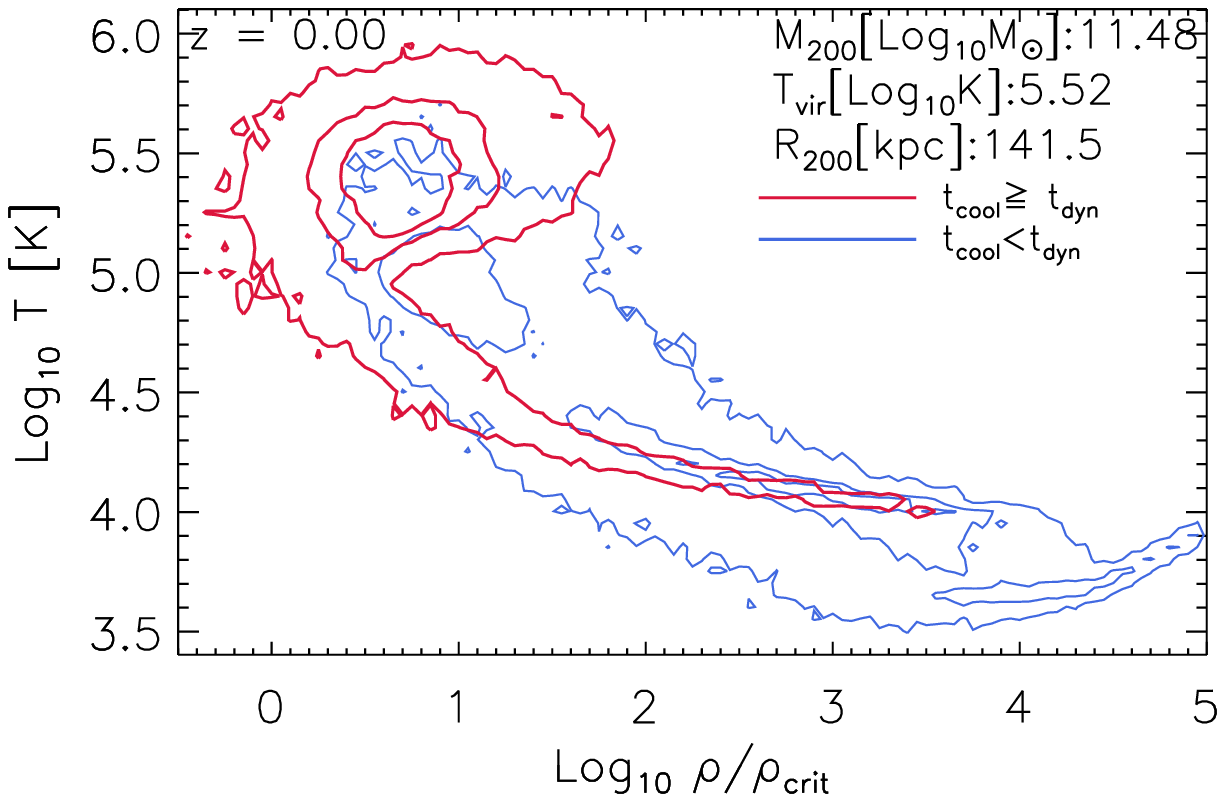}}\\
  \subfloat{\includegraphics[angle=0,width=0.35\textwidth]{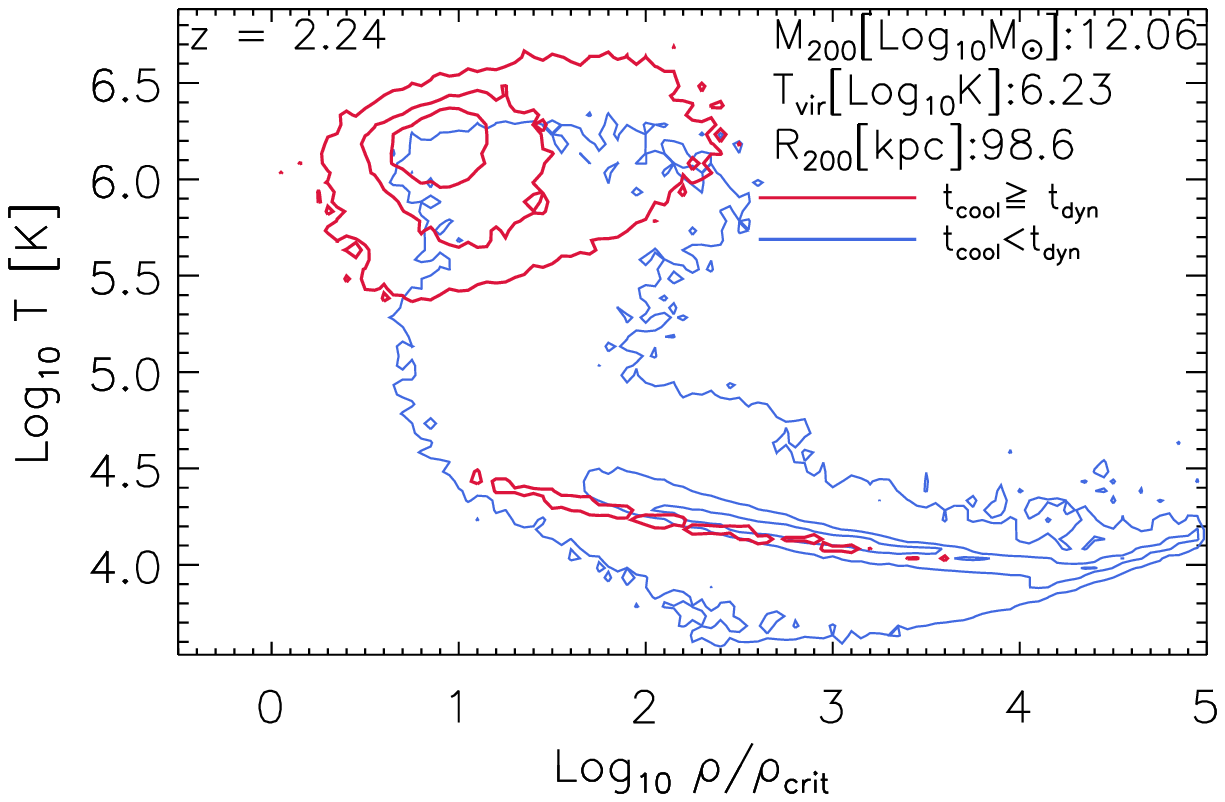}}
  \hspace{1cm}
  \subfloat{\includegraphics[angle=0,width=0.35\textwidth]{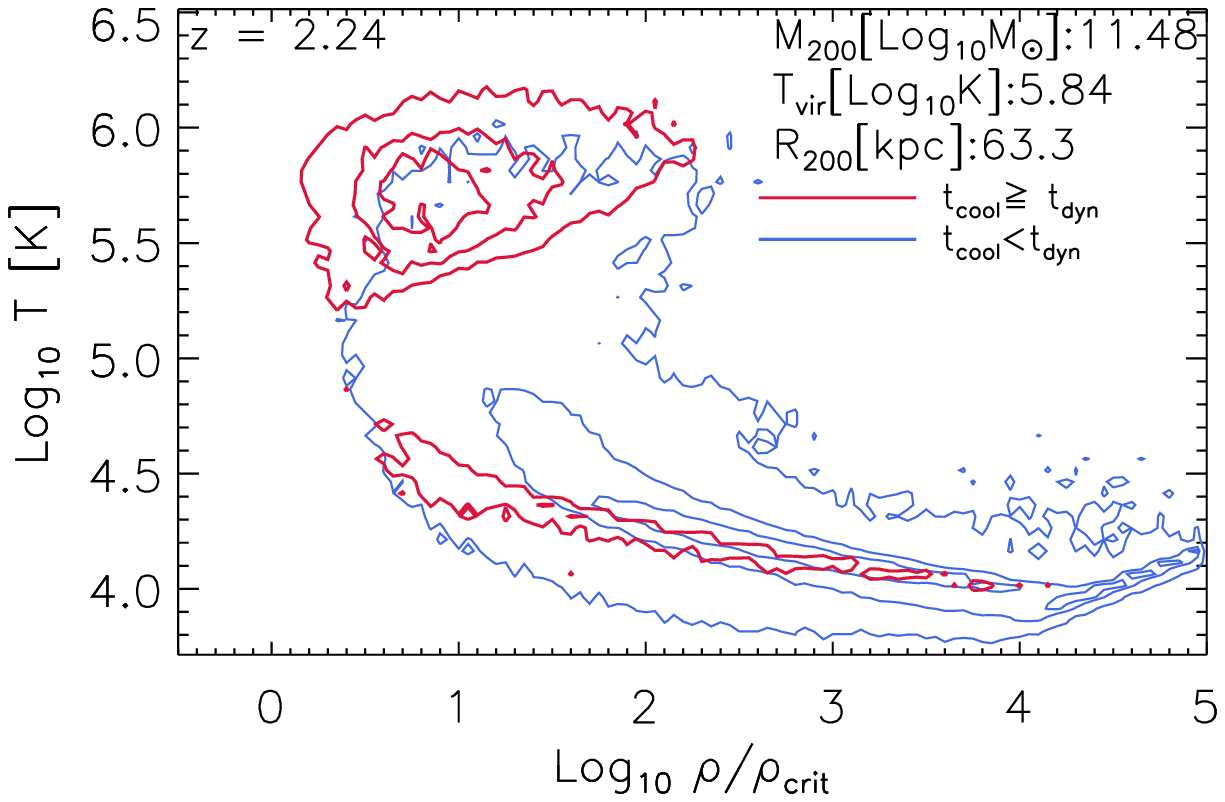}}\\
  \caption{Temperature vs. density of gas in $10^{12}\Msun$ (top-left panel) and $10^{11.5}\Msun$ haloes (top-right panel) at $z=0$, and in $10^{12}\Msun$ (bottom-left panel) and $10^{11.5}\Msun$ haloes (bottom-right panel) at $z=2.2$. The contours in the panels enclose 25, 50 and 75$\%$ of the distribution and the different colors correspond to the distribution of hot gas (i.e. with $t_{\rm{cool}}\ge t_{\rm{dyn}}$, red contours) and cold gas (blue contours). Note that the cooling time is computed from the net cooling rate, i.e. the difference between the photoheating and radiative cooling rate, and depends on metallicity.}
\label{density_plot}
\end{figure*}

In Section \ref{Toymodel} we derived an analytic model for hot halo formation, which considers the heating and cooling rates of gas in the halo. When calculating the cooling rates, we assumed that the hot gas density is about $10^{0.6}\rho_{\rm{crit}}$. In this section we analyse the density and temperature of the hot gas in $10^{11.5}-10^{12}\Msun$ haloes. 

Fig.~\ref{density_plot} shows the temperature vs. density of gas in $10^{12}\Msun$ (top-left panel) and $10^{11.5}\Msun$ haloes (top-right panel) at $z=0$, and in $10^{12}\Msun$ (bottom-left panel) and $10^{11.5}\Msun$ haloes (bottom-right panel) at $z=2.2$. The red contours indicate the distribution of the hot gas (i.e. with $t_{\rm{cool}}\ge t_{\rm{dyn}}$) while the blue contours indicate the cold gas (i.e. $t_{\rm{cool}}<t_{\rm{dyn}}$). It is interesting to see that there is gas with temperatures below $10^{5}$ K (and a large spread in density) that has a net cooling time longer than the local dynamical time. This gas is close to the equilibirum temperature obtained when photoheating balances radiative cooling rate. We define the hot halo gas as all gas with densities below $10^{2}\rho_{\rm{crit}}$ and temperature above $10^{5}$ K. At $z=0$, we find that the mean density of the hot gas in the halo is around $10^{0.6}\rho_{\rm{crit}}$ in both $10^{12}\Msun$ and $10^{11.5}\Msun$ haloes. At $z=2.2$ the mean density increases slightly to $10^{0.8}\rho_{\rm{crit}}$ for both halo masses. We find that the hot gas density does not change significantly with the host halo mass, nor with redshift (we also did the analysis for $z=3$ and 4), we then assume that the mean density of the hot gas is $10^{0.6}\rho_{\rm{crit}}$ and use this value to calculate of the mass scale for the hot halo formation.

\section{Comparison with the Dekel \& Birnboim model}\label{comparison_Dekel}

DB06 derived a post-shock stability criterion based on the interplay between the cooling time and the compression time. In their derivation, DB06 began by defining the adiabatic index

\begin{equation}\label{Dekel1}
\gamma_{\rm{eff}}\equiv \gamma-\rho q/(\dot{\rho}\mathscr{E}),
\end{equation}

\noindent which they rewrote in terms of the compression time, defined as $t_{\rm{comp}}\equiv \Gamma\rho/\dot{\rho}$, with $\Gamma=(3\gamma+2)/[\gamma(3\gamma-4)]$ and $\rho=N/V$, and the cooling time, $t_{\rm{cool}}=\mathscr{E}/q$ with $q$ the cooling rate, as follows

\begin{equation}\label{Dekel2}
\gamma_{\rm{eff}}=\gamma-\Gamma^{-1}t_{\rm{comp}}/t_{\rm{cool}}.
\end{equation}

\noindent They found that the shock is stable if $\gamma_{\rm{eff}}>\gamma_{\rm{crit}}=2\gamma/(\gamma+2/3)$, which is equivalent to $t_{\rm{cool}}>t_{\rm{comp}}$. Once the cooling time is larger, the pressure gained by compression can balance the loss by radiative cooling, and thus support the shock. In their calculation, $t_{\rm{comp}}\propto \frac{r_{\rm{s}}}{u}\propto \frac{R_{200}}{V_{\rm{vir}}}$, with $r_{\rm{s}}\approx R_{200}$, the radius where the spherical shock occurs, and $u\propto V_{\rm{vir}}$, the post-shock radial velocity. Thus obtaining that $t_{\rm{comp}}$ is comparable to the Hubble time at the corresponding epoch (but at inner radii becomes significantly shorter).

We compare our condition for hot halo formation with that of DB06. We begin by writing eq. (\ref{energy}) in terms of $t_{\rm{cool}}$ and $t_{\rm{heat}}$.

\begin{eqnarray}\label{energy2}
\frac{\dot{\mathscr{E}}}{\mathscr{E}} &=& \frac{\Gamma_{\rm{heat}}}{\mathscr{E}}-\frac{\Gamma_{\rm{cool}}}{\mathscr{E}},\\
&=& \frac{\frac{{\rm{d}}}{{\rm{d}}t}(\frac{3}{2}k_{\rm{B}}TN_{\rm{hot}})}{\frac{3}{2}k_{\rm{B}}TN_{\rm{hot}}}-\frac{M_{\rm{hot}}\Lambda/\rho_{\rm{hot}}}{\frac{3}{2}k_{\rm{B}}TN_{\rm{hot}}},\\
&=& \frac{\dot{M}_{200}}{M_{200}}(2/3+f_{\rm{acc,hot}}/f_{\rm{hot}})-\frac{\Lambda}{\frac{3}{2}n_{\rm{hot}}k_{\rm{B}}T},\\
&=& t_{\rm{heat}}^{-1}-t_{\rm{cool}}^{-1},
\end{eqnarray}

\noindent where in eq. (\ref{energy2}) we divided by $\mathscr{E}=\frac{3}{2}kT_{\rm{vir}}N_{\rm{hot}}$.

We find that our model is not equivalent to that of DB06 due to the different redshift dependence of $t_{\rm{comp}}$ and $t_{\rm{heat}}$. While $t^{-1}_{\rm{comp}}\propto [\Omega_{\rm{m}}(1+z)^{3}+\Omega_{\Lambda}]^{1/2}$, $t^{-1}_{\rm{heat}}\propto (1+z)[\Omega_{\rm{m}}(1+z)^{3}+\Omega_{\Lambda}]^{1/2}$. However, we obtain similar results at all redshifts when comparing to DB06's critical mass for the shock at $0.1R_{200}$ ($6\times 10^{11}\Msun$). We believe that we have improved upon the DB06 model by introducing a dependence on the amount of shock-heated gas, which we find to decrease with increasing redshift at fixed halo mass due to the presence of cold filaments (see panels in Fig.~\ref{HotFraction_plot_1}). To include the impact of cold filaments in their calculations, DB06 had to modified the gas density and assume $\rho_{\rm{stream}}/\rho_{\rm{vir}}\sim (3M_{*}/M)^{-2/3}$ (with $\rho_{\rm{stream}}$ the filamentary gas density, $\rho_{\rm{vir}}$ the gas density at the virial radius and $M_{*}$ the non-linear clustering mass scale). By doing so they obtained an upper limit for cold streams which increases with increasing redshift, in agreement with our results, but reaches $10^{14}\Msun$ at $z=3.5$.





\end{document}